\documentclass[pdflatex,sn-mathphys-num]{sn-jnl}

\usepackage{times}
\usepackage{siunitx}
\usepackage{graphicx}%
\usepackage{lscape}
\usepackage{booktabs}%
\usepackage{multirow}%

\usepackage[title]{appendix}%

\usepackage{amsthm}%
\usepackage{mathrsfs}%
\usepackage{xcolor}%
\usepackage{textcomp}%
\usepackage{manyfoot}%

\usepackage[T1]{fontenc}

\usepackage{algorithm}%
\usepackage{algorithmicx}%
\usepackage{algpseudocode}%
\usepackage{listings}%
\usepackage[version=4]{mhchem}
\usepackage{amsmath,amssymb,amsfonts}%

\theoremstyle{thmstyleone}%
\theoremstyle{thmstyletwo}%

\theoremstyle{thmstylethree}%

\begin{document}

\title[Detections of interstellar 2-cyanopyrene and 4-cyanopyrene]{Detections of interstellar 2-cyanopyrene and 4-cyanopyrene in TMC-1}



\author*[1]{\fnm{Gabi} \sur{Wenzel}}\email{gwenzel@mit.edu}

\author[2]{\fnm{Thomas H.} \sur{Speak}}

\author[3]{\fnm{P. Bryan} \sur{Changala}}

\author[2]{\fnm{Reace H. J.} \sur{Willis}}

\author[4]{\fnm{Andrew M.} \sur{Burkhardt}}

\author[1]{\fnm{Shuo} \sur{Zhang}}

\author[5]{\fnm{Edwin A.} \sur{Bergin}}

\author[1]{\fnm{Alex N.} \sur{Byrne}}

\author[6]{\fnm{Steven B.} \sur{Charnley}}

\author[1]{\fnm{Zachary T. P.} \sur{Fried}}

\author[3,7]{\fnm{Harshal} \sur{Gupta}}

\author[8]{\fnm{Eric} \sur{Herbst}}

\author[1]{\fnm{Martin S.} \sur{Holdren}}

\author[9]{\fnm{Andrew} \sur{Lipnicky}}

\author[9]{\fnm{Ryan A.} \sur{Loomis}}

\author[10]{\fnm{Christopher N.} \sur{Shingledecker}}

\author[1]{\fnm{Ci} \sur{Xue}}

\author[9]{\fnm{Anthony J.} \sur{Remijan}}

\author[1]{\fnm{Alison E.} \sur{Wendlandt}}

\author[3]{\fnm{Michael C.} \sur{McCarthy}}

\author*[2]{\fnm{Ilsa R.} \sur{Cooke}}\email{icooke@chem.ubc.ca}

\author*[1,9]{\fnm{Brett A.} \sur{McGuire}}\email{brettmc@mit.edu}

\affil*[1]{\orgdiv{Department of Chemistry}, \orgname{Massachusetts Institute of Technology}, \orgaddress{
\city{Cambridge}, \postcode{02139}, \state{MA}, \country{USA}}}

\affil[2]{\orgdiv{Department of Chemistry}, \orgname{University of British Columbia}, \orgaddress{
\city{Vancouver}, \postcode{V6T 1Z1}, \state{BC}, \country{Canada}}}

\affil[3]{
\orgname{Center for Astrophysics | Harvard \& Smithsonian}, \orgaddress{
\city{Cambridge}, \postcode{02138}, \state{MA}, \country{USA}}}

\affil[4]{\orgdiv{Department of Earth, Environment, and Physics}, \orgname{Worcester State University}, \orgaddress{
\city{Worcester}, \state{MA} \postcode{01602}, \country{USA}}}

\affil[5]{\orgdiv{Department of Astronomy}, \orgname{University of Michigan}, \orgaddress{
\city{Ann Arbor}, \postcode{48109}, \state{MI}, \country{USA}}}

\affil[6]{\orgdiv{Astrochemistry Laboratory}, \orgname{NASA Goddard Space Flight Center}, \orgaddress{
\city{Greenbelt}, \postcode{20771}, \state{MD}, \country{USA}}}

\affil[7]{\orgdiv{Division of Astronomical Sciences}, \orgname{National Science Foundation}, \orgaddress{
\city{Alexandria}, \postcode{22314}, \state{VA}, \country{USA}}}

\affil[8]{\orgdiv{Departments of Chemistry and Astronomy}, \orgname{University of Virginia}, \orgaddress{
\city{Charlottesville}, \postcode{22904}, \state{VA}, \country{USA}}}

\affil[9]{
\orgname{National Radio Astronomy Observatory}, \orgaddress{
\city{Charlottesville}, \postcode{22903}, \state{VA}, \country{USA}}}

\affil[10]{\orgdiv{Department of Chemistry}, \orgname{Virginia Military Institute}, \orgaddress{
\city{Lexington}, \postcode{24450}, \state{VA}, \country{USA}}}


\abstract{Polycyclic aromatic hydrocarbons (PAHs) are among the most ubiquitous compounds in the universe, accounting for up to ${\sim}$25\,\% of all interstellar carbon. Since most unsubstituted PAHs do not possess permanent dipole moments, they are invisible to radio astronomy. Constraining their abundances relies on the detection of polar chemical proxies, such as aromatic nitriles. We report the detection of 2- and 4-cyanopyrene, isomers of the recently detected 1-cyanopyrene. We find that these isomers are present in an abundance ratio of ${\sim}$2:1:2, which mirrors the number of equivalent sites available for CN addition. We conclude that there is evidence that the cyanopyrene isomers formed by direct CN addition to pyrene under kinetic control in hydrogen-rich gas at 10 K and discuss constraints on the H:CN ratio for PAHs in TMC-1.}  





\maketitle

\section*{Main}\label{Main}


Polycyclic aromatic hydrocarbons (PAHs) are the likely carriers of the unidentified infrared (UIR) bands that dominate the spectra of most galactic and extragalactic objects~\cite{tielens2008}. These bright features, specifically at 3.3, 6.2, 7.7, 8.6, 11.2, and $12.7\,\mathrm{\mu m}$, are generally associated with vibrational modes of PAHs that undergo infrared (IR) fluorescence after having been electronically excited by absorbing far-ultraviolet photons~\cite{leger1984,allamandola1985}. In the astronomical objects where UIR bands are observed, PAHs are highly abundant (${\sim}10^{-7}$ relative to hydrogen~\cite{tielens2008}) and therefore significantly impact the physics and chemistry of the interstellar medium (ISM). In particular, they play a key role in determining the ionization balance in molecular clouds, thus influencing ion-molecule chemistry~\cite{bakes1998} and contributing to the neutral gas heating due to the photoelectric effect~\cite{berne2022}. Despite their perceived importance, little is known about the nature of individual PAH molecules in the ISM. While the presence and abundance of PAHs in space is strongly supported by IR observations using, e.g., the Infrared Space Observatory (ISO)~\cite{cernicharo2001}, the Spitzer Space Observatory~\cite{li2020}, and the recently launched James Webb Space Telescope (JWST)~\cite{chown2024}, the spectra obtained in the mid-IR are a convolution of many different hot PAH molecules that all contain similar functional groups. Due to this spectral congestion, identification of individual PAHs in the ISM has not yet been achieved using their vibrational fingerprints. However, significant efforts to compare spectral variations across multiple astronomical objects have constrained the PAH families present in these astrophysical environments~\cite{peeters2002,galliano2008,peeters2011}, including recently with unprecedented spatial resolution using JWST~\cite{chown2024}.

While extraterrestrial PAHs have been found in carbonaceous chondrites such as Murchison and Orgueil~\cite{sabbah2017,lecasble2022} and in samples returned from comet 81P/Wild 2 during the Stardust mission~\cite{clemett2010}, 
their recent discovery in return samples from asteroid Ryugu shines new light on potential formation pathways~\cite{aponte2023,zeichner2023}. Carbon-13 isotopic analysis of the PAHs found in Ryugu showed that the 3-ring species such as anthracene and phenanthrene were formed at high temperatures (${>}1000\,\mathrm{K}$). Meanwhile, the 2- and 4-ring PAHs naphthalene, fluoranthene, and pyrene (the most abundant PAH in Ryugu) 
must have formed via a kinetically controlled route at low temperatures (${\sim}10\,\mathrm{K}$). Indeed, 2- and 4-ring PAHs have been unambiguously detected in the cold, dark molecular cloud TMC-1 by radio astronomical observations~\cite{mcguire2021,burkhardt2021,cernicharo2021,sita2022,wenzelinreview}.

In contrast to the hot, broad UIR bands, each molecule possessing a permanent dipole moment has a distinct rotational spectrum with narrow emission lines that can be observed using radio astronomy. 
Most PAHs considered in the literature are large (more than 30 carbon atoms), highly symmetric, and unsubstituted (“pure” hydrocarbons) for which models predict a viable chance of survival under the harsh interstellar conditions~\cite{montillaud2013}. However, due to their high symmetry, these PAHs often possess only a small or null dipole moment. Thus, despite their ubiquity, only five individual PAHs have been detected by radio astronomy to date~\cite{mcguire2021,burkhardt2021,cernicharo2021,sita2022,wenzelinreview}. 
The rotational emission from these unambiguously detected PAHs has been observed toward TMC-1 and originates from CN-functionalized PAHs (nitriles), with the exception of the asymmetric, pure PAH indene. It has been proposed that, owing to their large dipole moments, nitrile-substituted PAHs can be used as observational proxies for pure PAHs ~\cite{messinger2020,cooke2020}. Extracting quantitative abundances of unsubstituted PAHs from these proxies, however, relies on knowledge of the kinetics of their dominant formation and destruction pathways~\cite{balucani2000}.

Here, we present the interstellar detection of two additional CN-functionalized PAHs, 2- and 4-cyanopyrene, isomers of the recently discovered 1-cyanopyrene in TMC-1, using broadband radio astronomical observations and enhancing the statistical evidence for their detections with a stacking and matched filtering analysis. The discovery of 2- and 4-cyanopyrene completes the set of all possible singly CN-substituted pyrene isomers, allowing us to explore their potential formation routes by comparing their abundances to each other, and to further constrain the abundance of pure pyrene in TMC-1.


\begin{figure}[!htb]
    \centering
    \includegraphics[width=0.6\linewidth]{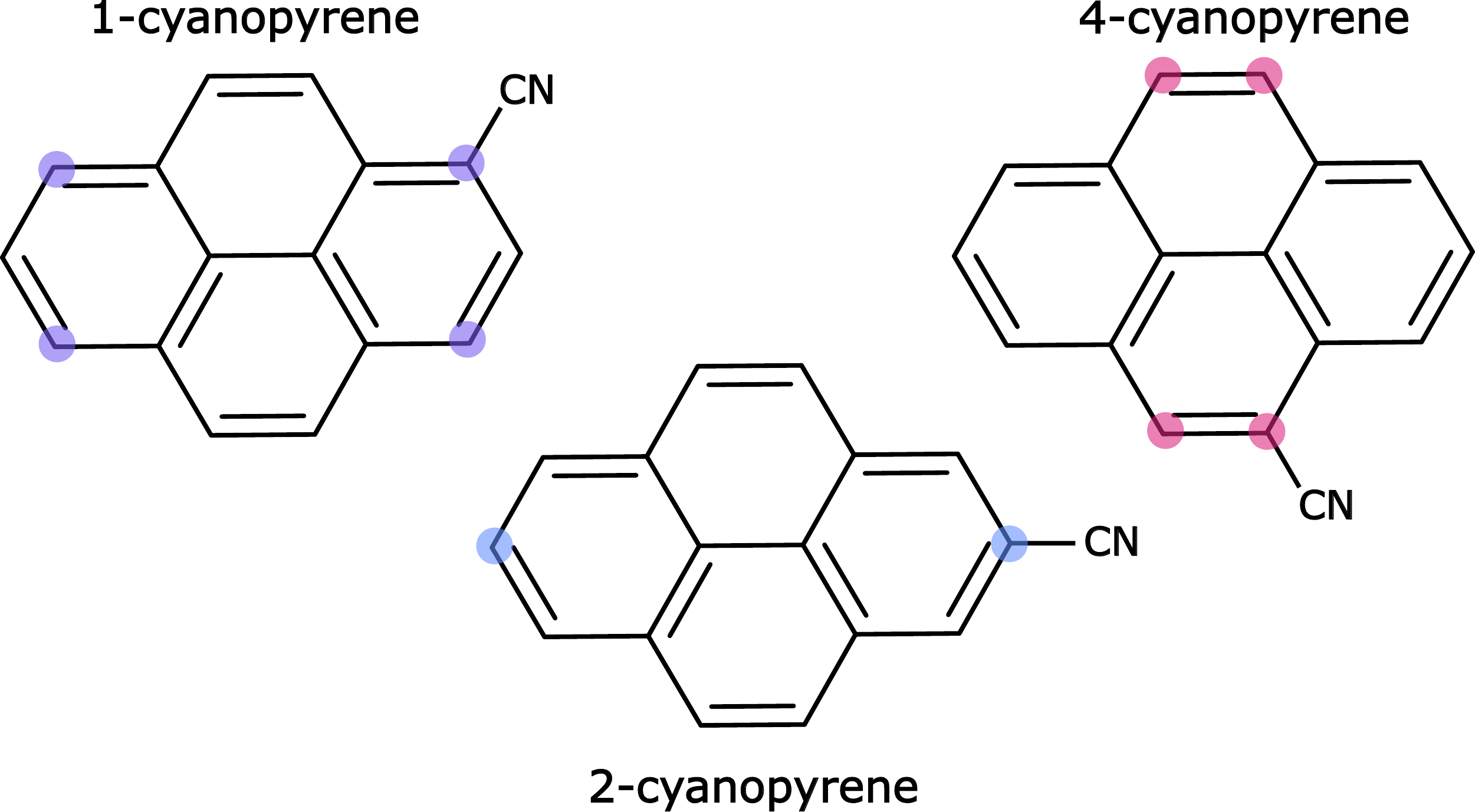}
    \caption{\textbf{Structures of the cyanopyrene isomers.} CN-functionalization of pyrene (\ce{C17H9N}) forms three possible isomers, namely 1-cyanopyrene (\ce{1-CN}--\ce{C16H9}), 2-cyanopyrene (\ce{2-CN}--\ce{C16H9}), and 4-cyanopyrene (\ce{4-CN}--\ce{C16H9}). Equivalent sites are shown with colored circles.}
    \label{fig:molecular_structures}
\end{figure}

\subsection*{Discovery of 2- and 4-cyanopyrene in GOTHAM observations toward TMC-1}

The GOTHAM -- GBT Observations of TMC-1: Hunting Aromatic Molecules -- project is a high-sensitivity, high-spectral resolution broadband line survey of TMC-1 with near-continuous coverage from approximately $8$ to $36\,\mathrm{GHz}$~\cite{mcguire2020}. The data were collected with the 100-m Robert C. Byrd Green Bank Telescope from 2018 to 2022, and the dataset used in this work was described previously~\cite{sita2022,cooke2023,wenzelinreview}.

First, the laboratory rotational spectra of pure samples of 2- and 4-cyanopyrene (see Fig.~\ref{fig:molecular_structures} for their structures) were measured between approximately $7$ and $18\,\mathrm{GHz}$ using a cavity-enhanced Fourier transform microwave spectrometer. To determine the rotational constants, 762 and 318 (individual or partly blended) transitions of 2- and 4-cyanopyrene, respectively, were fit to a standard asymmetric top rotational Hamiltonian (see Methods). The derived spectroscopic constants, which are reported in Supplementary Table~\ref{tab:rotconst}, allowed us to calculate the rotational rest transition frequencies up to ${\sim}25\,\mathrm{GHz}$ with an accuracy of ${\sim}2\,\mathrm{kHz}$. Searches for the radio emission features of 2- and 4-cyanopyrene toward TMC-1 were performed by simulating their rotational spectra under TMC-1 conditions (${\sim}5.8\,\mathrm{km\,s^{-1}}$, $5-10\,\mathrm{K}$) and comparing them to the GOTHAM data (see Supplementary Fig.~\ref{fig:coverage}). Comparing the simulated rotational spectra to the root-mean-square (RMS) noise of our GOTHAM data depicted in Supplementary Fig.~\ref{fig:coverage}, we identify only a few spectral windows in which the 4-cyanopyrene features might be $>1\,\sigma$ (above the noise, but $\leq 3\,\sigma$), while the 2-cyanopyrene lines are even weaker. These spectral windows are plotted in Supplementary Figs.~\ref{fig:lines_2-cyanopyrene} and \ref{fig:lines_4-cyanopyrene} for 2- and 4-cyanopyrene, respectively.

\begin{figure}[!htb]
    \centering
    \includegraphics[width=1\linewidth]{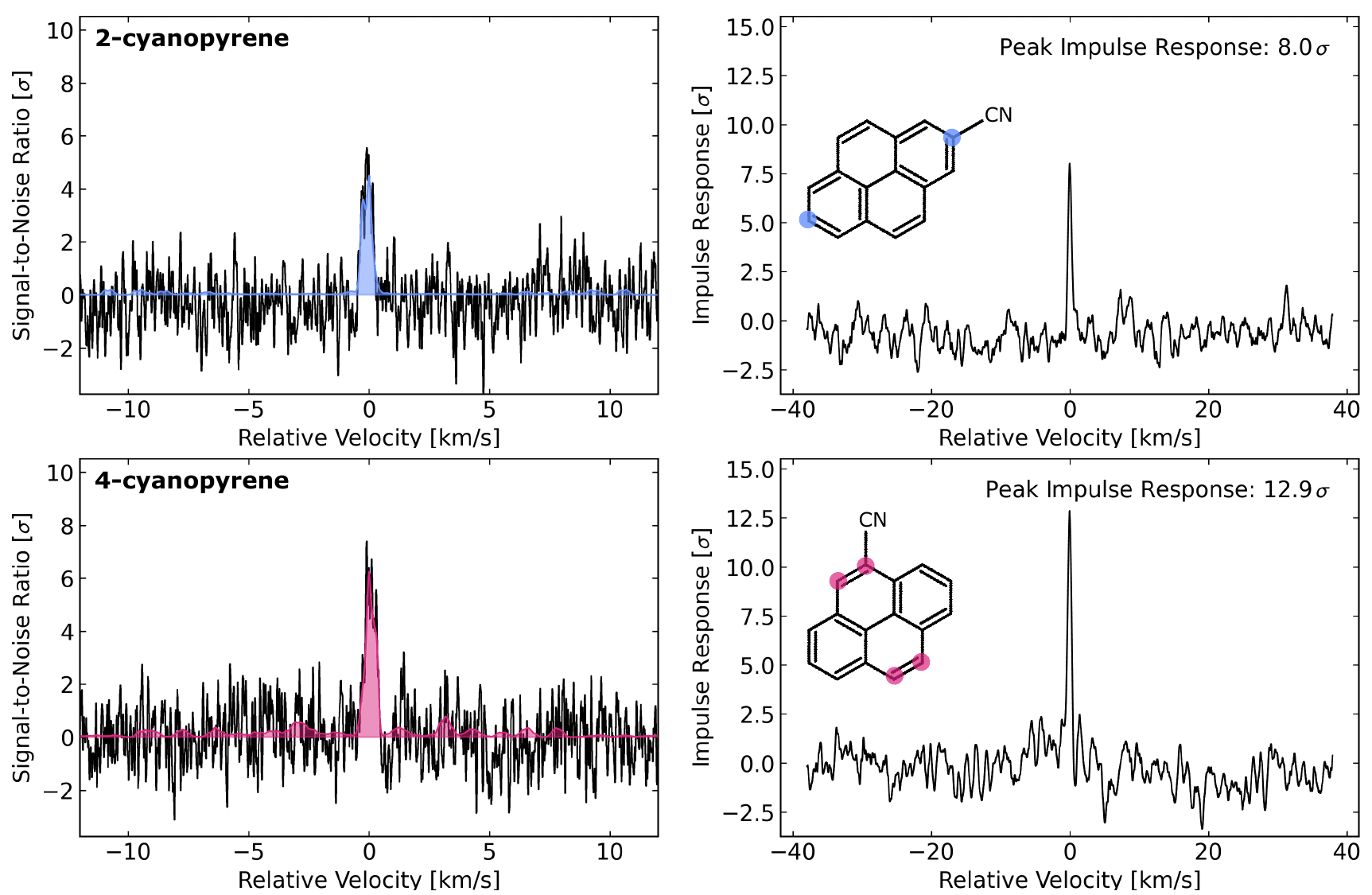}
    \caption{\textbf{Velocity-stacked spectra and matched filter responses of 2- and 4-cyanopyrene.} The stacked GOTHAM observations (black) are overlaid with the simulated stacked spectrum of 2- and 4-cyanopyrene (blue and pink, respectively), each consisting of the 150 brightest SNR lines (left panels). Marginalized posterior parameters were used in both simulations, as reported in Supplementary Table~\ref{tab:MCMCresults_isomers}. The corresponding impulse response for the matched filtering analysis is shown, yielding a significance of $8.0\,\sigma$ and $12.9\,\sigma$ for the 2- and 4-cyanopyrene detections, respectively (right panels). The small features in the stacked simulated spectrum (pink) 
    in the bottom left panel result from the densely populated 4-cyanopyrene lines (see Fig.~\ref{fig:lines_4-cyanopyrene}) that add up in the stack.}
    \label{fig:stack+mf}
\end{figure}

A Markov chain Monte Carlo (MCMC) analysis was used to derive the marginalized posterior parameters for the 2- and 4-cyanopyrene emission. These included the velocity in the local standard of rest, $v_\mathrm{lsr}$, and column densities, $N_T$, in all four spatially separated velocity components of TMC-1~\cite{loomis2021}, a single excitation temperature, $T_\mathrm{ex}$, and linewidth, $\Delta V$. From this analysis, using the 1-cyanopyrene marginalized posterior parameters as priors (see Methods, Supplementary Table~\ref{tab:MCMCpriors} and ref.~\cite{wenzelinreview}), we derived a column density of $0.84^{+0.09}_{-0.09} \times 10^{12}\,\mathrm{cm^{-2}}$ at an excitation temperature of $7.90^{+0.53}_{-0.48}\,\mathrm{K}$ for 2-cyanopyrene, and $1.33^{+0.10}_{-0.09} \times 10^{12}\,\mathrm{cm^{-2}}$ at $8.27^{+0.46}_{-0.44}\,\mathrm{K}$ for 4-cyanopyrene (see Supplementary Figs.~\ref{fig:corner2} and~\ref{fig:corner4} and Table~\ref{tab:MCMCresults_isomers}). The partition functions used for the MCMC analysis are listed in Table~\ref{tab:partfunc}.

To further explore the significance of our detections, we performed a velocity-stack and matched filtering analysis on both species with the same methodology described previously~\cite{loomis2021,mcguire2021}. For this purpose, spectral windows centered around the 150 brightest signal-to-noise ratio (SNR) lines of the simulated cyanopyrene spectra and the corresponding frequencies in the GOTHAM data were extracted in frequency space and collapsed into one SNR-weighted line in velocity space. Cross-correlating the latter stack (black in the left panel of Fig.~\ref{fig:stack+mf}) with the former (color in the left panel of Fig.~\ref{fig:stack+mf}) yielded an impulse response for the two detections. The statistical significance of the 2- and 4-cyanopyrene detections were $8.0\,\sigma$ and $12.9\,\sigma$, respectively.







\section*{Discussion}\label{sec:discussion}

\subsection*{Abundances of the cyanopyrene isomers}

We extracted column densities of $1.52^{+0.18}_{-0.16}$, $0.84^{+0.09}_{-0.09}$, and $1.33^{+0.10}_{-0.09} \times 10^{12}\,\mathrm{cm^{-2}}$ for the 1-, 2- and 4-cyanopyrene isomers, respectively, \textit{i.e.}, a 1.8:1:1.6 abundance ratio (with an uncertainty of approximately $\pm 0.3$) or roughly 2:1:2 \cite{wenzelinreview}.
On the assumption that CN addition to a double bond in pyrene is the major formation pathway for the cyanopyrene isomers, the observed ratio is consistent with two equivalent sites yielding 2-cyanopyrene, and four equivalent sites yielding 1- and 4-cyanopyrene (marked with colored circles in Fig.~\ref{fig:molecular_structures}).



The use of cyanopyrene abundance to estimate the abundance of pyrene relies on knowledge of the formation pathway of cyanopyrene. If cyanopyrene is formed predominantly from CN addition to pyrene under kinetic control, then the relative cyanopyrene/pyrene abundance would be determined by the ratio of the CN addition rate and the total cyanopyrene destruction rate. In this case, it is reasonable to expect that the 1-, 2-, and 4-cyanopyrene product branching ratios would be proportional to the number of equivalent sites available for CN addition. Calculations of the CN addition rate coefficients were therefore carried out to evaluate if the observed isomer distribution agrees with kinetic control for the addition of CN to pyrene (see Methods and Supplementary Table~\ref{tab:MESMER}). 

The EP3 corrected $\omega$B97X-D4/def2-TZVPP site-specific bimolecular rate coefficients for CN addition to pyrene at $10\,\mathrm{K}$ were predicted to be 
$k_1 = 2.02 \times 10^{-10}\,\mathrm{cm^3\,s^{-1}}$,
$k_2 = 1.01 \times 10^{-10}\,\mathrm{cm^3\,s^{-1}}$, 
$k_4 = 2.02 \times 10^{-10}\,\mathrm{cm^3\,s^{-1}}$,
consistent with the observation of the cyanopyrene isomers in a ${\sim}$2:1:2 ratio. We used these isomer-specific rate coefficients to place tighter constraints on the CN/H ratio for cyanopyrene:pyrene by comparing them to those obtained with the \texttt{nautilus} modeling code~\cite{ruaud2016} using typical cold dense cloud conditions, an adapted astrochemical network, and the observed ratio of 2-cyanoindene:indene in TMC-1~\cite{sita2022}. 

\begin{figure}
    \centering
    \includegraphics[width=0.95\linewidth]{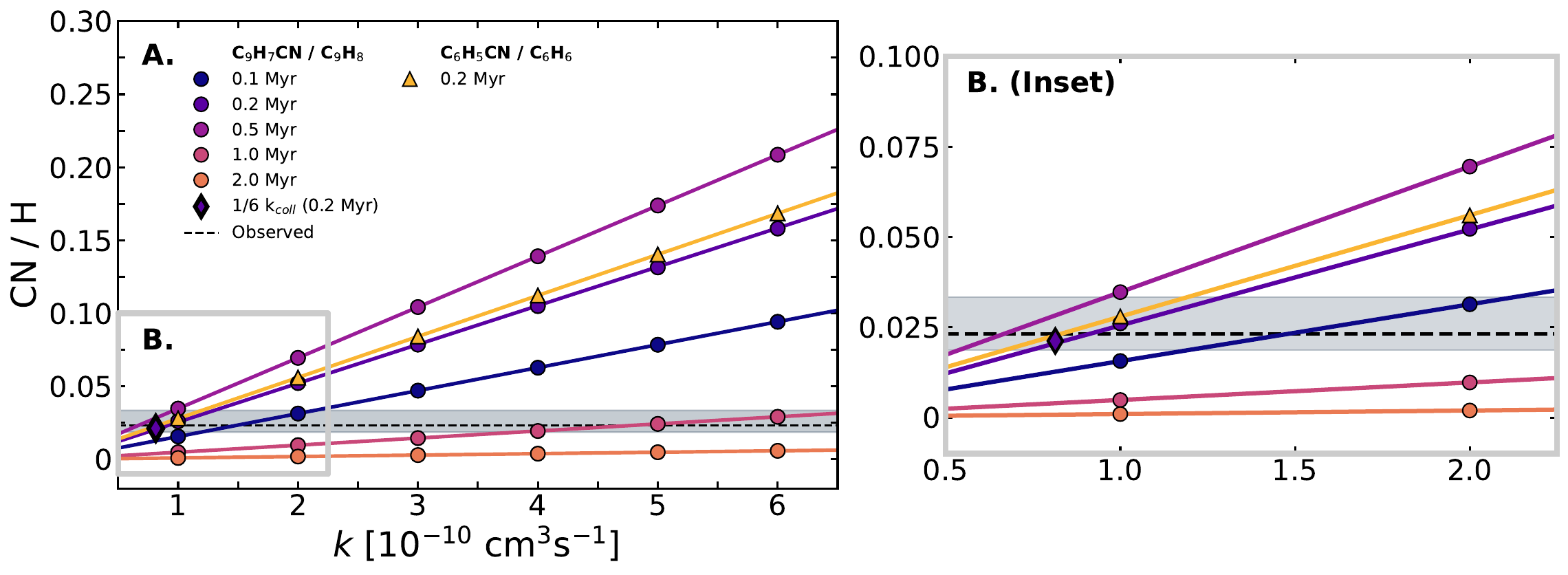}
    \caption{The CN/H ratio as a function of the CN + aromatic rate coefficient, $k$. The filled circles show ratios obtained from the astrochemical models for \ce{C9H7CN}/\ce{C9H8} at different simulation ages. The isomer-specific CN/H ratio obtained at 0.2\,Myrs by taking 1/6 of the collision rate coefficient for the reaction of CN with indene (i.e., $8.14 \times 10^{-11}\,\mathrm{cm^{-3}}$) is shown by the purple diamond. The observed \ce{C9H7CN}/\ce{C9H8} ratio is shown by the horizontal dashed line with uncertainties shown in the grey horizontal bar. The yellow filled triangles show \ce{C6H5CN}/\ce{C6H6} at 0.2\,Myrs as a function of the CN + \ce{C6H6} rate coefficient.}
    \label{fig:proxy}
\end{figure}

Figure~\ref{fig:proxy} shows the linear dependence of the CN/H ratio on the rate coefficient for the reaction of CN with an aromatic hydrocarbon, illustrated using 2-cyanoindene/indene, \ce{C9H7CN}/\ce{C9H8}, and benzonitrile/benzene, \ce{C6H5CN}/\ce{C6H6}. We explored this dependence for different chemical ages between 0.1 and 2\,Myrs, since a range of values have been reported for different molecules and model parameters~\cite{loomis2021,loison2014}. The linear dependence was observed for all chemical ages, albeit with different slopes. This trend is due to the increased efficiency of simultaneously destroying the pure aromatic and producing the nitrile-functionalized aromatic. Both benzene and indene display similar trends with their CN/H ratios nearly identical at 0.2\,Myr. Assuming similar production and destruction mechanisms, these models can be extended to the CN/H ratio for other aromatic molecules and their nitriles, such as naphthalene and the cyanonaphthalene isomers.

It is important to note that while CN addition to benzene produces only one isomer (benzonitrile), CN addition to indene can occur via 6 distinct sites, therefore the product-branching ratio must be taken into account when determining the appropriate CN + aromatic rate coefficient. To explore this in more detail, we estimated the CN + indene collision rate coefficient, $k_\mathrm{coll}$ (see Methods and Equation \ref{eq:kcoll}). Using the simple assumption that each product channel is equally likely, $k_\mathrm{coll}/6 = 8.14 \times 10^{-10}\,\mathrm{cm^3\,s^{-1}}$ is used to approximate the rate coefficient for CN + indene to form 2-cyanoindene in the collision limit. To compare to observations, we choose an age of 0.2\,Myr based on the approximate chemical age of TMC-1 that was derived previously from modelling of carbon chains ~\cite{mcguire2018,Siebert2022}). As shown in Fig.~\ref{fig:proxy}, this rate coefficient is consistent with both the modeled and observed CN/H ratio for 2-cyanoindene/indene of 0.023. We therefore used these results to constrain the CN/H ratio for cyanopyrene/pyrene. We calculated the channel-specific rate coefficients by incorporating our EP3//$\omega$B97X-D4/def2-TZVPP surface into the energy-grained master equation calculator MESMER 7.0 (see Supplementary Section~\ref{sec:MESMER}). The product-branching ratios are sensitive to the heights of the submerged barriers for H elimination. Using the EP3 corrected $\omega$B97X-D4/def2-TZVPP values, 
the abundance of pyrene can be estimated from Fig.~\ref{fig:proxy} as ${\sim}20\times$ 1- or 4-cyanopyrene or ${\sim}40\times$ the abundance of 2-cyanopyrene, \textit{i.e.}, a column density of ${\sim}3 \times10^{13}\,\mathrm{cm^{-2}}$.

However, when EP3 corrections are applied, the heights of the barriers are raised such that within the accuracy of the calculations the barriers could begin to impact the product distribution. Further computational work to evaluate the significance of the exit barriers will be important in elucidating whether the observed ratio of the cyanopyrene isomers agrees with a kinetically controlled addition of CN to pyrene.

\subsection*{Bottom-up versus top-down formation pathways}

Measurements of $^{13}$C isotopic substitutions in samples from asteroid Ryugu suggest that the two- and four-ring PAHs naphthalene, fluoranthene, and pyrene likely formed in low-temperature interstellar environments~\cite{zeichner2023}. However, the observed doubly-$^{13}$C substitutions for 3-ring PAHs phenanthrene and anthracene, as well as pyrene in the carbonaceous chondrite Murchison, suggest they may have formed in high-temperature circumstellar envelopes of evolved stars. Alternatively, the values could be explained by a scenario where these species formed or were altered on the parent body of asteroid Ryugu and hence after solar system formation. If these small PAHs are formed in circumstellar envelopes, they must survive destruction by radiation and shocks in the diffuse interstellar medium~\cite{micelotta2010,micelotta2011}. However, if this is the case, it is difficult to understand why only the 3-ring PAHs survived passage through the diffuse ISM. Indeed, cursory searches for the 3-ring PAHs 9-cyanoanthracene and 9-cyanophenanthrene, whose rotational spectra are known~\cite{mcnaughton2018}, in GOTHAM observations of TMC-1 have not yet been successful.

Various bottom-up mechanisms have been proposed to explain the formation pathways of PAHs in high-temperature environments, such as combustion systems~\cite{reizer2022} and circumstellar envelopes~\cite{zhao2018}. These mechanisms generally occur in two stages~\cite{reizer2022}: first-ring closure followed by growth through subsequent addition of aromatic rings. The most prominent of the latter is the Hydrogen Abstraction -- Acetylene (\ce{C2H2}) Addition (HACA) mechanism~\cite{reizer2022}. In this mechanism, hydrogen is first abstracted from an aromatic molecule (typically by a hydrogen atom), followed by addition to the radical site. A bottom-up HACA mechanism to pyrene has been proposed by~\citet{zhao2018} involving the reaction of the 4-phenanthrenyl radical (\ce{C14H9}$^{\bullet}$) with \ce{C2H2}. However, the addition requires overcoming a barrier on the order of $10-20\,\mathrm{kJ\,mol^{-1}}$ ($1200-2400\,\mathrm{K}$ equivalent temperature), prohibiting the HACA mechanism at the low temperatures of TMC-1
~\cite{kaiser2021}.

An alternative low-temperature mechanism for the bottom-up ring growth of PAHs has been proposed, the so-called HAVA (Hydrogen Abstraction -- Vinylacetylene (\ce{C4H4}) Addition) mechanism. Phenanthrene has been shown to form 
by the HAVA mechanism via naphthyl radical~\cite{zhao2018b}. Since the HAVA mechanism generally involves only submerged barriers, it can operate at low temperatures like those in dense molecular clouds. Recently, the reaction between the phenylethynyl radical (\ce{C6H5CC}$^{\bullet}$) and benzene has been proposed as another viable bottom-up mechanism to form phenanthrene at low temperature~\cite{goettl2024a}. However, vinylacetylene addition to the 4-phenanthrenyl radical is not expected to form pyrene~\cite{zhao2019}. While this mechanism may form vinylpyrene, it is one of the few HAVA mechanisms studied that possesses a barrier in the entrance channel, due to steric hindrance at the reaction site. Thus, a low-temperature bottom-up mechanism to pyrene has yet to be unveiled.

Top-down routes to smaller PAHs have also been explored, including the fragmentation of bulk amorphous carbon or graphite by collisions of dust grains or interstellar shocks~\cite{scott1997,merino2014}. The relative importance of top-down versus bottom-up chemistry is difficult to quantify and most astrochemical models focus on the latter. A comparison between the abundances of pyrene and smaller aromatic molecules detected in TMC-1 thus far can provide clues about its formation pathways.
The column density of benzonitrile is $1.73^{+0.85}_{-0.10} \times 10^{12}\,\mathrm{cm^{-2}}$, therefore, assuming a CN + benzene rate coefficient of ${\sim}4\times10^{-10}\,\mathrm{cm^{-3}\,s^{-1}}$ \cite{cooke2020}, we predict a benzene abundance of ${\sim}1.4\times 10^{13}\,\mathrm{cm^{-2}}$. Since this is approximately half the abundance that we derive above for pyrene, it is difficult to envision a bottom-up route to pyrene from benzene, unless benzene is destroyed much more efficiently than pyrene. Observational constraints on the abundance of larger PAHs would help to constrain the relative importance of top-down versus bottom-up formation pathways. If pyrene does form top-down, further work is required to reconcile this mechanism with the isotopic results from Ryugu. 



\section*{Conclusions}

We report the first interstellar detections of the two CN-functionalized pyrene isomers, 2- and 4-cyanopyrene, in our GOTHAM observations toward the dark molecular cloud TMC-1. Together with the previously detected 1-cyanopyrene, they form a family of the largest interstellar molecules identified by radio astronomy to date. New theoretical calculations of the CN + pyrene rate coefficients, their column densities of $1.52^{+0.18}_{-0.16}$, $0.84^{+0.09}_{-0.09}$, and $1.33^{+0.10}_{-0.09} \times 10^{12}\,\mathrm{cm^{-2}}$ for 1-, 2-, and 4-cyanopyrene, respectively, forming an approximate abundance ratio of 2:1:2, help better constrain the CN/H ratio of all aromatic species present in TMC-1 and hence the abundance of pure pyrene. We estimate a column density for gas-phase pyrene of ${\sim}3 \times10^{13}\,\mathrm{cm^{-2}}$ corresponding to an abundance of ${\sim}3 \times10^{-9}$ with respect to \ce{H2}. In light of this, PAH formation mechanisms should be revisited to help explain the origin and abundance of pyrene in TMC-1. 

\section*{Methods}\label{methods}

\subsection*{Rotational spectroscopy}

The synthetic routes to form the two cyanopyrene isomers are described in Supplementary Section~\ref{sec:synthesis}. Rotational spectra of 2- and 4-cyanopyrene were predicted using the open-source quantum chemical package Psi4~\cite{smith2020}. Their geometries were initially optimized at the B3LYP/6-311++G(d,p)~\cite{becke1993a} level of theory and basis set and subsequently using M06-2X/6-31+G(d)~\cite{zhao2008,lee2020}, determining their rotational constants $A$, $B$, and $C$ as presented in Supplementary Table~\ref{tab:rotconst}. They agree well when compared to the constants derived by the `Lego brick' approach~\cite{ye2022}. The $^{14}$N nuclear electric quadrupole hyperfine coupling constants, $\chi_{aa}$ and $\chi_{bb}$, were estimated by rotating the $\chi$ tensor of benzonitrile (cyanobenzene), which has been accurately measured experimentally~\cite{wohlfart2008}, to the principal axis coordinate systems of 2- and 4-cyanopyrene, assuming that the electric field gradients remain identical with respect to the local CN bond axis and molecular plane.

The laboratory rotational transition rest frequencies were measured using a cavity-enhanced Fourier transform microwave (FTMW) spectrometer~\cite{grabow2005, crabtree2016}. A laser ablation source was employed for solid sample introduction. The molecule of interest was either mixed with anthracene (Sigma Aldrich, purity $\geq 97\,\%$) as a binder material in a 1:1 ratio to produce a homogeneous mixture (4-cyanopyrene) or used as is (2-cyanopyrene) and pressed with $3\,\mathrm{metric\;ton}$ of press force in a hydraulic press into a 0.25''-diameter cylindrical sample rod. The sample rod was mounted $8\,\mathrm{mm}$ downstream of a pulsed solenoid valve which was backed with $2.5\,\mathrm{kTorr}$ of neon as carrier gas. Sample ablation was performed using the second harmonic of a Continuum Surelite (SLI-10) Nd:YAG laser at a wavelength of $532\,\mathrm{nm}$ with a pulse energy of $50\,\mathrm{mJ}$ synchronized to operate during the $\leq1\,\mathrm{ms}$ opening time of the solenoid valve. The ablated cyanopyrenes were carried into the FTMW spectrometer and the supersonic expansion cooled them to a rotational temperature of ${\sim}2\,\mathrm{K}$. After a brief search near the predicted frequencies of the strongest transitions, we observed several lines that could be successfully assigned to 2- and 4-cyanopyrene and were used to iteratively refine the spectroscopic predictions to search for additional transitions. Ultimately, we observed 762 and 318 individual or partly blended rotational transitions over the $7 - 16\,\mathrm{GHz}$ frequency range for 2- and 4-cyanopyrene, respectively. The rest frequencies were least-squares fit to a standard rotational Hamiltonian using SPCAT/SPFIT in Pickett’s CALPGM suite of programs~\cite{pickett1991} (A-reduced, $I^r$ representation) including quartic centrifugal distortion and $^{14}$N nuclear electric quadrupole hyperfine coupling constants, which are reported in Supplementary Table~\ref{tab:rotconst} in Supplementary Section~\ref{sec:rotconst}.

\subsection*{Observations and analysis}

The GOTHAM project has been described in previous publications~\cite{mcguire2020,mcguire2021} and a detailed description of the data used in this study can be found in refs.~\cite{sita2022,cooke2023}. These data include observations up until May 2022.

Briefly, spectra were collected using the VErsatile GBT Astronomical Spectrometer (VEGAS) on the 100-m Robert C. Byrd Green Bank Telescope (GBT). Project codes for the observations used in the data set are GBT17A\nobreakdash-164, GBT17A\nobreakdash-434, GBT18A\nobreakdash-333, GBT18B\nobreakdash-007, GBT19B\nobreakdash-047, AGBT20A\nobreakdash-516, AGBT21A\nobreakdash-414, and AGBT21B\nobreakdash-210. Data were recorded with a uniform frequency resolution of $1.4\,\mathrm{kHz}$, or $0.05 - 0.01\,\mathrm{km\,s^{-1}}$ in velocity space.

The data span the X-, Ku-, K-, and most of the Ka-receiver bands, with nearly continuous coverage from $7.9-11.6\,\mathrm{GHz}$, $12.7-15.6\,\mathrm{GHz}$, and $18.0-36.4\,\mathrm{GHz}$. The total bandwidth covered is $24.9\,\mathrm{GHz}$. To visualize the spectral coverage, the GOTHAM dataset is overlaid in Supplementary Fig.~\ref{fig:coverage} with the simulated rotational spectra of the cyanopyrene isomers at ${\sim}8\,\mathrm{K}$. The cyanopyrene lines covered by the GOTHAM dataset are shown in violet (1-cyanopyrene), blue (2-cyanopyrene), and pink (4-cyanopyrene). The shaded grey boxes represent the averaged root mean square (RMS) noise level in each data chunk, typically ${\sim}2-20\,\mathrm{mK}$. The RMS noise increase at higher frequencies is due to the shorter total integration times in those frequency ranges.

Pointing was performed on the cyanopolyyne peak (CP) in TMC-1 at (J2000) $\alpha$~=~04$^h$41$^m$42.50$^s$ $\delta$~=~+25$^{\circ}$41$^{\prime}$26.8$^{\prime\prime}$. Spectra (ON/OFF source) were collected using position-switching between the source and an emission-free position offset by $1^{\circ}$. Re-pointing and focusing were generally carried out every $1-2$\,hours, primarily on the calibrator J0530+1331. Flux calibration was performed using an internal noise diode and Karl G. Jansky Very Large Array (VLA) observations of J0530+1331, \textit{i.e.}, the same source used for pointing. The flux uncertainty is estimated to be ${\sim}20\,\%$~\cite{sita2022,cooke2023}. 

\subsection*{MCMC Analysis}

Prior observations of TMC-1 have shown that most cm-emission can be separated into contributions from four different velocity components~\cite{dobashi2018, dobashi2019, xue2020}. Thus, we consider four different Doppler components, each with independent source size, velocity (in the local standard of rest, $v_\mathrm{lsr}$) and column densities ($N_T$) and a shared uniform excitation temperature ($T_\mathrm{ex}$) and linewidth ($\Delta V$), resulting in 14 modelling parameters for each molecule.

To account for covariance between the model parameters, we use an Affine-invariant Markov Chain Monte Carlo (MCMC) sampling analysis, which has previously been applied to complex probability distributions in many components, including for previous observations of molecules (see refs.~\cite{loomis2021,mcguire2021}). 

The priors we adopted for the MCMC analysis of the two newly detected cyanopyrene isomers are listed in Table~\ref{tab:MCMCpriors}. A uniform distribution (\textit{i.e.}, unconstrained within the minima and maxima) was chosen for the $N_T$ and $T_\mathrm{ex}$ priors, whereas the remaining parameters were set to have more tightly constrained Gaussian distributed priors centered at the values determined by prior observations of chemically similar species 
benzonitrile, cyanonaphthalenes~\cite{mcguire2021}, and 1-cyanopyrene~\cite{wenzelinreview}. 

Posterior probability distributions for each of the model parameters, along with their covariances, were generated using 100 walkers with 10,000 samples. The resulting source-dependent molecular parameters for the cyanopyrene isomers are reported in Supplementary Table~\ref{tab:MCMCresults_isomers}. The covariance plots of the 14 parameters resulting from our MCMC analysis for the 2- and 4-cyanopyrene isomers are presented in Supplementary Figs.~\ref{fig:corner2} and~\ref{fig:corner4}. The covariance plot for 1-cyanopyrene can be found in the Supplementary Material for ref.~\cite{wenzelinreview}.


\subsection*{Density functional theory (DFT) computations}

Initial structures for the adducts and separated reagents were optimized with the RI-BP86 DFT functional~\cite{lee1988} using the def2-SVP basis set~\cite{weigend2005} and including D3(BJ) empirical dispersion corrections~\cite{grimme2011} (RI-BP86-D3(BJ)/def2-SVP will be referred to as DFT-Cheap). The presence of barriers to subsequent transfer and elimination from the adducts was evaluated using relaxed surfaces scans with DFT-Cheap; where peaks in the scans were observed these were used as the initial guess for transition state optimizations. All of the structures calculated previously were then further refined and harmonic vibrations were calculated with the $\omega$B97X functional~\cite{chai2008a} with the triple zeta basis set def2-TZVPP~\cite{weigend2006} and D4 empirical dispersion corrections~\cite{caldeweyher2017} (henceforth DFT-2). Single point energy corrections were then carried out using the EP3 approximation of the CCSD(T) complete basis set limit~\cite{jurecka2006,liakos2012}. This approach can typically yield accuracy on the order of $5-15\,\mathrm{kJ\,mol^{-1}}$ with the accuracy of the relative energies of similar species (the 3 isomers formed, for example) likely to be much higher on the order of $\pm1\,\mathrm{kJ\,mol^{-1}}$. All calculations were performed using the open-source quantum chemical package ORCA 5.0.4~\cite{neese2022}. As no barriers at the DFT-Cheap level had been observed for the addition of CN to the ring in the 1-, 2- or 4-position, additional relaxed scans with the more reliable hybrid functionals and triple zeta basis sets DFT-2~\cite{chai2008a, weigend2006, caldeweyher2017} and M06-2X/def2-TZVPP~\cite{zhao2008, weigend2006} were carried out to verify this result. 


The results of this work show that the approach of CN to pyrene is barrierless for the addition of carbon to the 1-, 2- or 4-positions, leading to the formation of deeply bound adducts $121-182\,\mathrm{kJ\,mol^{-1}}$ below the entrance energies. The barriers to the subsequent H atom elimination are submerged by $2-17\,\mathrm{kJ\,mol^{-1}}$.\\


This EP3//DFT-2 surface was then incorporated into the energy-grained master equation calculator MESMER 7.0~\cite{glowacki2012}, which allowed the reaction to be simulated over a range of pressures and temperatures. There are eight equivalent pathways to addition at the 1- and 4-positions and four equivalent pathways to addition at the 2-position. The overall temperature-dependent collision rate coefficient for CN + pyrene was estimated using classical capture theory (CCT)~\cite{georgievskii2005}:

\begin{equation}\label{eq:kcoll}
    k_{\textrm{coll}}(T) = \sigma_{\textrm{coll}}\langle v(T) \rangle = \left[\pi \left(\frac{2C_6}{k_BT}\right)^{1/3} \Gamma \left( \frac{2}{3}\right)\right] \left[ \left( \frac{8k_BT}{\pi \mu}\right)^{1/2} \right],\vspace{12pt}
\end{equation}
where $k_B$ is the Boltzmann constant, $\Gamma(x)$ is the gamma function such that $\Gamma(2/3) = 1.353$, $\mu$ is the reduced mass of the collision, and $C_6$ is the sum of coefficients describing the magnitude of the attractive forces between collision partners (see~\citet{west2019a}). This approach is generally accurate within a factor of 2 for the prediction of rate coefficients for neutral--neutral barrierless reactions \cite{west2019a}. 
This was then treated with the inverse Laplace transformation  (ILT) methodology~\cite{davies1986} in MESMER along with Rice–Ramsperger–Kassel–Marcus (RRKM) treatment~\cite{Robertson_1996,baer1996,lourderaj2009} for the H atom elimination reactions. The A factor for the ILT into the individual adduct wells was set to the fraction of the total number of pathways available that led to the formation of that isomer.

The results of the MESMER simulations have been summarized in Supplementary Table~\ref{tab:MESMER}. The primary route to product formation is via well-skipping. An estimate for the uncertainties in the barrier heights was accounted for by concertedly raising and lowering all three barriers by $10\,\mathrm{kJ\,mol^{-1}}$. The predicted branching ratios are 2:1:2 for 1-, 2-, and 4-cyanopyrene formation with an overall removal rate coefficient of $5.05 \times 10^{-10}\,\mathrm{cm^3\,s^{-1}}$. 



\subsection*{Astrochemical models}

To gain insights into the CN:H ratio in TMC-1, we adapted the three-phase chemical network model \texttt{nautilus} v1.1 code~\citep{ruaud2016}. The reaction network is that of \citet{Siebert2022}, with modifications to the ion--neutral reaction rate coefficients as described below. We use the same initial conditions as described in~\citet{burkhardt2021a}, \textit{i.e.}, gas and dust temperatures of 10\,K, gas densities of $2\times10^4\,\mathrm{cm^{-3}}$, and a cosmic-ray ionization rate of $1.3\times10^{-17}\,\mathrm{s^{-1}}$.

Initial elemental abundances were taken from~\cite{hincelin2011} with the exception of atomic oxygen, where we utilize a slightly carbon-rich C/O $\approx1.1$ and $X_\mathrm{O}(t=0)\approx1.5\times10^{-4}$, as described in refs.~\cite{loomis2021} and ~\cite{burkhardt2021a}.

Rate coefficients for the destruction of benzonitrile via ion--neutral reactions were updated (see Supplementary Table~\ref{tab:ionrxns}), estimated assuming a Su-Chesnavich capture model to account for long-range Coulombic attractions~\cite{su1982, woon2009}. In addition, the reaction of carbon atoms with benzene was added to the network for consistency with the other aromatics (indene and naphthalene) currently in the network. 

\begin{equation}
    \ce{C + C6H6 -> C7H5 + H}
\end{equation}

The rate coefficient was initially set to $5\times10^{-10}\,\mathrm{cm^3\,s^{-1}}$, the room temperature value~\cite{haider1993}. Modifying the rate coefficient of C + aromatic was found to strongly influence the absolute abundance of benzene, benzonitrile, indene, and cyanoindene; however, it had negligible influence on the observed abundance ratios ($\leq 4\,\%$ for both benzonitrile/benzene and 2-cyanoindene/indene).

The rate coefficient for CN and indene was recalculated using CCT~\cite{georgievskii2005,west2019a} to be $4.88 \times 10^{-10}\,\mathrm{cm^3\,s^{-1}}$; assuming equal formation of all isomers this leads to a site-specific rate coefficient of $8.14 \times 10^{-10}\,\mathrm{cm^3\,s^{-1}}$ for 2-cyanoindene.

To evaluate the sensitivity of the model predictions to the CN + aromatic rate coefficient, the model was run for a range of $k_{\textrm{CN+benzene}}$ between $0.5 - 6 \times 10^{-10}\,\mathrm{cm^3\,s^{-1}}$ and for a range of $k_{\textrm{CN+indene}}$ between $0.81 - 10\times 10^{-10}\,\mathrm{cm^3\,s^{-1}}$. The observed ratios of benzonitrile/benzene and cyanoindene/indene depend linearly on their formation rate coefficient ($k_{\textrm{CN+aromatic}}$).
As can be seen in Figure~\ref{fig:proxy}, there is excellent agreement between the predicted ratio of cyanoindene to indene and benzonitrile to benzene using our updated reaction network. Furthermore, there is also a strong agreement to the observed isomer-specific ratio for 2-cyanoindene to indene. 


\section*{Data availability}

All data from our observing program GOTHAM are now available to the scientific community through the NRAO and GBO archives at  https://data.nrao.edu/portal/ under project codes GBT17A\nobreakdash-164, GBT17A\nobreakdash-434, GBT18A\nobreakdash-333, GBT18B\nobreakdash-007, GBT19B\nobreakdash-047, AGBT20A\nobreakdash-516, AGBT21A\nobreakdash-414, and AGBT21B\nobreakdash-210. Calibrated and reduced observational data windowed around the reported transitions; the full catalogs of 2- and 4-cyanopyrene, including quantum numbers of each transition; and the partition functions used in the MCMC analysis are available in a Harvard Dataverse repository OR zenodo.\\



\backmatter

\bmhead{Additional information}

\textbf{Supplementary information} is available for this paper at XYZ.com.

\bmhead{Acknowledgements}


The National Radio Astronomy Observatory is a facility of the National Science Foundation operated under cooperative agreement by Associated Universities, Inc. The Green Bank Observatory is a facility of the National Science Foundation operated under cooperative agreement by Associated Universities, Inc.\\
 
\textbf{Funding:} G.W.
and B.A.M. acknowledge the support of the Arnold and Mabel Beckman Foundation Beckman Young Investigator Award. 
Z.T.P.F., M.S.H., and B.A.M. acknowledge support from the Schmidt Family Futures Foundation. A.M.B. was supported in part by the Aisiku Summer Research Fellowship. A.N.B. acknowledges the support of NSF Graduate Research Fellowship grant 2141064. C.X. and B.A.M. acknowledge support of National Science Foundation grant AST-2205126. I.R.C. acknowledges support from the University of British Columbia, the Natural Sciences and Engineering Research Council of Canada (RGPIN-2022-04684), the Canada Foundation for Innovation and the B.C. Knowledge Development Fund (BCKDF). P.B.C. and M.C.M. are supported by the National Science Foundation award No. AST-2307137. 
S.B.C. is supported by the Goddard Center for Astrobiology and by the NASA Planetary Science Division Internal Scientist Funding Program through the Fundamental Laboratory Research work package (FLaRe). H.G. acknowledges support from the National Science Foundation for participation in this work as part of his independent research and development plan. Any opinions, findings, and conclusions expressed in this material are those of the authors and do not necessarily reflect the views of the National Science Foundation.

\section*{Author contributions}

All authors edited and reviewed the manuscript. In addition, G.W. performed spectroscopic experiments, analyzed observational data, conducted quantum chemical calculations, and wrote the manuscript. T.S. performed quantum chemical calculations. H.G. performed spectroscopic experiments. I.R.C. analyzed observational data and wrote the manuscript. P.B.C. performed spectroscopic experiments. E.A.B. analyzed observational results. S.Z. performed the synthesis. A.M.B. performed observations. A.J.R. performed observations. C.X. performed observations. M.C.M. supervised laboratory experiments. A.E.W. supervised laboratory experiments. B.A.M. performed observations, analyzed observational data, wrote the manuscript, and designed the project.

\section*{Competing interests}
The authors declare no competing interests.







\begin{thebibliography}{81}
\ifx \bisbn   \undefined \def \bisbn  #1{ISBN #1}\fi
\ifx \binits  \undefined \def \binits#1{#1}\fi
\ifx \bauthor  \undefined \def \bauthor#1{#1}\fi
\ifx \batitle  \undefined \def \batitle#1{#1}\fi
\ifx \bjtitle  \undefined \def \bjtitle#1{#1}\fi
\ifx \bvolume  \undefined \def \bvolume#1{\textbf{#1}}\fi
\ifx \byear  \undefined \def \byear#1{#1}\fi
\ifx \bissue  \undefined \def \bissue#1{#1}\fi
\ifx \bfpage  \undefined \def \bfpage#1{#1}\fi
\ifx \blpage  \undefined \def \blpage #1{#1}\fi
\ifx \burl  \undefined \def \burl#1{\textsf{#1}}\fi
\ifx \doiurl  \undefined \def \doiurl#1{\url{https://doi.org/#1}}\fi
\ifx \betal  \undefined \def \betal{\textit{et al.}}\fi
\ifx \binstitute  \undefined \def \binstitute#1{#1}\fi
\ifx \binstitutionaled  \undefined \def \binstitutionaled#1{#1}\fi
\ifx \bctitle  \undefined \def \bctitle#1{#1}\fi
\ifx \beditor  \undefined \def \beditor#1{#1}\fi
\ifx \bpublisher  \undefined \def \bpublisher#1{#1}\fi
\ifx \bbtitle  \undefined \def \bbtitle#1{#1}\fi
\ifx \bedition  \undefined \def \bedition#1{#1}\fi
\ifx \bseriesno  \undefined \def \bseriesno#1{#1}\fi
\ifx \blocation  \undefined \def \blocation#1{#1}\fi
\ifx \bsertitle  \undefined \def \bsertitle#1{#1}\fi
\ifx \bsnm \undefined \def \bsnm#1{#1}\fi
\ifx \bsuffix \undefined \def \bsuffix#1{#1}\fi
\ifx \bparticle \undefined \def \bparticle#1{#1}\fi
\ifx \barticle \undefined \def \barticle#1{#1}\fi
\bibcommenthead
\ifx \bconfdate \undefined \def \bconfdate #1{#1}\fi
\ifx \botherref \undefined \def \botherref #1{#1}\fi
\ifx \url \undefined \def \url#1{\textsf{#1}}\fi
\ifx \bchapter \undefined \def \bchapter#1{#1}\fi
\ifx \bbook \undefined \def \bbook#1{#1}\fi
\ifx \bcomment \undefined \def \bcomment#1{#1}\fi
\ifx \oauthor \undefined \def \oauthor#1{#1}\fi
\ifx \citeauthoryear \undefined \def \citeauthoryear#1{#1}\fi
\ifx \endbibitem  \undefined \def \endbibitem {}\fi
\ifx \bconflocation  \undefined \def \bconflocation#1{#1}\fi
\ifx \arxivurl  \undefined \def \arxivurl#1{\textsf{#1}}\fi
\csname PreBibitemsHook\endcsname

\bibitem[\protect\citeauthoryear{Tielens}{2008}]{tielens2008}
\begin{barticle}
\bauthor{\bsnm{Tielens}, \binits{A.G.G.M.}}:
\batitle{Interstellar {{Polycyclic Aromatic Hydrocarbon Molecules}}}.
\bjtitle{Annual Review of Astronomy and Astrophysics}
\bvolume{46}(\bissue{1}),
\bfpage{289}--\blpage{337}
(\byear{2008})
\doiurl{10.1146/annurev.astro.46.060407.145211}
\end{barticle}
\endbibitem

\bibitem[\protect\citeauthoryear{L{\'e}ger and Puget}{1984}]{leger1984}
\begin{barticle}
\bauthor{\bsnm{L{\'e}ger}, \binits{A.}},
\bauthor{\bsnm{Puget}, \binits{J.L.}}:
\batitle{Identification of the ``{{Unidentified}}'' {{IR Emission Features}} of
  {{Interstellar Dust}}?}
\bjtitle{Astronomy \& Astrophysics}
\bvolume{500},
\bfpage{279}
(\byear{1984})
\end{barticle}
\endbibitem

\bibitem[\protect\citeauthoryear{Allamandola et~al.}{1985}]{allamandola1985}
\begin{barticle}
\bauthor{\bsnm{Allamandola}, \binits{L.J.}},
\bauthor{\bsnm{Tielens}, \binits{A.G.G.M.}},
\bauthor{\bsnm{Barker}, \binits{J.R.}}:
\batitle{Polycyclic aromatic hydrocarbons and the unidentified infrared
  emission bands: Auto exhaust along the milky way.}
\bjtitle{The Astrophysical Journal}
\bvolume{290},
\bfpage{25}
(\byear{1985})
\doiurl{10.1086/184435}
\end{barticle}
\endbibitem

\bibitem[\protect\citeauthoryear{Bakes and Tielens}{1998}]{bakes1998}
\begin{barticle}
\bauthor{\bsnm{Bakes}, \binits{E.L.O.}},
\bauthor{\bsnm{Tielens}, \binits{A.G.G.M.}}:
\batitle{The {{Effects}} of {{Polycyclic Aromatic Hydrocarbons}} on the
  {{Chemistry}} of {{Photodissociation Regions}}}.
\bjtitle{The Astrophysical Journal}
\bvolume{499}(\bissue{1}),
\bfpage{258}
(\byear{1998})
\doiurl{10.1086/305625}
\end{barticle}
\endbibitem

\bibitem[\protect\citeauthoryear{Bern{\'e} et~al.}{2022}]{berne2022}
\begin{barticle}
\bauthor{\bsnm{Bern{\'e}}, \binits{O.}},
\bauthor{\bsnm{Foschino}, \binits{S.}},
\bauthor{\bsnm{Jalabert}, \binits{F.}},
\bauthor{\bsnm{Joblin}, \binits{C.}}:
\batitle{Contribution of polycyclic aromatic hydrocarbon ionization to neutral
  gas heating in galaxies: Model versus observations}.
\bjtitle{Astronomy \& Astrophysics}
\bvolume{667},
\bfpage{159}
(\byear{2022})
\doiurl{10.1051/0004-6361/202243171}
\end{barticle}
\endbibitem

\bibitem[\protect\citeauthoryear{Cernicharo et~al.}{2001}]{cernicharo2001}
\begin{barticle}
\bauthor{\bsnm{Cernicharo}, \binits{J.}},
\bauthor{\bsnm{Heras}, \binits{A.M.}},
\bauthor{\bsnm{Tielens}, \binits{A.G.G.M.}},
\bauthor{\bsnm{Pardo}, \binits{J.R.}},
\bauthor{\bsnm{Herpin}, \binits{F.}},
\bauthor{\bsnm{Gu{\'e}lin}, \binits{M.}},
\bauthor{\bsnm{Waters}, \binits{L.B.F.M.}}:
\batitle{Infrared {{Space Observatory}}'s {{Discovery}} of {{C4H2}}, {{C6H2}},
  and {{Benzene}} in {{CRL}} 618}.
\bjtitle{The Astrophysical Journal}
\bvolume{546}(\bissue{2}),
\bfpage{123}
(\byear{2001})
\doiurl{10.1086/318871}
\end{barticle}
\endbibitem

\bibitem[\protect\citeauthoryear{Li}{2020}]{li2020}
\begin{barticle}
\bauthor{\bsnm{Li}, \binits{A.}}:
\batitle{Spitzer's perspective of polycyclic aromatic hydrocarbons in
  galaxies}.
\bjtitle{Nature Astronomy}
\bvolume{4}(\bissue{4}),
\bfpage{339}--\blpage{351}
(\byear{2020})
\doiurl{10.1038/s41550-020-1051-1}
\end{barticle}
\endbibitem

\bibitem[\protect\citeauthoryear{Chown et~al.}{2024}]{chown2024}
\begin{barticle}
\bauthor{\bsnm{Chown}, \binits{R.}},
\bauthor{\bsnm{Sidhu}, \binits{A.}},
\bauthor{\bsnm{Peeters}, \binits{E.}},
\bauthor{\bsnm{Tielens}, \binits{A.G.G.M.}},
\bauthor{\bsnm{Cami}, \binits{J.}},
\bauthor{\bsnm{Bern{\'e}}, \binits{O.}},
\bauthor{\bsnm{Habart}, \binits{E.}},
\bauthor{\bsnm{Alarc{\'o}n}, \binits{F.}},
\bauthor{\bsnm{Canin}, \binits{A.}},
\bauthor{\bsnm{Schroetter}, \binits{I.}},
\bauthor{\bsnm{Trahin}, \binits{B.}},
\bauthor{\bsnm{Putte}, \binits{D.V.D.}},
\bauthor{\bsnm{Abergel}, \binits{A.}},
\bauthor{\bsnm{Bergin}, \binits{E.A.}},
\bauthor{\bsnm{{Bernard-Salas}}, \binits{J.}},
\bauthor{\bsnm{Boersma}, \binits{C.}},
\bauthor{\bsnm{Bron}, \binits{E.}},
\bauthor{\bsnm{Cuadrado}, \binits{S.}},
\bauthor{\bsnm{Dartois}, \binits{E.}},
\bauthor{\bsnm{Dicken}, \binits{D.}},
\bauthor{\bsnm{{El-Yajouri}}, \binits{M.}},
\bauthor{\bsnm{Fuente}, \binits{A.}},
\bauthor{\bsnm{Goicoechea}, \binits{J.R.}},
\bauthor{\bsnm{Gordon}, \binits{K.D.}},
\bauthor{\bsnm{Issa}, \binits{L.}},
\bauthor{\bsnm{Joblin}, \binits{C.}},
\bauthor{\bsnm{Kannavou}, \binits{O.}},
\bauthor{\bsnm{Khan}, \binits{B.}},
\bauthor{\bsnm{Lacinbala}, \binits{O.}},
\bauthor{\bsnm{Languignon}, \binits{D.}},
\bauthor{\bsnm{Gal}, \binits{R.L.}},
\bauthor{\bsnm{Maragkoudakis}, \binits{A.}},
\bauthor{\bsnm{Meshaka}, \binits{R.}},
\bauthor{\bsnm{Okada}, \binits{Y.}},
\bauthor{\bsnm{Onaka}, \binits{T.}},
\bauthor{\bsnm{Pasquini}, \binits{S.}},
\bauthor{\bsnm{Pound}, \binits{M.W.}},
\bauthor{\bsnm{Robberto}, \binits{M.}},
\bauthor{\bsnm{R{\"o}llig}, \binits{M.}},
\bauthor{\bsnm{Schefter}, \binits{B.}},
\bauthor{\bsnm{Schirmer}, \binits{T.}},
\bauthor{\bsnm{Vicente}, \binits{S.}},
\bauthor{\bsnm{Wolfire}, \binits{M.G.}},
\bauthor{\bsnm{Zannese}, \binits{M.}},
\bauthor{\bsnm{Aleman}, \binits{I.}},
\bauthor{\bsnm{Allamandola}, \binits{L.}},
\bauthor{\bsnm{Auchettl}, \binits{R.}},
\bauthor{\bsnm{Baratta}, \binits{G.A.}},
\bauthor{\bsnm{Bejaoui}, \binits{S.}},
\bauthor{\bsnm{Bera}, \binits{P.P.}},
\bauthor{\bsnm{Black}, \binits{J.H.}},
\bauthor{\bsnm{Boulanger}, \binits{F.}},
\bauthor{\bsnm{Bouwman}, \binits{J.}},
\bauthor{\bsnm{Brandl}, \binits{B.}},
\bauthor{\bsnm{Brechignac}, \binits{P.}},
\bauthor{\bsnm{Br{\"u}nken}, \binits{S.}},
\bauthor{\bsnm{Buragohain}, \binits{M.}},
\bauthor{\bsnm{Burkhardt}, \binits{A.}},
\bauthor{\bsnm{Candian}, \binits{A.}},
\bauthor{\bsnm{Cazaux}, \binits{S.}},
\bauthor{\bsnm{Cernicharo}, \binits{J.}},
\bauthor{\bsnm{Chabot}, \binits{M.}},
\bauthor{\bsnm{Chakraborty}, \binits{S.}},
\bauthor{\bsnm{Champion}, \binits{J.}},
\bauthor{\bsnm{Colgan}, \binits{S.W.J.}},
\bauthor{\bsnm{Cooke}, \binits{I.R.}},
\bauthor{\bsnm{Coutens}, \binits{A.}},
\bauthor{\bsnm{Cox}, \binits{N.L.J.}},
\bauthor{\bsnm{Demyk}, \binits{K.}},
\bauthor{\bsnm{Meyer}, \binits{J.D.}},
\bauthor{\bsnm{Foschino}, \binits{S.}},
\bauthor{\bsnm{{Garc{\'i}a-Lario}}, \binits{P.}},
\bauthor{\bsnm{Gavilan}, \binits{L.}},
\bauthor{\bsnm{Gerin}, \binits{M.}},
\bauthor{\bsnm{Gottlieb}, \binits{C.A.}},
\bauthor{\bsnm{Guillard}, \binits{P.}},
\bauthor{\bsnm{Gusdorf}, \binits{A.}},
\bauthor{\bsnm{Hartigan}, \binits{P.}},
\bauthor{\bsnm{He}, \binits{J.}},
\bauthor{\bsnm{Herbst}, \binits{E.}},
\bauthor{\bsnm{Hornekaer}, \binits{L.}},
\bauthor{\bsnm{J{\"a}ger}, \binits{C.}},
\bauthor{\bsnm{{Janot-Pacheco}}, \binits{E.}},
\bauthor{\bsnm{Kaufman}, \binits{M.}},
\bauthor{\bsnm{Kemper}, \binits{F.}},
\bauthor{\bsnm{Kendrew}, \binits{S.}},
\bauthor{\bsnm{Kirsanova}, \binits{M.S.}},
\bauthor{\bsnm{Klaassen}, \binits{P.}},
\bauthor{\bsnm{Kwok}, \binits{S.}},
\bauthor{\bsnm{Labiano}, \binits{{\'A}.}},
\bauthor{\bsnm{Lai}, \binits{T.S.-Y.}},
\bauthor{\bsnm{Lee}, \binits{T.J.}},
\bauthor{\bsnm{Lefloch}, \binits{B.}},
\bauthor{\bsnm{Petit}, \binits{F.L.}},
\bauthor{\bsnm{Li}, \binits{A.}},
\bauthor{\bsnm{Linz}, \binits{H.}},
\bauthor{\bsnm{Mackie}, \binits{C.J.}},
\bauthor{\bsnm{Madden}, \binits{S.C.}},
\bauthor{\bsnm{Mascetti}, \binits{J.}},
\bauthor{\bsnm{McGuire}, \binits{B.A.}},
\bauthor{\bsnm{Merino}, \binits{P.}},
\bauthor{\bsnm{Micelotta}, \binits{E.R.}},
\bauthor{\bsnm{Misselt}, \binits{K.}},
\bauthor{\bsnm{Morse}, \binits{J.A.}},
\bauthor{\bsnm{Mulas}, \binits{G.}},
\bauthor{\bsnm{Neelamkodan}, \binits{N.}},
\bauthor{\bsnm{Ohsawa}, \binits{R.}},
\bauthor{\bsnm{Omont}, \binits{A.}},
\bauthor{\bsnm{Paladini}, \binits{R.}},
\bauthor{\bsnm{Palumbo}, \binits{M.E.}},
\bauthor{\bsnm{Pathak}, \binits{A.}},
\bauthor{\bsnm{Pendleton}, \binits{Y.J.}},
\bauthor{\bsnm{Petrignani}, \binits{A.}},
\bauthor{\bsnm{Pino}, \binits{T.}},
\bauthor{\bsnm{Puga}, \binits{E.}},
\bauthor{\bsnm{Rangwala}, \binits{N.}},
\bauthor{\bsnm{Rapacioli}, \binits{M.}},
\bauthor{\bsnm{Ricca}, \binits{A.}},
\bauthor{\bsnm{{Roman-Duval}}, \binits{J.}},
\bauthor{\bsnm{Roser}, \binits{J.}},
\bauthor{\bsnm{Roueff}, \binits{E.}},
\bauthor{\bsnm{Rouill{\'e}}, \binits{G.}},
\bauthor{\bsnm{Salama}, \binits{F.}},
\bauthor{\bsnm{Sales}, \binits{D.A.}},
\bauthor{\bsnm{Sandstrom}, \binits{K.}},
\bauthor{\bsnm{Sarre}, \binits{P.}},
\bauthor{\bsnm{{Sciamma-O'Brien}}, \binits{E.}},
\bauthor{\bsnm{Sellgren}, \binits{K.}},
\bauthor{\bsnm{Shenoy}, \binits{S.S.}},
\bauthor{\bsnm{Teyssier}, \binits{D.}},
\bauthor{\bsnm{Thomas}, \binits{R.D.}},
\bauthor{\bsnm{Togi}, \binits{A.}},
\bauthor{\bsnm{Verstraete}, \binits{L.}},
\bauthor{\bsnm{Witt}, \binits{A.N.}},
\bauthor{\bsnm{Wootten}, \binits{A.}},
\bauthor{\bsnm{Zettergren}, \binits{H.}},
\bauthor{\bsnm{Zhang}, \binits{Y.}},
\bauthor{\bsnm{Zhang}, \binits{Z.E.}},
\bauthor{\bsnm{Zhen}, \binits{J.}}:
\batitle{{{PDRs4All}} - {{IV}}. {{An}} embarrassment of riches: {{Aromatic}}
  infrared bands in the {{Orion Bar}}}.
\bjtitle{Astronomy \& Astrophysics}
\bvolume{685},
\bfpage{75}
(\byear{2024})
\doiurl{10.1051/0004-6361/202346662}
\end{barticle}
\endbibitem

\bibitem[\protect\citeauthoryear{Peeters et~al.}{2002}]{peeters2002}
\begin{barticle}
\bauthor{\bsnm{Peeters}, \binits{E.}},
\bauthor{\bsnm{Hony}, \binits{S.}},
\bauthor{\bsnm{Kerckhoven}, \binits{C.V.}},
\bauthor{\bsnm{Tielens}, \binits{A.G.G.M.}},
\bauthor{\bsnm{Allamandola}, \binits{L.J.}},
\bauthor{\bsnm{Hudgins}, \binits{D.M.}},
\bauthor{\bsnm{Bauschlicher}, \binits{C.W.}}:
\batitle{The rich 6 to 9 m spectrum of interstellar {{PAHs}}}.
\bjtitle{Astronomy \& Astrophysics}
\bvolume{390}(\bissue{3}),
\bfpage{1089}--\blpage{1113}
(\byear{2002})
\doiurl{10.1051/0004-6361:20020773}
\end{barticle}
\endbibitem

\bibitem[\protect\citeauthoryear{Galliano et~al.}{2008}]{galliano2008}
\begin{barticle}
\bauthor{\bsnm{Galliano}, \binits{F.}},
\bauthor{\bsnm{Madden}, \binits{S.C.}},
\bauthor{\bsnm{Tielens}, \binits{A.G.G.M.}},
\bauthor{\bsnm{Peeters}, \binits{E.}},
\bauthor{\bsnm{Jones}, \binits{A.P.}}:
\batitle{Variations of the {{Mid-IR Aromatic Features}} inside and among
  {{Galaxies}}}.
\bjtitle{The Astrophysical Journal}
\bvolume{679}(\bissue{1}),
\bfpage{310}--\blpage{345}
(\byear{2008})
\doiurl{10.1086/587051}
\end{barticle}
\endbibitem

\bibitem[\protect\citeauthoryear{Peeters}{2011}]{peeters2011}
\begin{barticle}
\bauthor{\bsnm{Peeters}, \binits{E.}}:
\batitle{Astronomical observations of the {{PAH}} emission bands}.
\bjtitle{European Astronomical Society Publications Series}
\bvolume{46},
\bfpage{13}--\blpage{27}
(\byear{2011})
\doiurl{10.1051/eas/1146002}
\end{barticle}
\endbibitem

\bibitem[\protect\citeauthoryear{Sabbah et~al.}{2017}]{sabbah2017}
\begin{barticle}
\bauthor{\bsnm{Sabbah}, \binits{H.}},
\bauthor{\bsnm{Bonnamy}, \binits{A.}},
\bauthor{\bsnm{Papanastasiou}, \binits{D.}},
\bauthor{\bsnm{Cernicharo}, \binits{J.}},
\bauthor{\bsnm{{Mart{\'i}n-Gago}}, \binits{J.-A.}},
\bauthor{\bsnm{Joblin}, \binits{C.}}:
\batitle{Identification of {{PAH Isomeric Structure}} in {{Cosmic Dust
  Analogs}}: {{The AROMA Setup}}}.
\bjtitle{The Astrophysical Journal}
\bvolume{843}(\bissue{1}),
\bfpage{34}
(\byear{2017})
\doiurl{10.3847/1538-4357/aa73dd}
\end{barticle}
\endbibitem

\bibitem[\protect\citeauthoryear{Lecasble et~al.}{2022}]{lecasble2022}
\begin{barticle}
\bauthor{\bsnm{Lecasble}, \binits{M.}},
\bauthor{\bsnm{Remusat}, \binits{L.}},
\bauthor{\bsnm{Viennet}, \binits{J.-C.}},
\bauthor{\bsnm{Laurent}, \binits{B.}},
\bauthor{\bsnm{Bernard}, \binits{S.}}:
\batitle{Polycyclic aromatic hydrocarbons in carbonaceous chondrites can be
  used as tracers of both pre-accretion and secondary processes}.
\bjtitle{Geochimica et Cosmochimica Acta}
\bvolume{335},
\bfpage{243}--\blpage{255}
(\byear{2022})
\doiurl{10.1016/j.gca.2022.08.039}
\end{barticle}
\endbibitem

\bibitem[\protect\citeauthoryear{Clemett et~al.}{2010}]{clemett2010}
\begin{barticle}
\bauthor{\bsnm{Clemett}, \binits{S.J.}},
\bauthor{\bsnm{Sandford}, \binits{S.A.}},
\bauthor{\bsnm{{Nakamura-Messenger}}, \binits{K.}},
\bauthor{\bsnm{H{\"o}rz}, \binits{F.}},
\bauthor{\bsnm{McKAY}, \binits{D.S.}}:
\batitle{Complex aromatic hydrocarbons in {{Stardust}} samples collected from
  comet {{81P}}/{{Wild}} 2}.
\bjtitle{Meteoritics \& Planetary Science}
\bvolume{45}(\bissue{5}),
\bfpage{701}--\blpage{722}
(\byear{2010})
\doiurl{10.1111/j.1945-5100.2010.01062.x}
\end{barticle}
\endbibitem

\bibitem[\protect\citeauthoryear{Aponte et~al.}{2023}]{aponte2023}
\begin{barticle}
\bauthor{\bsnm{Aponte}, \binits{J.C.}},
\bauthor{\bsnm{Dworkin}, \binits{J.P.}},
\bauthor{\bsnm{Glavin}, \binits{D.P.}},
\bauthor{\bsnm{Elsila}, \binits{J.E.}},
\bauthor{\bsnm{Parker}, \binits{E.T.}},
\bauthor{\bsnm{McLain}, \binits{H.L.}},
\bauthor{\bsnm{Naraoka}, \binits{H.}},
\bauthor{\bsnm{Okazaki}, \binits{R.}},
\bauthor{\bsnm{Takano}, \binits{Y.}},
\bauthor{\bsnm{Tachibana}, \binits{S.}},
\bauthor{\bsnm{Dong}, \binits{G.}},
\bauthor{\bsnm{Zeichner}, \binits{S.S.}},
\bauthor{\bsnm{Eiler}, \binits{J.M.}},
\bauthor{\bsnm{Yurimoto}, \binits{H.}},
\bauthor{\bsnm{Nakamura}, \binits{T.}},
\bauthor{\bsnm{Yabuta}, \binits{H.}},
\bauthor{\bsnm{Terui}, \binits{F.}},
\bauthor{\bsnm{Noguchi}, \binits{T.}},
\bauthor{\bsnm{Sakamoto}, \binits{K.}},
\bauthor{\bsnm{Yada}, \binits{T.}},
\bauthor{\bsnm{Nishimura}, \binits{M.}},
\bauthor{\bsnm{Nakato}, \binits{A.}},
\bauthor{\bsnm{Miyazaki}, \binits{A.}},
\bauthor{\bsnm{Yogata}, \binits{K.}},
\bauthor{\bsnm{Abe}, \binits{M.}},
\bauthor{\bsnm{Okada}, \binits{T.}},
\bauthor{\bsnm{Usui}, \binits{T.}},
\bauthor{\bsnm{Yoshikawa}, \binits{M.}},
\bauthor{\bsnm{Saiki}, \binits{T.}},
\bauthor{\bsnm{Tanaka}, \binits{S.}},
\bauthor{\bsnm{Nakazawa}, \binits{S.}},
\bauthor{\bsnm{Tsuda}, \binits{Y.}},
\bauthor{\bsnm{Watanabe}, \binits{S.-i.}},
\bauthor{\bsnm{{The Hayabusa2-initial-analysis SOM team}}},
\bauthor{\bsnm{{The Hayabusa2-initial-analysis core team}}}:
\batitle{{{PAHs}}, hydrocarbons, and dimethylsulfides in {{Asteroid Ryugu}}
  samples {{A0106}} and {{C0107}} and the {{Orgueil}} ({{CI1}}) meteorite}.
\bjtitle{Earth, Planets and Space}
\bvolume{75}(\bissue{1}),
\bfpage{28}
(\byear{2023})
\doiurl{10.1186/s40623-022-01758-4}
\end{barticle}
\endbibitem

\bibitem[\protect\citeauthoryear{Zeichner et~al.}{2023}]{zeichner2023}
\begin{barticle}
\bauthor{\bsnm{Zeichner}, \binits{S.S.}},
\bauthor{\bsnm{Aponte}, \binits{J.C.}},
\bauthor{\bsnm{Bhattacharjee}, \binits{S.}},
\bauthor{\bsnm{Dong}, \binits{G.}},
\bauthor{\bsnm{Hofmann}, \binits{A.E.}},
\bauthor{\bsnm{Dworkin}, \binits{J.P.}},
\bauthor{\bsnm{Glavin}, \binits{D.P.}},
\bauthor{\bsnm{Elsila}, \binits{J.E.}},
\bauthor{\bsnm{Graham}, \binits{H.V.}},
\bauthor{\bsnm{Naraoka}, \binits{H.}},
\bauthor{\bsnm{Takano}, \binits{Y.}},
\bauthor{\bsnm{Tachibana}, \binits{S.}},
\bauthor{\bsnm{Karp}, \binits{A.T.}},
\bauthor{\bsnm{Grice}, \binits{K.}},
\bauthor{\bsnm{Holman}, \binits{A.I.}},
\bauthor{\bsnm{Freeman}, \binits{K.H.}},
\bauthor{\bsnm{Yurimoto}, \binits{H.}},
\bauthor{\bsnm{Nakamura}, \binits{T.}},
\bauthor{\bsnm{Noguchi}, \binits{T.}},
\bauthor{\bsnm{Okazaki}, \binits{R.}},
\bauthor{\bsnm{Yabuta}, \binits{H.}},
\bauthor{\bsnm{Sakamoto}, \binits{K.}},
\bauthor{\bsnm{Yada}, \binits{T.}},
\bauthor{\bsnm{Nishimura}, \binits{M.}},
\bauthor{\bsnm{Nakato}, \binits{A.}},
\bauthor{\bsnm{Miyazaki}, \binits{A.}},
\bauthor{\bsnm{Yogata}, \binits{K.}},
\bauthor{\bsnm{Abe}, \binits{M.}},
\bauthor{\bsnm{Okada}, \binits{T.}},
\bauthor{\bsnm{Usui}, \binits{T.}},
\bauthor{\bsnm{Yoshikawa}, \binits{M.}},
\bauthor{\bsnm{Saiki}, \binits{T.}},
\bauthor{\bsnm{Tanaka}, \binits{S.}},
\bauthor{\bsnm{Terui}, \binits{F.}},
\bauthor{\bsnm{Nakazawa}, \binits{S.}},
\bauthor{\bsnm{Watanabe}, \binits{S.-i.}},
\bauthor{\bsnm{Tsuda}, \binits{Y.}},
\bauthor{\bsnm{Hamase}, \binits{K.}},
\bauthor{\bsnm{Fukushima}, \binits{K.}},
\bauthor{\bsnm{Aoki}, \binits{D.}},
\bauthor{\bsnm{Hashiguchi}, \binits{M.}},
\bauthor{\bsnm{Mita}, \binits{H.}},
\bauthor{\bsnm{Chikaraishi}, \binits{Y.}},
\bauthor{\bsnm{Ohkouchi}, \binits{N.}},
\bauthor{\bsnm{Ogawa}, \binits{N.O.}},
\bauthor{\bsnm{Sakai}, \binits{S.}},
\bauthor{\bsnm{Parker}, \binits{E.T.}},
\bauthor{\bsnm{McLain}, \binits{H.L.}},
\bauthor{\bsnm{{Orthous-Daunay}}, \binits{F.-R.}},
\bauthor{\bsnm{Vuitton}, \binits{V.}},
\bauthor{\bsnm{Wolters}, \binits{C.}},
\bauthor{\bsnm{{Schmitt-Kopplin}}, \binits{P.}},
\bauthor{\bsnm{Hertkorn}, \binits{N.}},
\bauthor{\bsnm{Thissen}, \binits{R.}},
\bauthor{\bsnm{Ruf}, \binits{A.}},
\bauthor{\bsnm{Isa}, \binits{J.}},
\bauthor{\bsnm{Oba}, \binits{Y.}},
\bauthor{\bsnm{Koga}, \binits{T.}},
\bauthor{\bsnm{Yoshimura}, \binits{T.}},
\bauthor{\bsnm{Araoka}, \binits{D.}},
\bauthor{\bsnm{Sugahara}, \binits{H.}},
\bauthor{\bsnm{Furusho}, \binits{A.}},
\bauthor{\bsnm{Furukawa}, \binits{Y.}},
\bauthor{\bsnm{Aoki}, \binits{J.}},
\bauthor{\bsnm{Kano}, \binits{K.}},
\bauthor{\bsnm{Nomura}, \binits{S.-i.M.}},
\bauthor{\bsnm{Sasaki}, \binits{K.}},
\bauthor{\bsnm{Sato}, \binits{H.}},
\bauthor{\bsnm{Yoshikawa}, \binits{T.}},
\bauthor{\bsnm{Tanaka}, \binits{S.}},
\bauthor{\bsnm{Morita}, \binits{M.}},
\bauthor{\bsnm{Onose}, \binits{M.}},
\bauthor{\bsnm{Kabashima}, \binits{F.}},
\bauthor{\bsnm{Fujishima}, \binits{K.}},
\bauthor{\bsnm{Yamazaki}, \binits{T.}},
\bauthor{\bsnm{Kimura}, \binits{Y.}},
\bauthor{\bsnm{Eiler}, \binits{J.M.}}:
\batitle{Polycyclic aromatic hydrocarbons in samples of {{Ryugu}} formed in the
  interstellar medium}.
\bjtitle{Science}
\bvolume{382}(\bissue{6677}),
\bfpage{1411}--\blpage{1416}
(\byear{2023})
\doiurl{10.1126/science.adg6304}
\end{barticle}
\endbibitem

\bibitem[\protect\citeauthoryear{McGuire et~al.}{2021}]{mcguire2021}
\begin{barticle}
\bauthor{\bsnm{McGuire}, \binits{B.A.}},
\bauthor{\bsnm{Loomis}, \binits{R.A.}},
\bauthor{\bsnm{Burkhardt}, \binits{A.M.}},
\bauthor{\bsnm{Lee}, \binits{K.L.K.}},
\bauthor{\bsnm{Shingledecker}, \binits{C.N.}},
\bauthor{\bsnm{Charnley}, \binits{S.B.}},
\bauthor{\bsnm{Cooke}, \binits{I.R.}},
\bauthor{\bsnm{Cordiner}, \binits{M.A.}},
\bauthor{\bsnm{Herbst}, \binits{E.}},
\bauthor{\bsnm{Kalenskii}, \binits{S.}},
\bauthor{\bsnm{Siebert}, \binits{M.A.}},
\bauthor{\bsnm{Willis}, \binits{E.R.}},
\bauthor{\bsnm{Xue}, \binits{C.}},
\bauthor{\bsnm{Remijan}, \binits{A.J.}},
\bauthor{\bsnm{McCarthy}, \binits{M.C.}}:
\batitle{Detection of two interstellar polycyclic aromatic hydrocarbons via
  spectral matched filtering}.
\bjtitle{Science}
\bvolume{371}(\bissue{6535}),
\bfpage{1265}--\blpage{1269}
(\byear{2021})
\doiurl{10.1126/science.abb7535} .
\bcomment{Chap. Report}
\end{barticle}
\endbibitem

\bibitem[\protect\citeauthoryear{Burkhardt et~al.}{2021}]{burkhardt2021}
\begin{barticle}
\bauthor{\bsnm{Burkhardt}, \binits{A.M.}},
\bauthor{\bsnm{Loomis}, \binits{R.A.}},
\bauthor{\bsnm{Shingledecker}, \binits{C.N.}},
\bauthor{\bsnm{Lee}, \binits{K.L.K.}},
\bauthor{\bsnm{Remijan}, \binits{A.J.}},
\bauthor{\bsnm{McCarthy}, \binits{M.C.}},
\bauthor{\bsnm{McGuire}, \binits{B.A.}}:
\batitle{Ubiquitous aromatic carbon chemistry at the earliest stages of star
  formation}.
\bjtitle{Nature Astronomy}
\bvolume{5}(\bissue{2}),
\bfpage{181}--\blpage{187}
(\byear{2021})
\doiurl{10.1038/s41550-020-01253-4}
\end{barticle}
\endbibitem

\bibitem[\protect\citeauthoryear{Cernicharo et~al.}{2021}]{cernicharo2021}
\begin{barticle}
\bauthor{\bsnm{Cernicharo}, \binits{J.}},
\bauthor{\bsnm{Ag{\'u}ndez}, \binits{M.}},
\bauthor{\bsnm{Cabezas}, \binits{C.}},
\bauthor{\bsnm{Tercero}, \binits{B.}},
\bauthor{\bsnm{Marcelino}, \binits{N.}},
\bauthor{\bsnm{Pardo}, \binits{J.R.}},
\bauthor{\bsnm{{de Vicente}}, \binits{P.}}:
\batitle{Pure hydrocarbon cycles in {{TMC-1}}: {{Discovery}} of ethynyl
  cyclopropenylidene, cyclopentadiene and indene.}
\bjtitle{Astronomy and astrophysics}
\bvolume{649},
\bfpage{15}
(\byear{2021})
\doiurl{10.1051/0004-6361/202141156}
\end{barticle}
\endbibitem

\bibitem[\protect\citeauthoryear{Sita et~al.}{2022}]{sita2022}
\begin{barticle}
\bauthor{\bsnm{Sita}, \binits{M.L.}},
\bauthor{\bsnm{Changala}, \binits{P.B.}},
\bauthor{\bsnm{Xue}, \binits{C.}},
\bauthor{\bsnm{Burkhardt}, \binits{A.M.}},
\bauthor{\bsnm{Shingledecker}, \binits{C.N.}},
\bauthor{\bsnm{Lee}, \binits{K.L.K.}},
\bauthor{\bsnm{Loomis}, \binits{R.A.}},
\bauthor{\bsnm{Momjian}, \binits{E.}},
\bauthor{\bsnm{Siebert}, \binits{M.A.}},
\bauthor{\bsnm{Gupta}, \binits{D.}},
\bauthor{\bsnm{Herbst}, \binits{E.}},
\bauthor{\bsnm{Remijan}, \binits{A.J.}},
\bauthor{\bsnm{McCarthy}, \binits{M.C.}},
\bauthor{\bsnm{Cooke}, \binits{I.R.}},
\bauthor{\bsnm{McGuire}, \binits{B.A.}}:
\batitle{Discovery of {{Interstellar}} 2-{{Cyanoindene}} (2-{{C9H7CN}}) in
  {{GOTHAM Observations}} of {{TMC-1}}}.
\bjtitle{The Astrophysical Journal Letters}
\bvolume{938}(\bissue{2}),
\bfpage{12}
(\byear{2022})
\doiurl{10.3847/2041-8213/ac92f4}
\end{barticle}
\endbibitem

\bibitem[\protect\citeauthoryear{Wenzel et~al.}{in review}]{wenzelinreview}
\begin{botherref}
\oauthor{\bsnm{Wenzel}, \binits{G.}},
\oauthor{\bsnm{Cooke}, \binits{I.R.}},
\oauthor{\bsnm{Changala}, \binits{P.B.}},
\oauthor{\bsnm{Bergin}, \binits{E.A.}},
\oauthor{\bsnm{Zhang}, \binits{S.}},
\oauthor{\bsnm{Burkhardt}, \binits{A.M.}},
\oauthor{\bsnm{Byrne}, \binits{A.N.}},
\oauthor{\bsnm{Charnley}, \binits{S.B.}},
\oauthor{\bsnm{Cordiner}, \binits{M.A.}},
\oauthor{\bsnm{Duffy}, \binits{M.}},
\oauthor{\bsnm{Fried}, \binits{Z.T.P.}},
\oauthor{\bsnm{Gupta}, \binits{H.}},
\oauthor{\bsnm{Holdren}, \binits{M.S.}},
\oauthor{\bsnm{Lipnicky}, \binits{A.}},
\oauthor{\bsnm{Loomis}, \binits{R.A.}},
\oauthor{\bsnm{Toru~Shay}, \binits{H.}},
\oauthor{\bsnm{Shingledecker}, \binits{C.N.}},
\oauthor{\bsnm{Siebert}, \binits{M.A.}},
\oauthor{\bsnm{Stewart}, \binits{D.A.}},
\oauthor{\bsnm{Willis}, \binits{R.H.J.}},
\oauthor{\bsnm{Xue}, \binits{C.}},
\oauthor{\bsnm{Remijan}, \binits{A.J.}},
\oauthor{\bsnm{Wendlandt}, \binits{A.E.}},
\oauthor{\bsnm{McCarthy}, \binits{M.C.}},
\oauthor{\bsnm{McGuire}, \binits{B.A.}}:
Discovery of interstellar 1-cyanopyrene: A four-ring polycyclic aromatic
  hydrocarbon in {{TMC-1}}
(in review)
\end{botherref}
\endbibitem

\bibitem[\protect\citeauthoryear{Montillaud et~al.}{2013}]{montillaud2013}
\begin{barticle}
\bauthor{\bsnm{Montillaud}, \binits{J.}},
\bauthor{\bsnm{Joblin}, \binits{C.}},
\bauthor{\bsnm{Toublanc}, \binits{D.}}:
\batitle{Evolution of polycyclic aromatic hydrocarbons in photodissociation
  regions - {{Hydrogenation}} and charge states}.
\bjtitle{Astronomy \& Astrophysics}
\bvolume{552},
\bfpage{15}
(\byear{2013})
\doiurl{10.1051/0004-6361/201220757}
\end{barticle}
\endbibitem

\bibitem[\protect\citeauthoryear{Messinger et~al.}{2020}]{messinger2020}
\begin{barticle}
\bauthor{\bsnm{Messinger}, \binits{J.P.}},
\bauthor{\bsnm{Gupta}, \binits{D.}},
\bauthor{\bsnm{Cooke}, \binits{I.R.}},
\bauthor{\bsnm{Okumura}, \binits{M.}},
\bauthor{\bsnm{Sims}, \binits{I.R.}}:
\batitle{Rate {{Constants}} of the {{CN}} + {{Toluene Reaction}} from 15 to 294
  {{K}} and {{Interstellar Implications}}}.
\bjtitle{The Journal of Physical Chemistry A}
\bvolume{124}(\bissue{39}),
\bfpage{7950}--\blpage{7958}
(\byear{2020})
\doiurl{10.1021/acs.jpca.0c06900}
\end{barticle}
\endbibitem

\bibitem[\protect\citeauthoryear{Cooke et~al.}{2020}]{cooke2020}
\begin{barticle}
\bauthor{\bsnm{Cooke}, \binits{I.R.}},
\bauthor{\bsnm{Gupta}, \binits{D.}},
\bauthor{\bsnm{Messinger}, \binits{J.P.}},
\bauthor{\bsnm{Sims}, \binits{I.R.}}:
\batitle{Benzonitrile as a {{Proxy}} for {{Benzene}} in the {{Cold ISM}}:
  {{Low-temperature Rate Coefficients}} for {{CN}} + {{C6H6}}}.
\bjtitle{The Astrophysical Journal Letters}
\bvolume{891}(\bissue{2}),
\bfpage{41}
(\byear{2020})
\doiurl{10.3847/2041-8213/ab7a9c}
\end{barticle}
\endbibitem

\bibitem[\protect\citeauthoryear{Balucani et~al.}{2000}]{balucani2000}
\begin{barticle}
\bauthor{\bsnm{Balucani}, \binits{N.}},
\bauthor{\bsnm{Asvany}, \binits{O.}},
\bauthor{\bsnm{Huang}, \binits{L.C.L.}},
\bauthor{\bsnm{Lee}, \binits{Y.T.}},
\bauthor{\bsnm{Kaiser}, \binits{R.I.}},
\bauthor{\bsnm{Osamura}, \binits{Y.}},
\bauthor{\bsnm{Bettinger}, \binits{H.F.}}:
\batitle{Formation of {{Nitriles}} in the {{Interstellar Medium}} via
  {{Reactions ofCyano Radicals}},{{CN}}({{X 2$\Sigma$}}+), {{withUnsaturated
  Hydrocarbons}}}.
\bjtitle{The Astrophysical Journal}
\bvolume{545}(\bissue{2}),
\bfpage{892}
(\byear{2000})
\doiurl{10.1086/317848}
\end{barticle}
\endbibitem

\bibitem[\protect\citeauthoryear{McGuire et~al.}{2020}]{mcguire2020}
\begin{barticle}
\bauthor{\bsnm{McGuire}, \binits{B.A.}},
\bauthor{\bsnm{Burkhardt}, \binits{A.M.}},
\bauthor{\bsnm{Loomis}, \binits{R.A.}},
\bauthor{\bsnm{Shingledecker}, \binits{C.N.}},
\bauthor{\bsnm{Lee}, \binits{K.L.K.}},
\bauthor{\bsnm{Charnley}, \binits{S.B.}},
\bauthor{\bsnm{Cordiner}, \binits{M.A.}},
\bauthor{\bsnm{Herbst}, \binits{E.}},
\bauthor{\bsnm{Kalenskii}, \binits{S.}},
\bauthor{\bsnm{Momjian}, \binits{E.}},
\bauthor{\bsnm{Willis}, \binits{E.R.}},
\bauthor{\bsnm{Xue}, \binits{C.}},
\bauthor{\bsnm{Remijan}, \binits{A.J.}},
\bauthor{\bsnm{McCarthy}, \binits{M.C.}}:
\batitle{Early {{Science}} from {{GOTHAM}}: {{Project Overview}}, {{Methods}},
  and the {{Detection}} of {{Interstellar Propargyl Cyanide}} ({{HCCCH2CN}}) in
  {{TMC-1}}}.
\bjtitle{The Astrophysical Journal Letters}
\bvolume{900}(\bissue{1}),
\bfpage{10}
(\byear{2020})
\doiurl{10.3847/2041-8213/aba632}
\end{barticle}
\endbibitem

\bibitem[\protect\citeauthoryear{Cooke et~al.}{2023}]{cooke2023}
\begin{barticle}
\bauthor{\bsnm{Cooke}, \binits{I.R.}},
\bauthor{\bsnm{Xue}, \binits{C.}},
\bauthor{\bsnm{Changala}, \binits{P.B.}},
\bauthor{\bsnm{Shay}, \binits{H.T.}},
\bauthor{\bsnm{Byrne}, \binits{A.N.}},
\bauthor{\bsnm{Tang}, \binits{Q.Y.}},
\bauthor{\bsnm{Fried}, \binits{Z.T.P.}},
\bauthor{\bsnm{Lee}, \binits{K.L.K.}},
\bauthor{\bsnm{Loomis}, \binits{R.A.}},
\bauthor{\bsnm{Lamberts}, \binits{T.}},
\bauthor{\bsnm{Remijan}, \binits{A.}},
\bauthor{\bsnm{Burkhardt}, \binits{A.M.}},
\bauthor{\bsnm{Herbst}, \binits{E.}},
\bauthor{\bsnm{McCarthy}, \binits{M.C.}},
\bauthor{\bsnm{McGuire}, \binits{B.A.}}:
\batitle{Detection of {{Interstellar E-1-cyano-1}},3-butadiene in {{GOTHAM
  Observations}} of {{TMC-1}}}.
\bjtitle{The Astrophysical Journal}
\bvolume{948}(\bissue{2}),
\bfpage{133}
(\byear{2023})
\doiurl{10.3847/1538-4357/acc584}
\end{barticle}
\endbibitem

\bibitem[\protect\citeauthoryear{Loomis et~al.}{2021}]{loomis2021}
\begin{barticle}
\bauthor{\bsnm{Loomis}, \binits{R.A.}},
\bauthor{\bsnm{Burkhardt}, \binits{A.M.}},
\bauthor{\bsnm{Shingledecker}, \binits{C.N.}},
\bauthor{\bsnm{Charnley}, \binits{S.B.}},
\bauthor{\bsnm{Cordiner}, \binits{M.A.}},
\bauthor{\bsnm{Herbst}, \binits{E.}},
\bauthor{\bsnm{Kalenskii}, \binits{S.}},
\bauthor{\bsnm{Lee}, \binits{K.L.K.}},
\bauthor{\bsnm{Willis}, \binits{E.R.}},
\bauthor{\bsnm{Xue}, \binits{C.}},
\bauthor{\bsnm{Remijan}, \binits{A.J.}},
\bauthor{\bsnm{McCarthy}, \binits{M.C.}},
\bauthor{\bsnm{McGuire}, \binits{B.A.}}:
\batitle{An investigation of spectral line stacking techniques and application
  to the detection of {{HC11N}}}.
\bjtitle{Nature Astronomy}
\bvolume{5}(\bissue{2}),
\bfpage{188}--\blpage{196}
(\byear{2021})
\doiurl{10.1038/s41550-020-01261-4}
\end{barticle}
\endbibitem

\bibitem[\protect\citeauthoryear{Ruaud et~al.}{2016}]{ruaud2016}
\begin{barticle}
\bauthor{\bsnm{Ruaud}, \binits{M.}},
\bauthor{\bsnm{Wakelam}, \binits{V.}},
\bauthor{\bsnm{Hersant}, \binits{F.}}:
\batitle{Gas and grain chemical composition in cold cores as predicted by the
  {{Nautilus}} three-phase model}.
\bjtitle{Monthly Notices of the Royal Astronomical Society}
\bvolume{459}(\bissue{4}),
\bfpage{3756}--\blpage{3767}
(\byear{2016})
\doiurl{10.1093/mnras/stw887}
\end{barticle}
\endbibitem

\bibitem[\protect\citeauthoryear{Loison et~al.}{2014}]{loison2014}
\begin{barticle}
\bauthor{\bsnm{Loison}, \binits{J.-C.}},
\bauthor{\bsnm{Wakelam}, \binits{V.}},
\bauthor{\bsnm{Hickson}, \binits{K.M.}},
\bauthor{\bsnm{Bergeat}, \binits{A.}},
\bauthor{\bsnm{Mereau}, \binits{R.}}:
\batitle{The gas-phase chemistry of carbon chains in dark cloud chemical
  models}.
\bjtitle{Monthly Notices of the Royal Astronomical Society}
\bvolume{437}(\bissue{1}),
\bfpage{930}--\blpage{945}
(\byear{2014})
\doiurl{10.1093/mnras/stt1956}
\end{barticle}
\endbibitem

\bibitem[\protect\citeauthoryear{McGuire et~al.}{2018}]{mcguire2018}
\begin{barticle}
\bauthor{\bsnm{McGuire}, \binits{B.A.}},
\bauthor{\bsnm{Burkhardt}, \binits{A.M.}},
\bauthor{\bsnm{Kalenskii}, \binits{S.}},
\bauthor{\bsnm{Shingledecker}, \binits{C.N.}},
\bauthor{\bsnm{Remijan}, \binits{A.J.}},
\bauthor{\bsnm{Herbst}, \binits{E.}},
\bauthor{\bsnm{McCarthy}, \binits{M.C.}}:
\batitle{Detection of the aromatic molecule benzonitrile (c-c$_6$h$_5$cn) in
  the interstellar medium}.
\bjtitle{Science}
\bvolume{359}(\bissue{6372}),
\bfpage{202}--\blpage{205}
(\byear{2018})
\doiurl{10.1126/science.aao4890}
\end{barticle}
\endbibitem

\bibitem[\protect\citeauthoryear{Micelotta et~al.}{2010}]{micelotta2010}
\begin{barticle}
\bauthor{\bsnm{Micelotta}, \binits{E.R.}},
\bauthor{\bsnm{Jones}, \binits{A.P.}},
\bauthor{\bsnm{Tielens}, \binits{A.G.G.M.}}:
\batitle{Polycyclic aromatic hydrocarbon processing in interstellar shocks}.
\bjtitle{Astronomy \& Astrophysics}
\bvolume{510},
\bfpage{36}
(\byear{2010})
\doiurl{10.1051/0004-6361/200911682}
\end{barticle}
\endbibitem

\bibitem[\protect\citeauthoryear{Micelotta et~al.}{2011}]{micelotta2011}
\begin{barticle}
\bauthor{\bsnm{Micelotta}, \binits{E.R.}},
\bauthor{\bsnm{Jones}, \binits{A.P.}},
\bauthor{\bsnm{Tielens}, \binits{A.G.G.M.}}:
\batitle{Polycyclic aromatic hydrocarbon processing by cosmic rays}.
\bjtitle{Astronomy \& Astrophysics}
\bvolume{526},
\bfpage{52}
(\byear{2011})
\doiurl{10.1051/0004-6361/201015741}
\end{barticle}
\endbibitem

\bibitem[\protect\citeauthoryear{McNaughton et~al.}{2018}]{mcnaughton2018}
\begin{barticle}
\bauthor{\bsnm{McNaughton}, \binits{D.}},
\bauthor{\bsnm{Jahn}, \binits{M.K.}},
\bauthor{\bsnm{Travers}, \binits{M.J.}},
\bauthor{\bsnm{Wachsmuth}, \binits{D.}},
\bauthor{\bsnm{Godfrey}, \binits{P.D.}},
\bauthor{\bsnm{Grabow}, \binits{J.-U.}}:
\batitle{Laboratory rotational spectroscopy of cyano substituted polycyclic
  aromatic hydrocarbons}.
\bjtitle{Monthly Notices of the Royal Astronomical Society}
\bvolume{476}(\bissue{4}),
\bfpage{5268}--\blpage{5273}
(\byear{2018})
\doiurl{10.1093/mnras/sty557}
\end{barticle}
\endbibitem

\bibitem[\protect\citeauthoryear{Reizer et~al.}{2022}]{reizer2022}
\begin{barticle}
\bauthor{\bsnm{Reizer}, \binits{E.}},
\bauthor{\bsnm{Viskolcz}, \binits{B.}},
\bauthor{\bsnm{Fiser}, \binits{B.}}:
\batitle{Formation and growth mechanisms of polycyclic aromatic hydrocarbons:
  {{A}} mini-review}.
\bjtitle{Chemosphere}
\bvolume{291},
\bfpage{132793}
(\byear{2022})
\doiurl{10.1016/j.chemosphere.2021.132793}
\end{barticle}
\endbibitem

\bibitem[\protect\citeauthoryear{Zhao et~al.}{2018}]{zhao2018}
\begin{barticle}
\bauthor{\bsnm{Zhao}, \binits{L.}},
\bauthor{\bsnm{Kaiser}, \binits{R.I.}},
\bauthor{\bsnm{Xu}, \binits{B.}},
\bauthor{\bsnm{Ablikim}, \binits{U.}},
\bauthor{\bsnm{Ahmed}, \binits{M.}},
\bauthor{\bsnm{Joshi}, \binits{D.}},
\bauthor{\bsnm{Veber}, \binits{G.}},
\bauthor{\bsnm{Fischer}, \binits{F.R.}},
\bauthor{\bsnm{Mebel}, \binits{A.M.}}:
\batitle{Pyrene synthesis in circumstellar envelopes and its role in the
  formation of {{2D}} nanostructures}.
\bjtitle{Nature Astronomy}
\bvolume{2}(\bissue{5}),
\bfpage{413}--\blpage{419}
(\byear{2018})
\doiurl{10.1038/s41550-018-0399-y}
\end{barticle}
\endbibitem

\bibitem[\protect\citeauthoryear{Kaiser and Hansen}{2021}]{kaiser2021}
\begin{barticle}
\bauthor{\bsnm{Kaiser}, \binits{R.I.}},
\bauthor{\bsnm{Hansen}, \binits{N.}}:
\batitle{An {{Aromatic Universe}}--{{A Physical Chemistry Perspective}}}.
\bjtitle{The Journal of Physical Chemistry A}
\bvolume{125}(\bissue{18}),
\bfpage{3826}--\blpage{3840}
(\byear{2021})
\doiurl{10.1021/acs.jpca.1c00606}
\end{barticle}
\endbibitem

\bibitem[\protect\citeauthoryear{Zhao et~al.}{2018}]{zhao2018b}
\begin{barticle}
\bauthor{\bsnm{Zhao}, \binits{L.}},
\bauthor{\bsnm{Kaiser}, \binits{R.I.}},
\bauthor{\bsnm{Xu}, \binits{B.}},
\bauthor{\bsnm{Ablikim}, \binits{U.}},
\bauthor{\bsnm{Ahmed}, \binits{M.}},
\bauthor{\bsnm{Evseev}, \binits{M.M.}},
\bauthor{\bsnm{Bashkirov}, \binits{E.K.}},
\bauthor{\bsnm{Azyazov}, \binits{V.N.}},
\bauthor{\bsnm{Mebel}, \binits{A.M.}}:
\batitle{Low-temperature formation of polycyclic aromatic hydrocarbons in
  {{Titan}}'s atmosphere}.
\bjtitle{Nature Astronomy}
\bvolume{2}(\bissue{12}),
\bfpage{973}--\blpage{979}
(\byear{2018})
\doiurl{10.1038/s41550-018-0585-y}
\end{barticle}
\endbibitem

\bibitem[\protect\citeauthoryear{Goettl et~al.}{2024}]{goettl2024a}
\begin{barticle}
\bauthor{\bsnm{Goettl}, \binits{S.J.}},
\bauthor{\bsnm{Yang}, \binits{Z.}},
\bauthor{\bsnm{He}, \binits{C.}},
\bauthor{\bsnm{Somani}, \binits{A.}},
\bauthor{\bsnm{{Portela-Gonzalez}}, \binits{A.}},
\bauthor{\bsnm{Sander}, \binits{W.}},
\bauthor{\bsnm{Mebel}, \binits{A.M.}},
\bauthor{\bsnm{Kaiser}, \binits{R.I.}}:
\batitle{Exploring the chemical dynamics of phenanthrene ({{C14H10}}) formation
  via the bimolecular gas-phase reaction of the phenylethynyl radical
  ({{C6H5CC}}) with benzene ({{C6H6}})}.
\bjtitle{Faraday Discussions}
(\byear{2024})
\doiurl{10.1039/D3FD00159H}
\end{barticle}
\endbibitem

\bibitem[\protect\citeauthoryear{Zhao et~al.}{2019}]{zhao2019}
\begin{barticle}
\bauthor{\bsnm{Zhao}, \binits{L.}},
\bauthor{\bsnm{Kaiser}, \binits{R.I.}},
\bauthor{\bsnm{Xu}, \binits{B.}},
\bauthor{\bsnm{Ablikim}, \binits{U.}},
\bauthor{\bsnm{Lu}, \binits{W.}},
\bauthor{\bsnm{Ahmed}, \binits{M.}},
\bauthor{\bsnm{Evseev}, \binits{M.M.}},
\bauthor{\bsnm{Bashkirov}, \binits{E.K.}},
\bauthor{\bsnm{Azyazov}, \binits{V.N.}},
\bauthor{\bsnm{Zagidullin}, \binits{M.V.}},
\bauthor{\bsnm{Morozov}, \binits{A.N.}},
\bauthor{\bsnm{Howlader}, \binits{A.H.}},
\bauthor{\bsnm{Wnuk}, \binits{S.F.}},
\bauthor{\bsnm{Mebel}, \binits{A.M.}},
\bauthor{\bsnm{Joshi}, \binits{D.}},
\bauthor{\bsnm{Veber}, \binits{G.}},
\bauthor{\bsnm{Fischer}, \binits{F.R.}}:
\batitle{Gas phase synthesis of [4]-helicene}.
\bjtitle{Nature Communications}
\bvolume{10}(\bissue{1}),
\bfpage{1510}
(\byear{2019})
\doiurl{10.1038/s41467-019-09224-8}
\end{barticle}
\endbibitem

\bibitem[\protect\citeauthoryear{Scott et~al.}{1997}]{scott1997}
\begin{barticle}
\bauthor{\bsnm{Scott}, \binits{A.}},
\bauthor{\bsnm{Duley}, \binits{W.W.}},
\bauthor{\bsnm{Pinho}, \binits{G.P.}}:
\batitle{Polycyclic {{Aromatic Hydrocarbons}} and {{Fullerenes}} as
  {{Decomposition Products}} of {{Hydrogenated Amorphous Carbon}}}.
\bjtitle{The Astrophysical Journal}
\bvolume{489}(\bissue{2}),
\bfpage{193}
(\byear{1997})
\doiurl{10.1086/316789}
\end{barticle}
\endbibitem

\bibitem[\protect\citeauthoryear{Merino et~al.}{2014}]{merino2014}
\begin{barticle}
\bauthor{\bsnm{Merino}, \binits{P.}},
\bauthor{\bsnm{{\v S}vec}, \binits{M.}},
\bauthor{\bsnm{Martinez}, \binits{J.I.}},
\bauthor{\bsnm{Jelinek}, \binits{P.}},
\bauthor{\bsnm{Lacovig}, \binits{P.}},
\bauthor{\bsnm{Dalmiglio}, \binits{M.}},
\bauthor{\bsnm{Lizzit}, \binits{S.}},
\bauthor{\bsnm{Soukiassian}, \binits{P.}},
\bauthor{\bsnm{Cernicharo}, \binits{J.}},
\bauthor{\bsnm{{Martin-Gago}}, \binits{J.A.}}:
\batitle{Graphene etching on {{SiC}} grains as a path to interstellar
  polycyclic aromatic hydrocarbons formation}.
\bjtitle{Nature Communications}
\bvolume{5}(\bissue{1}),
\bfpage{3054}
(\byear{2014})
\doiurl{10.1038/ncomms4054}
\end{barticle}
\endbibitem

\bibitem[\protect\citeauthoryear{Smith et~al.}{2020}]{smith2020}
\begin{barticle}
\bauthor{\bsnm{Smith}, \binits{D.G.A.}},
\bauthor{\bsnm{Burns}, \binits{L.A.}},
\bauthor{\bsnm{Simmonett}, \binits{A.C.}},
\bauthor{\bsnm{Parrish}, \binits{R.M.}},
\bauthor{\bsnm{Schieber}, \binits{M.C.}},
\bauthor{\bsnm{Galvelis}, \binits{R.}},
\bauthor{\bsnm{Kraus}, \binits{P.}},
\bauthor{\bsnm{Kruse}, \binits{H.}},
\bauthor{\bsnm{Di~Remigio}, \binits{R.}},
\bauthor{\bsnm{Alenaizan}, \binits{A.}},
\bauthor{\bsnm{James}, \binits{A.M.}},
\bauthor{\bsnm{Lehtola}, \binits{S.}},
\bauthor{\bsnm{Misiewicz}, \binits{J.P.}},
\bauthor{\bsnm{Scheurer}, \binits{M.}},
\bauthor{\bsnm{Shaw}, \binits{R.A.}},
\bauthor{\bsnm{Schriber}, \binits{J.B.}},
\bauthor{\bsnm{Xie}, \binits{Y.}},
\bauthor{\bsnm{Glick}, \binits{Z.L.}},
\bauthor{\bsnm{Sirianni}, \binits{D.A.}},
\bauthor{\bsnm{O'Brien}, \binits{J.S.}},
\bauthor{\bsnm{Waldrop}, \binits{J.M.}},
\bauthor{\bsnm{Kumar}, \binits{A.}},
\bauthor{\bsnm{Hohenstein}, \binits{E.G.}},
\bauthor{\bsnm{Pritchard}, \binits{B.P.}},
\bauthor{\bsnm{Brooks}, \binits{B.R.}},
\bauthor{\bsnm{Schaefer}, \binits{H.F.}},
\bauthor{\bsnm{Sokolov}, \binits{A.Y.}},
\bauthor{\bsnm{Patkowski}, \binits{K.}},
\bauthor{\bsnm{DePrince}, \binits{A.E.}},
\bauthor{\bsnm{Bozkaya}, \binits{U.}},
\bauthor{\bsnm{King}, \binits{R.A.}},
\bauthor{\bsnm{Evangelista}, \binits{F.A.}},
\bauthor{\bsnm{Turney}, \binits{J.M.}},
\bauthor{\bsnm{Crawford}, \binits{T.D.}},
\bauthor{\bsnm{Sherrill}, \binits{C.D.}}:
\batitle{Psi4 1.4: {{Open-source}} software for high-throughput quantum
  chemistry}.
\bjtitle{The Journal of Chemical Physics}
\bvolume{152}(\bissue{18}),
\bfpage{184108}
(\byear{2020})
\doiurl{10.1063/5.0006002}
\end{barticle}
\endbibitem

\bibitem[\protect\citeauthoryear{Becke}{1993}]{becke1993a}
\begin{barticle}
\bauthor{\bsnm{Becke}, \binits{A.D.}}:
\batitle{Density-functional thermochemistry. {{III}}. {{The}} role of exact
  exchange}.
\bjtitle{The Journal of Chemical Physics}
\bvolume{98}(\bissue{7}),
\bfpage{5648}--\blpage{5652}
(\byear{1993})
\doiurl{10.1063/1.464913}
\end{barticle}
\endbibitem

\bibitem[\protect\citeauthoryear{Zhao and Truhlar}{2008}]{zhao2008}
\begin{barticle}
\bauthor{\bsnm{Zhao}, \binits{Y.}},
\bauthor{\bsnm{Truhlar}, \binits{D.G.}}:
\batitle{The {{M06}} suite of density functionals for main group
  thermochemistry, thermochemical kinetics, noncovalent interactions, excited
  states, and transition elements: Two new functionals and systematic testing
  of four {{M06-class}} functionals and 12 other functionals}.
\bjtitle{Theoretical Chemistry Accounts}
\bvolume{120}(\bissue{1}),
\bfpage{215}--\blpage{241}
(\byear{2008})
\doiurl{10.1007/s00214-007-0310-x}
\end{barticle}
\endbibitem

\bibitem[\protect\citeauthoryear{Lee and McCarthy}{2020}]{lee2020}
\begin{barticle}
\bauthor{\bsnm{Lee}, \binits{K.L.K.}},
\bauthor{\bsnm{McCarthy}, \binits{M.}}:
\batitle{Bayesian {{Analysis}} of {{Theoretical Rotational Constants}} from
  {{Low-Cost Electronic Structure Methods}}}.
\bjtitle{The Journal of Physical Chemistry A}
\bvolume{124}(\bissue{5}),
\bfpage{898}--\blpage{910}
(\byear{2020})
\doiurl{10.1021/acs.jpca.9b09982}
\end{barticle}
\endbibitem

\bibitem[\protect\citeauthoryear{Ye et~al.}{2022}]{ye2022}
\begin{barticle}
\bauthor{\bsnm{Ye}, \binits{H.}},
\bauthor{\bsnm{Alessandrini}, \binits{S.}},
\bauthor{\bsnm{Melosso}, \binits{M.}},
\bauthor{\bsnm{Puzzarini}, \binits{C.}}:
\batitle{Exploiting the ``{{Lego}} brick'' approach to predict accurate
  molecular structures of {{PAHs}} and {{PANHs}}}.
\bjtitle{Physical Chemistry Chemical Physics}
\bvolume{24}(\bissue{38}),
\bfpage{23254}--\blpage{23264}
(\byear{2022})
\doiurl{10.1039/D2CP03294E}
\end{barticle}
\endbibitem

\bibitem[\protect\citeauthoryear{Wohlfart et~al.}{2008}]{wohlfart2008}
\begin{barticle}
\bauthor{\bsnm{Wohlfart}, \binits{K.}},
\bauthor{\bsnm{Schnell}, \binits{M.}},
\bauthor{\bsnm{Grabow}, \binits{J.-U.}},
\bauthor{\bsnm{K{\"u}pper}, \binits{J.}}:
\batitle{Precise dipole moment and quadrupole coupling constants of
  benzonitrile}.
\bjtitle{Journal of Molecular Spectroscopy}
\bvolume{247}(\bissue{1}),
\bfpage{119}--\blpage{121}
(\byear{2008})
\doiurl{10.1016/j.jms.2007.10.006}
\end{barticle}
\endbibitem

\bibitem[\protect\citeauthoryear{Grabow et~al.}{2005}]{grabow2005}
\begin{barticle}
\bauthor{\bsnm{Grabow}, \binits{J.-U.}},
\bauthor{\bsnm{Palmer}, \binits{E.S.}},
\bauthor{\bsnm{McCarthy}, \binits{M.C.}},
\bauthor{\bsnm{Thaddeus}, \binits{P.}}:
\batitle{Supersonic-jet cryogenic-resonator coaxially oriented beam-resonator
  arrangement {{Fourier}} transform microwave spectrometer}.
\bjtitle{Review of Scientific Instruments}
\bvolume{76}(\bissue{9}),
\bfpage{093106}
(\byear{2005})
\doiurl{10.1063/1.2039347}
\end{barticle}
\endbibitem

\bibitem[\protect\citeauthoryear{Crabtree et~al.}{2016}]{crabtree2016}
\begin{barticle}
\bauthor{\bsnm{Crabtree}, \binits{K.N.}},
\bauthor{\bsnm{{Martin-Drumel}}, \binits{M.-A.}},
\bauthor{\bsnm{Brown}, \binits{G.G.}},
\bauthor{\bsnm{Gaster}, \binits{S.A.}},
\bauthor{\bsnm{Hall}, \binits{T.M.}},
\bauthor{\bsnm{McCarthy}, \binits{M.C.}}:
\batitle{Microwave spectral taxonomy: {{A}} semi-automated combination of
  chirped-pulse and cavity {{Fourier-transform}} microwave spectroscopy}.
\bjtitle{The Journal of Chemical Physics}
\bvolume{144}(\bissue{12}),
\bfpage{124201}
(\byear{2016})
\doiurl{10.1063/1.4944072}
\end{barticle}
\endbibitem

\bibitem[\protect\citeauthoryear{Pickett}{1991}]{pickett1991}
\begin{barticle}
\bauthor{\bsnm{Pickett}, \binits{H.M.}}:
\batitle{The fitting and prediction of vibration-rotation spectra with spin
  interactions}.
\bjtitle{Journal of Molecular Spectroscopy}
\bvolume{148}(\bissue{2}),
\bfpage{371}--\blpage{377}
(\byear{1991})
\doiurl{10.1016/0022-2852(91)90393-O}
\end{barticle}
\endbibitem

\bibitem[\protect\citeauthoryear{Dobashi et~al.}{2018}]{dobashi2018}
\begin{barticle}
\bauthor{\bsnm{Dobashi}, \binits{K.}},
\bauthor{\bsnm{Shimoikura}, \binits{T.}},
\bauthor{\bsnm{Nakamura}, \binits{F.}},
\bauthor{\bsnm{Kameno}, \binits{S.}},
\bauthor{\bsnm{Mizuno}, \binits{I.}},
\bauthor{\bsnm{Taniguchi}, \binits{K.}}:
\batitle{Spectral {{Tomography}} for the {{Line-of-sight Structures}} of the
  {{Taurus Molecular Cloud}} 1}.
\bjtitle{The Astrophysical Journal}
\bvolume{864}(\bissue{1}),
\bfpage{82}
(\byear{2018})
\doiurl{10.3847/1538-4357/aad62f}
\end{barticle}
\endbibitem

\bibitem[\protect\citeauthoryear{Dobashi et~al.}{2019}]{dobashi2019}
\begin{barticle}
\bauthor{\bsnm{Dobashi}, \binits{K.}},
\bauthor{\bsnm{Shimoikura}, \binits{T.}},
\bauthor{\bsnm{Ochiai}, \binits{T.}},
\bauthor{\bsnm{Nakamura}, \binits{F.}},
\bauthor{\bsnm{Kameno}, \binits{S.}},
\bauthor{\bsnm{Mizuno}, \binits{I.}},
\bauthor{\bsnm{Taniguchi}, \binits{K.}}:
\batitle{Discovery of {{CCS Velocity-coherent Substructures}} in the {{Taurus
  Molecular Cloud}} 1}.
\bjtitle{The Astrophysical Journal}
\bvolume{879}(\bissue{2}),
\bfpage{88}
(\byear{2019})
\doiurl{10.3847/1538-4357/ab25f0}
\end{barticle}
\endbibitem

\bibitem[\protect\citeauthoryear{Xue et~al.}{2020}]{xue2020}
\begin{barticle}
\bauthor{\bsnm{Xue}, \binits{C.}},
\bauthor{\bsnm{Willis}, \binits{E.R.}},
\bauthor{\bsnm{Loomis}, \binits{R.A.}},
\bauthor{\bsnm{Lee}, \binits{K.L.K.}},
\bauthor{\bsnm{Burkhardt}, \binits{A.M.}},
\bauthor{\bsnm{Shingledecker}, \binits{C.N.}},
\bauthor{\bsnm{Charnley}, \binits{S.B.}},
\bauthor{\bsnm{Cordiner}, \binits{M.A.}},
\bauthor{\bsnm{Kalenskii}, \binits{S.}},
\bauthor{\bsnm{McCarthy}, \binits{M.C.}},
\bauthor{\bsnm{Herbst}, \binits{E.}},
\bauthor{\bsnm{Remijan}, \binits{A.J.}},
\bauthor{\bsnm{McGuire}, \binits{B.A.}}:
\batitle{Detection of {{Interstellar HC4NC}} and an {{Investigation}} of
  {{Isocyanopolyyne Chemistry}} under {{TMC-1 Conditions}}}.
\bjtitle{The Astrophysical Journal Letters}
\bvolume{900}(\bissue{1}),
\bfpage{9}
(\byear{2020})
\doiurl{10.3847/2041-8213/aba631}
\end{barticle}
\endbibitem

\bibitem[\protect\citeauthoryear{Lee et~al.}{1988}]{lee1988}
\begin{barticle}
\bauthor{\bsnm{Lee}, \binits{C.}},
\bauthor{\bsnm{Yang}, \binits{W.}},
\bauthor{\bsnm{Parr}, \binits{R.G.}}:
\batitle{Development of the {{Colle-Salvetti}} correlation-energy formula into
  a functional of the electron density}.
\bjtitle{Physical Review B}
\bvolume{37}(\bissue{2}),
\bfpage{785}--\blpage{789}
(\byear{1988})
\doiurl{10.1103/PhysRevB.37.785}
\end{barticle}
\endbibitem

\bibitem[\protect\citeauthoryear{Weigend and Ahlrichs}{2005}]{weigend2005}
\begin{barticle}
\bauthor{\bsnm{Weigend}, \binits{F.}},
\bauthor{\bsnm{Ahlrichs}, \binits{R.}}:
\batitle{Balanced basis sets of split valence, triple zeta valence and
  quadruple zeta valence quality for {{H}} to {{Rn}}: {{Design}} and assessment
  of accuracy}.
\bjtitle{Physical Chemistry Chemical Physics}
\bvolume{7}(\bissue{18}),
\bfpage{3297}--\blpage{3305}
(\byear{2005})
\doiurl{10.1039/B508541A}
\end{barticle}
\endbibitem

\bibitem[\protect\citeauthoryear{Grimme et~al.}{2011}]{grimme2011}
\begin{barticle}
\bauthor{\bsnm{Grimme}, \binits{S.}},
\bauthor{\bsnm{Ehrlich}, \binits{S.}},
\bauthor{\bsnm{Goerigk}, \binits{L.}}:
\batitle{Effect of the damping function in dispersion corrected density
  functional theory}.
\bjtitle{Journal of Computational Chemistry}
\bvolume{32}(\bissue{7}),
\bfpage{1456}--\blpage{1465}
(\byear{2011})
\doiurl{10.1002/jcc.21759}
\end{barticle}
\endbibitem

\bibitem[\protect\citeauthoryear{Chai and {Head-Gordon}}{2008}]{chai2008a}
\begin{barticle}
\bauthor{\bsnm{Chai}, \binits{J.-D.}},
\bauthor{\bsnm{{Head-Gordon}}, \binits{M.}}:
\batitle{Systematic optimization of long-range corrected hybrid density
  functionals}.
\bjtitle{The Journal of Chemical Physics}
\bvolume{128}(\bissue{8}),
\bfpage{084106}
(\byear{2008})
\doiurl{10.1063/1.2834918}
\end{barticle}
\endbibitem

\bibitem[\protect\citeauthoryear{Weigend}{2006}]{weigend2006}
\begin{barticle}
\bauthor{\bsnm{Weigend}, \binits{F.}}:
\batitle{Accurate {{Coulomb-fitting}} basis sets for {{H}} to {{Rn}}}.
\bjtitle{Physical Chemistry Chemical Physics}
\bvolume{8}(\bissue{9}),
\bfpage{1057}
(\byear{2006})
\doiurl{10.1039/b515623h}
\end{barticle}
\endbibitem

\bibitem[\protect\citeauthoryear{Caldeweyher et~al.}{2017}]{caldeweyher2017}
\begin{barticle}
\bauthor{\bsnm{Caldeweyher}, \binits{E.}},
\bauthor{\bsnm{Bannwarth}, \binits{C.}},
\bauthor{\bsnm{Grimme}, \binits{S.}}:
\batitle{Extension of the {{D3}} dispersion coefficient model}.
\bjtitle{The Journal of Chemical Physics}
\bvolume{147}(\bissue{3}),
\bfpage{034112}
(\byear{2017})
\doiurl{10.1063/1.4993215}
\end{barticle}
\endbibitem

\bibitem[\protect\citeauthoryear{Jure{\v c}ka et~al.}{2006}]{jurecka2006}
\begin{barticle}
\bauthor{\bsnm{Jure{\v c}ka}, \binits{P.}},
\bauthor{\bsnm{{\v S}poner}, \binits{J.}},
\bauthor{\bsnm{{\v C}ern{\'y}}, \binits{J.}},
\bauthor{\bsnm{Hobza}, \binits{P.}}:
\batitle{Benchmark database of accurate ({{MP2}} and {{CCSD}}({{T}}) complete
  basis set limit) interaction energies of small model complexes, {{DNA}} base
  pairs, and amino acid pairs}.
\bjtitle{Physical Chemistry Chemical Physics}
\bvolume{8}(\bissue{17}),
\bfpage{1985}--\blpage{1993}
(\byear{2006})
\doiurl{10.1039/B600027D}
\end{barticle}
\endbibitem

\bibitem[\protect\citeauthoryear{Liakos and Neese}{2012}]{liakos2012}
\begin{barticle}
\bauthor{\bsnm{Liakos}, \binits{D.G.}},
\bauthor{\bsnm{Neese}, \binits{F.}}:
\batitle{Improved {{Correlation Energy Extrapolation Schemes Based}} on {{Local
  Pair Natural Orbital Methods}}}.
\bjtitle{The Journal of Physical Chemistry A}
\bvolume{116}(\bissue{19}),
\bfpage{4801}--\blpage{4816}
(\byear{2012})
\doiurl{10.1021/jp302096v}
\end{barticle}
\endbibitem

\bibitem[\protect\citeauthoryear{Neese}{2022}]{neese2022}
\begin{barticle}
\bauthor{\bsnm{Neese}, \binits{F.}}:
\batitle{Software update: {{The ORCA}} program system---{{Version}} 5.0}.
\bjtitle{WIREs Computational Molecular Science}
\bvolume{12}(\bissue{5}),
\bfpage{1606}
(\byear{2022})
\doiurl{10.1002/wcms.1606}
\end{barticle}
\endbibitem

\bibitem[\protect\citeauthoryear{Glowacki et~al.}{2012}]{glowacki2012}
\begin{barticle}
\bauthor{\bsnm{Glowacki}, \binits{D.R.}},
\bauthor{\bsnm{Liang}, \binits{C.-H.}},
\bauthor{\bsnm{Morley}, \binits{C.}},
\bauthor{\bsnm{Pilling}, \binits{M.J.}},
\bauthor{\bsnm{Robertson}, \binits{S.H.}}:
\batitle{{{MESMER}}: {{An Open-Source Master Equation Solver}} for
  {{Multi-Energy Well Reactions}}}.
\bjtitle{The Journal of Physical Chemistry A}
\bvolume{116}(\bissue{38}),
\bfpage{9545}--\blpage{9560}
(\byear{2012})
\doiurl{10.1021/jp3051033}
\end{barticle}
\endbibitem

\bibitem[\protect\citeauthoryear{Georgievskii and
  Klippenstein}{2005}]{georgievskii2005}
\begin{barticle}
\bauthor{\bsnm{Georgievskii}, \binits{Y.}},
\bauthor{\bsnm{Klippenstein}, \binits{S.J.}}:
\batitle{Long-range transition state theory}.
\bjtitle{The Journal of Chemical Physics}
\bvolume{122}(\bissue{19}),
\bfpage{194103}
(\byear{2005})
\doiurl{10.1063/1.1899603}
\end{barticle}
\endbibitem

\bibitem[\protect\citeauthoryear{West et~al.}{2019}]{west2019a}
\begin{barticle}
\bauthor{\bsnm{West}, \binits{N.A.}},
\bauthor{\bsnm{Millar}, \binits{T.J.}},
\bauthor{\bsnm{Sande}, \binits{M.V.}},
\bauthor{\bsnm{Rutter}, \binits{E.}},
\bauthor{\bsnm{Blitz}, \binits{M.A.}},
\bauthor{\bsnm{Decin}, \binits{L.}},
\bauthor{\bsnm{Heard}, \binits{D.E.}}:
\batitle{Measurements of {{Low Temperature Rate Coefficients}} for the
  {{Reaction}} of {{CH}} with {{CH2O}} and {{Application}} to {{Dark Cloud}}
  and {{AGB Stellar Wind Models}}}.
\bjtitle{The Astrophysical Journal}
\bvolume{885}(\bissue{2}),
\bfpage{134}
(\byear{2019})
\doiurl{10.3847/1538-4357/ab480e}
\end{barticle}
\endbibitem

\bibitem[\protect\citeauthoryear{Davies et~al.}{1986}]{davies1986}
\begin{barticle}
\bauthor{\bsnm{Davies}, \binits{J.W.}},
\bauthor{\bsnm{Green}, \binits{N.J.B.}},
\bauthor{\bsnm{Pilling}, \binits{M.J.}}:
\batitle{The testing of models for unimolecular decomposition via inverse
  laplace transformation of experimental recombination rate data}.
\bjtitle{Chemical Physics Letters}
\bvolume{126}(\bissue{3}),
\bfpage{373}--\blpage{379}
(\byear{1986})
\doiurl{10.1016/S0009-2614(86)80101-4}
\end{barticle}
\endbibitem

\bibitem[\protect\citeauthoryear{Holbrook et~al.}{1996}]{Robertson_1996}
\begin{bbook}
\bauthor{\bsnm{Holbrook}, \binits{K.A.}},
\bauthor{\bsnm{Pilling}, \binits{M.J.}},
\bauthor{\bsnm{Robertson}, \binits{S.H.}}:
\bbtitle{Unimolecular Reactions},
\bedition{2}nd edn.
\bpublisher{Wiley}, \blocation{???}
(\byear{1996})
\end{bbook}
\endbibitem

\bibitem[\protect\citeauthoryear{Baer and Hase}{1996}]{baer1996}
\begin{bbook}
\bauthor{\bsnm{Baer}, \binits{T.}},
\bauthor{\bsnm{Hase}, \binits{W.L.}}:
\bbtitle{Unimolecular {{Reaction Dynamics}}: {{Theory}} and {{Experiments}}},
(\byear{1996})
\end{bbook}
\endbibitem

\bibitem[\protect\citeauthoryear{Lourderaj and Hase}{2009}]{lourderaj2009}
\begin{barticle}
\bauthor{\bsnm{Lourderaj}, \binits{U.}},
\bauthor{\bsnm{Hase}, \binits{W.L.}}:
\batitle{Theoretical and {{Computational Studies}} of {{Non-RRKM Unimolecular
  Dynamics}}}.
\bjtitle{The Journal of Physical Chemistry A}
\bvolume{113}(\bissue{11}),
\bfpage{2236}--\blpage{2253}
(\byear{2009})
\doiurl{10.1021/jp806659f}
\end{barticle}
\endbibitem

\bibitem[\protect\citeauthoryear{Burkhardt et~al.}{2021}]{burkhardt2021a}
\begin{barticle}
\bauthor{\bsnm{Burkhardt}, \binits{A.M.}},
\bauthor{\bsnm{Lee}, \binits{K.L.K.}},
\bauthor{\bsnm{Changala}, \binits{P.B.}},
\bauthor{\bsnm{Shingledecker}, \binits{C.N.}},
\bauthor{\bsnm{Cooke}, \binits{I.R.}},
\bauthor{\bsnm{Loomis}, \binits{R.A.}},
\bauthor{\bsnm{Wei}, \binits{H.}},
\bauthor{\bsnm{Charnley}, \binits{S.B.}},
\bauthor{\bsnm{Herbst}, \binits{E.}},
\bauthor{\bsnm{McCarthy}, \binits{M.C.}},
\bauthor{\bsnm{McGuire}, \binits{B.A.}}:
\batitle{Discovery of the {{Pure Polycyclic Aromatic Hydrocarbon Indene}}
  (c-{{C9H8}}) with {{GOTHAM Observations}} of {{TMC-1}}}.
\bjtitle{The Astrophysical Journal Letters}
\bvolume{913}(\bissue{2}),
\bfpage{18}
(\byear{2021})
\doiurl{10.3847/2041-8213/abfd3a}
\end{barticle}
\endbibitem

\bibitem[\protect\citeauthoryear{Hincelin et~al.}{2011}]{hincelin2011}
\begin{barticle}
\bauthor{\bsnm{Hincelin}, \binits{U.}},
\bauthor{\bsnm{Wakelam}, \binits{V.}},
\bauthor{\bsnm{Hersant}, \binits{F.}},
\bauthor{\bsnm{Guilloteau}, \binits{S.}},
\bauthor{\bsnm{Loison}, \binits{J.C.}},
\bauthor{\bsnm{Honvault}, \binits{P.}},
\bauthor{\bsnm{Troe}, \binits{J.}}:
\batitle{Oxygen depletion in dense molecular clouds: A clue to a low {{O2}}
  abundance?}
\bjtitle{Astronomy \& Astrophysics}
\bvolume{530},
\bfpage{61}
(\byear{2011})
\doiurl{10.1051/0004-6361/201016328}
\end{barticle}
\endbibitem

\bibitem[\protect\citeauthoryear{Su and Chesnavich}{1982}]{su1982}
\begin{barticle}
\bauthor{\bsnm{Su}, \binits{T.}},
\bauthor{\bsnm{Chesnavich}, \binits{W.J.}}:
\batitle{Parametrization of the ion--polar molecule collision rate constant by
  trajectory calculations}.
\bjtitle{The Journal of Chemical Physics}
\bvolume{76}(\bissue{10}),
\bfpage{5183}--\blpage{5185}
(\byear{1982})
\doiurl{10.1063/1.442828}
\end{barticle}
\endbibitem

\bibitem[\protect\citeauthoryear{Woon and Herbst}{2009}]{woon2009}
\begin{barticle}
\bauthor{\bsnm{Woon}, \binits{D.E.}},
\bauthor{\bsnm{Herbst}, \binits{E.}}:
\batitle{{{QUANTUM CHEMICAL PREDICTIONS OF THE PROPERTIES OF KNOWN AND
  POSTULATED NEUTRAL INTERS}}tel{{LAR MOLECULES}}}.
\bjtitle{The Astrophysical Journal Supplement Series}
\bvolume{185}(\bissue{2}),
\bfpage{273}
(\byear{2009})
\doiurl{10.1088/0067-0049/185/2/273}
\end{barticle}
\endbibitem

\bibitem[\protect\citeauthoryear{Haider and Husain}{1993}]{haider1993}
\begin{barticle}
\bauthor{\bsnm{Haider}, \binits{N.}},
\bauthor{\bsnm{Husain}, \binits{D.}}:
\batitle{Kinetic investigation of the collisional behavior of ground state
  atomic carbon, {{C}}(2p2({{3PJ}})), with halogenated olefins and aromatic
  compounds studied by time-resolved atomic resonance absorption spectroscopy
  in the vacuum ultra-violet}.
\bjtitle{International Journal of Chemical Kinetics}
\bvolume{25}(\bissue{6}),
\bfpage{423}--\blpage{435}
(\byear{1993})
\doiurl{10.1002/kin.550250602}
\end{barticle}
\endbibitem

\bibitem[\protect\citeauthoryear{Wang et~al.}{2021}]{wang2021}
\begin{barticle}
\bauthor{\bsnm{Wang}, \binits{X.}},
\bauthor{\bsnm{Wang}, \binits{L.}},
\bauthor{\bsnm{Mao}, \binits{X.}},
\bauthor{\bsnm{Wang}, \binits{Q.}},
\bauthor{\bsnm{Mu}, \binits{Z.}},
\bauthor{\bsnm{An}, \binits{L.}},
\bauthor{\bsnm{Zhang}, \binits{W.}},
\bauthor{\bsnm{Feng}, \binits{X.}},
\bauthor{\bsnm{Redshaw}, \binits{C.}},
\bauthor{\bsnm{Cao}, \binits{C.}},
\bauthor{\bsnm{Qin}, \binits{A.}},
\bauthor{\bsnm{Tang}, \binits{B.Z.}}:
\batitle{Pyrene-based aggregation-induced emission luminogens ({{AIEgens}})
  with less colour migration for anti-counterfeiting applications}.
\bjtitle{Journal of Materials Chemistry C}
\bvolume{9}(\bissue{37}),
\bfpage{12828}--\blpage{12838}
(\byear{2021})
\doiurl{10.1039/D1TC03022A}
\end{barticle}
\endbibitem

\bibitem[\protect\citeauthoryear{{\DJ}or{\dj}evi{\'c}
  et~al.}{2020}]{dordevic2020}
\begin{barticle}
\bauthor{\bsnm{{\DJ}or{\dj}evi{\'c}}, \binits{L.}},
\bauthor{\bsnm{Milano}, \binits{D.}},
\bauthor{\bsnm{Demitri}, \binits{N.}},
\bauthor{\bsnm{Bonifazi}, \binits{D.}}:
\batitle{O-{{Annulation}} to {{Polycyclic Aromatic Hydrocarbons}}: {{A Tale}}
  of {{Optoelectronic Properties}} from {{Five-}} to {{Seven-Membered Rings}}}.
\bjtitle{Organic Letters}
\bvolume{22}(\bissue{11}),
\bfpage{4283}--\blpage{4288}
(\byear{2020})
\doiurl{10.1021/acs.orglett.0c01331}
\end{barticle}
\endbibitem

\bibitem[\protect\citeauthoryear{Biswas et~al.}{2024}]{biswas2024}
\begin{barticle}
\bauthor{\bsnm{Biswas}, \binits{K.}},
\bauthor{\bsnm{Chen}, \binits{Q.}},
\bauthor{\bsnm{Obermann}, \binits{S.}},
\bauthor{\bsnm{Ma}, \binits{J.}},
\bauthor{\bsnm{{Soler-Polo}}, \binits{D.}},
\bauthor{\bsnm{Melidonie}, \binits{J.}},
\bauthor{\bsnm{Barrag{\'a}n}, \binits{A.}},
\bauthor{\bsnm{{S{\'a}nchez-Grande}}, \binits{A.}},
\bauthor{\bsnm{Lauwaet}, \binits{K.}},
\bauthor{\bsnm{Gallego}, \binits{J.M.}},
\bauthor{\bsnm{Miranda}, \binits{R.}},
\bauthor{\bsnm{{\'E}cija}, \binits{D.}},
\bauthor{\bsnm{Jel{\'i}nek}, \binits{P.}},
\bauthor{\bsnm{Feng}, \binits{X.}},
\bauthor{\bsnm{Urgel}, \binits{J.I.}}:
\batitle{On-{{Surface Synthesis}} of {{Non-Benzenoid Nanographenes Embedding
  Azulene}} and {{Stone-Wales Topologies}}}.
\bjtitle{Angewandte Chemie International Edition}
\bvolume{63}(\bissue{13}),
\bfpage{202318185}
(\byear{2024})
\doiurl{10.1002/anie.202318185}
\end{barticle}
\endbibitem

\bibitem[\protect\citeauthoryear{Ji et~al.}{2015}]{ji2015}
\begin{barticle}
\bauthor{\bsnm{Ji}, \binits{L.}},
\bauthor{\bsnm{Lorbach}, \binits{A.}},
\bauthor{\bsnm{Edkins}, \binits{R.M.}},
\bauthor{\bsnm{Marder}, \binits{T.B.}}:
\batitle{Synthesis and {{Photophysics}} of a 2,7-{{Disubstituted
  Donor}}--{{Acceptor Pyrene Derivative}}: {{An Example}} of the
  {{Application}} of {{Sequential Ir-Catalyzed C}}--{{H Borylation}} and
  {{Substitution Chemistry}}}.
\bjtitle{The Journal of Organic Chemistry}
\bvolume{80}(\bissue{11}),
\bfpage{5658}--\blpage{5665}
(\byear{2015})
\doiurl{10.1021/acs.joc.5b00618}
\end{barticle}
\endbibitem

\bibitem[\protect\citeauthoryear{Lu et~al.}{2020}]{lu2020}
\begin{barticle}
\bauthor{\bsnm{Lu}, \binits{Q.}},
\bauthor{\bsnm{Kole}, \binits{G.K.}},
\bauthor{\bsnm{Friedrich}, \binits{A.}},
\bauthor{\bsnm{{M{\"u}ller-Buschbaum}}, \binits{K.}},
\bauthor{\bsnm{Liu}, \binits{Z.}},
\bauthor{\bsnm{Yu}, \binits{X.}},
\bauthor{\bsnm{Marder}, \binits{T.B.}}:
\batitle{Comparison {{Study}} of the {{Site-Effect}} on {{Regioisomeric
  Pyridyl-Pyrene Conjugates}}: {{Synthesis}}, {{Structures}}, and
  {{Photophysical Properties}}}.
\bjtitle{The Journal of Organic Chemistry}
\bvolume{85}(\bissue{6}),
\bfpage{4256}--\blpage{4266}
(\byear{2020})
\doiurl{10.1021/acs.joc.9b03421}
\end{barticle}
\endbibitem

\bibitem[\protect\citeauthoryear{{Bao-Xi} et~al.}{2019}]{bao-xi2019}
\begin{barticle}
\bauthor{\bsnm{{Bao-Xi}}, \binits{M.}},
\bauthor{\bsnm{{Li-Fang}}, \binits{Z.}},
\bauthor{\bsnm{Yun}, \binits{Z.}}:
\batitle{Crystal structure of pyrene-4-aldehyde, {{C17H10O}}}.
\bjtitle{Zeitschrift f{\"u}r Kristallographie - New Crystal Structures}
\bvolume{234}(\bissue{1}),
\bfpage{173}--\blpage{175}
(\byear{2019})
\doiurl{10.1515/ncrs-2018-0254}
\end{barticle}
\endbibitem

\end{thebibliography}

\begin{appendices}

\clearpage

\section{Supplementary Information}\label{SI}

\subsection{Spectroscopic constants for 2- and 4-cyanopyrene}
\label{sec:rotconst}

\begin{table}[htb!]
    \centering
    \caption{Spectroscopic constants of 2- and 4-cyanopyrene derived by least-squares fitting to a standard rotational Hamiltonian using SPCAT/SPFIT in Pickett’s CALPGM suite of programs~\cite{pickett1991} (A-reduced, $I^r$ representation). The number of lines in the fit, $N_\mathrm{lines}$, and the RMS value of the fit residuals, $\sigma_\mathrm{fit}$, are reported. Except for the dipole moments, $\mu_\mathrm{a}$ and $\mu_\mathrm{b}$, which are reported in Debye, all dimensional values are noted in MHz.}
    \begin{tabular}{lS[table-format=3.10]S[table-format=3.5]lS[table-format=3.5]c}
    \toprule
        \multicolumn{6}{c}{\textbf{2-cyanopyrene}} \\
    \midrule
        & \multicolumn{2}{c}{This work} & & \multicolumn{1}{c}{Ye et al. (2022)} \\
        
        {Parameter} & {Experimental$^a$} & {M06-2X/6-31+G(d)} & &{`Lego brick'~\cite{ye2022}} \\
    \cmidrule{1-3}
    \cmidrule{5-6}
        &  &  &  &  &  \\
        $A$ &  1009.19382(60) & 1010.2544 & & 1008.579  \\
        $B$ &  313.1345299(202) & 313.2383 & & 313.082  \\
        $C$ &  239.0427225(184) & 239.1025 & & 238.969  \\
        $\Delta_J \times 10^6$ &  0.7008(109) & &  \\
        $\Delta_{JK} \times 10^6$ &  5.814(94) & &  \\
        $\Delta_K \times 10^6$ & 15.3(114) &  &   \\
        $\delta_J \times 10^6$ & 0.1759(60) &  &  \\
        $\delta_K \times 10^6$ & 4.22(39) &  & \\
        $\chi_{aa}$ & -4.1981(128) & -4.2374$^b$ & & \\
        $\chi_{bb}$ & 2.3082(107) & 2.2886$^b$ & & \\
         &  &  &  &  & \\
        $N_\mathrm{lines}$ & 762 &  & \\
        $\sigma_\mathrm{fit} \times 10^3$ & 1.919 &  &  & \\
        $(J, K_\mathrm{a})_\mathrm{max}$ & {$(34, 11)$} &  &  & \\
        &  &  &  &  & \\
        $\mu_\mathrm{a}$ & & 5.617 & & \\

    \midrule 
    \\
    \multicolumn{6}{c}{\textbf{4-cyanopyrene}} \\
    \midrule
        & \multicolumn{2}{c}{This work} & & \multicolumn{1}{c}{Ye et al. (2022)} \\
        
        {Parameter} & {Experimental$^a$} & {M06-2X/6-31+G(d)} & &{`Lego brick'~\cite{ye2022}} \\
    \cmidrule{1-3}
    \cmidrule{5-6}
        &  &  &  &  &  \\
        $A$ &  651.383034(69) & 651.4828 & & 650.771  \\
        $B$ &  453.731352(45) & 454.1186 & & 453.662  \\
        $C$ &  267.5078168(224) & 267.5923 & & 267.368  \\
        $\Delta_J \times 10^6$ &  1.780(98) & &  \\
        $\Delta_{JK} \times 10^6$ &  0.43(58) & &  \\
        $\Delta_K \times 10^6$ & 10.89(73) &  &   \\
        $\delta_J \times 10^6$ & 0.677(49) &  &  \\
        $\delta_K \times 10^6$ & 2.56(33) &  & \\
        $\chi_{aa}$ & -3.3964(99) & -3.4668$^c$ & & \\
        $\chi_{bb}$ & 1.4704(69) & 1.5180$^c$ & & \\
         &  &  &  &  & \\
        $N_\mathrm{lines}$ & 318 &  & \\
        $\sigma_\mathrm{fit} \times 10^3$ & 2.121 &  &  & \\
        $(J, K_\mathrm{a})_\mathrm{max}$ & {$(25, 10)$} &  &  & \\
        &  &  &  &  & \\
        $\mu_\mathrm{a}$ & & 4.555 & & \\
        $\mu_\mathrm{b}$ & & 1.963 & & \\
        \bottomrule
        \end{tabular}
        {$^a$ $1\,\sigma$ uncertainties are given in parentheses in units of the last digit.}\\
        {$^b$ Due to symmetry, the experimental $\chi$ tensor of benzonitrile~\cite{wohlfart2008} was used.}\\
        {$^c$ Derived by rotating the experimental $\chi$ tensor of benzonitrile~\cite{wohlfart2008} to the local CN bond axis frame of the calculated 4-cyanopyrene equilibrium geometries.}
    \label{tab:rotconst}
\end{table}

\newpage
\subsection{Rotational spectra and GOTHAM coverage}

\begin{figure}[!htb]
    \centering
    \includegraphics[width=1\linewidth]{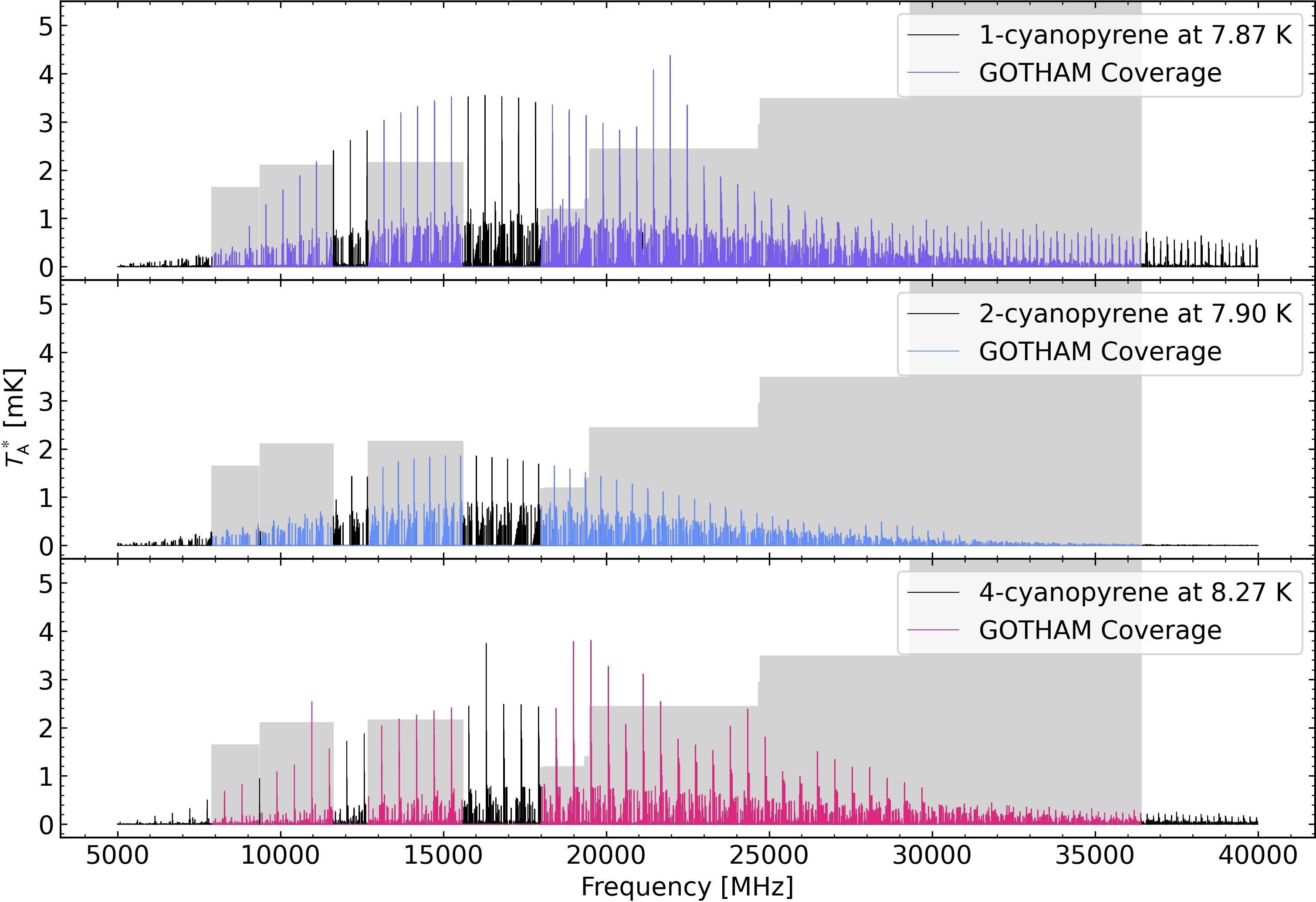}
    \caption{\textbf{Frequency coverage of TMC-1 in the GOTHAM data set.} Rotational spectra of 1-cyanopyrene (top panel), 2-cyanopyrene (center panel), and 4-cyanopyrene (bottom panel) simulated at the MCMC derived excitation temperatures of $7.87\,\mathrm{K}$, $7.90\,\mathrm{K}$, and $8.27\,\mathrm{K}$, respectively, are depicted in black. The lines that are covered by the GOTHAM data are depicted in color. The grey shaded areas indicate the averaged noise level of the observation in each data chunk.}
    \label{fig:coverage}
\end{figure}

\newpage
\subsection{Individual lines of 2-cyanopyrene}

\begin{figure}[!htb]
    \centering
    \includegraphics[width=1\linewidth]{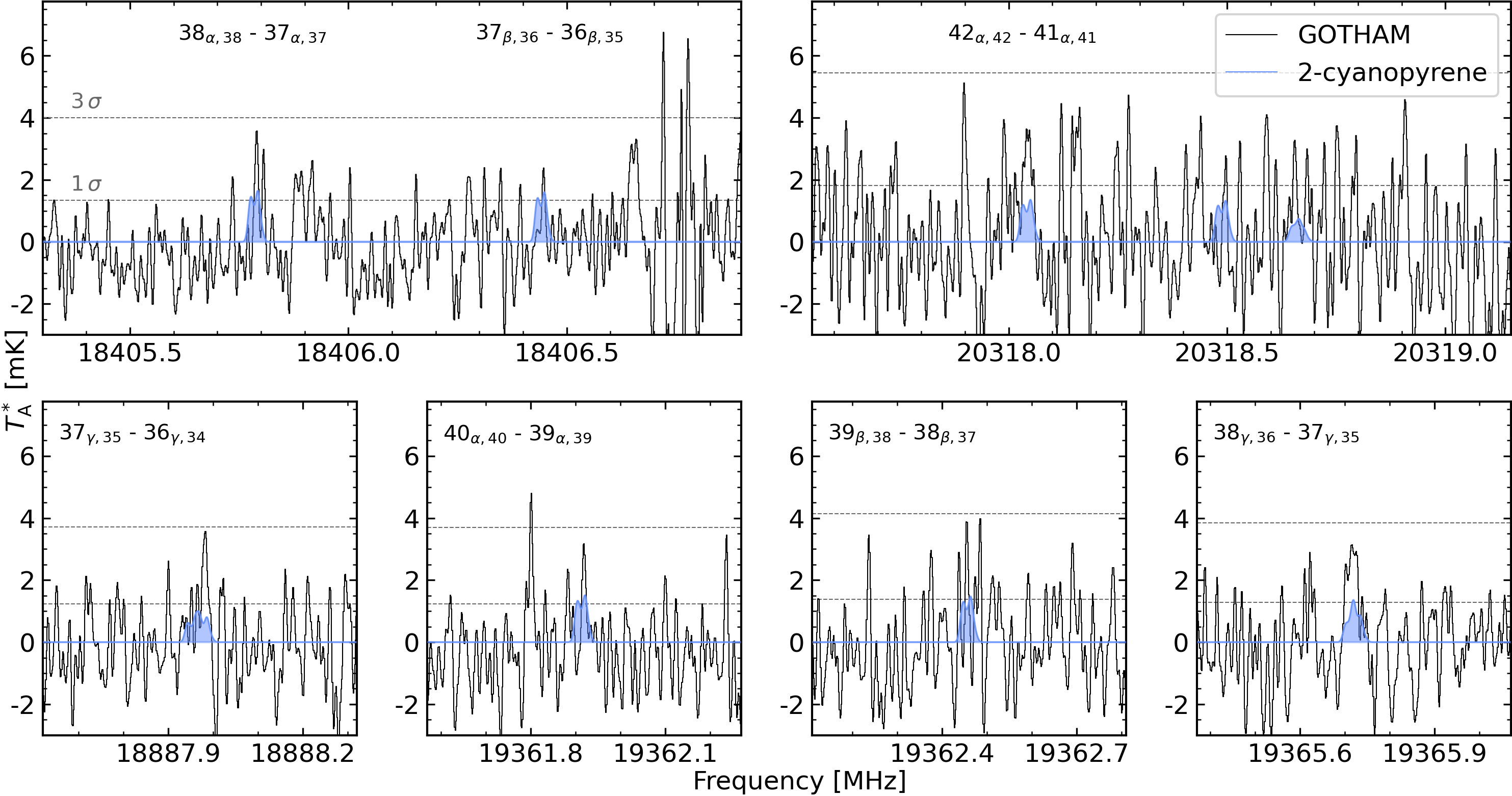}
    \caption{\textbf{GOTHAM spectra showing several 2-cyanopyrene lines under the noise of the TMC-1 observations.} The original GOTHAM observational data, for which one channel corresponds to $1.4\,\mathrm{kHz}$, were smoothed with a Hanning window to a resolution of $14\,\mathrm{kHz}$, depicted in black. The spectrum of 2-cyanopyrene is overplotted in blue using source-dependent molecular parameters as reported in Supplementary Table~\ref{tab:MCMCresults_isomers}. The horizontal dashed lines represent the RMS noise ($1\,\sigma$) and $3\,\times$ RMS noise ($3\,\sigma$) levels in each depicted spectral window. Quantum numbers of each transition, ignoring $^{14}$N nuclear electric quadrupole splitting, are reported. Each line contains multiple closely spaced $K$-components of each transition denoted $\alpha, \beta, \gamma, \delta = \{0,1\}, \{1,2\}, \{2,3\}, \{3,4\}$. The two weak lines not marked in the upper right panel correspond to transitions $41_{\beta,40} - 40_{\beta,39}$ and $35_{6,29} - 34_{6,28}$.
    }
    \label{fig:lines_2-cyanopyrene}
\end{figure}

\begin{figure}[!htb]
    \centering
    \includegraphics[width=1\linewidth]{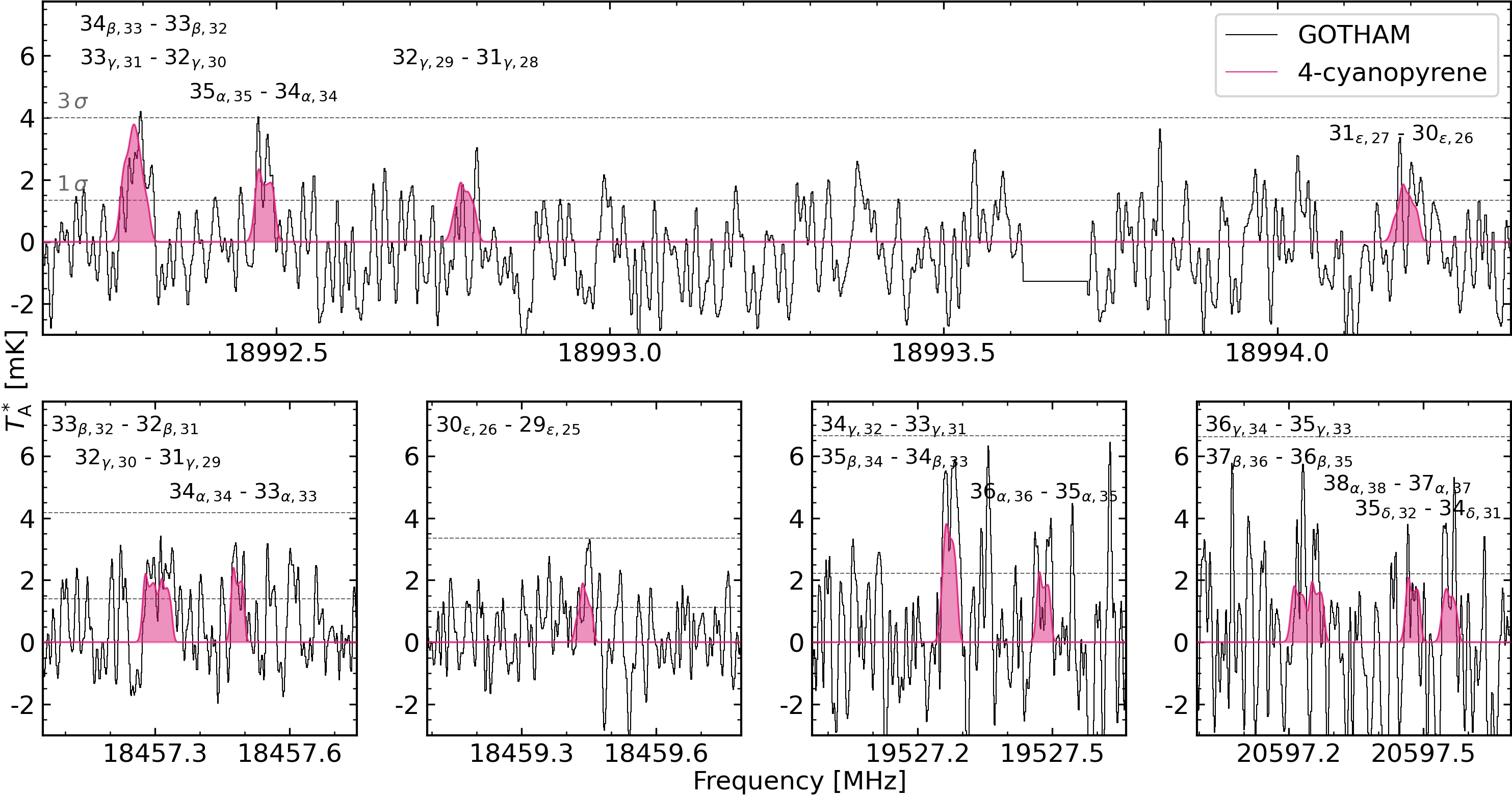}
    \caption{\textbf{GOTHAM data of TMC-1 compared to the brightest features of the rotational spectrum of 4-cyanopyrene.} GOTHAM observational data were collected at a spectral resolution of a $1.4\,\mathrm{kHz}$ and smoothed using a Hanning window to $14\,\mathrm{kHz}$ (black). The rotational spectrum of 4-cyanopyrene is overlaid in pink using source-dependent molecular parameters derived from the MCMC analysis as reported in Supplementary Table~\ref{tab:MCMCresults_isomers}. The horizontal dashed lines represent the RMS noise ($1\,\sigma$) and $3\,\times$ RMS noise ($3\,\sigma$) levels in each depicted spectral window. Quantum numbers of each transition, ignoring $^{14}$N nuclear electric quadrupole splitting, are reported. Each line contains multiple closely spaced $K$-components of each transition denoted $\alpha, \beta, \gamma, \delta, \varepsilon = \{0,1\}, \{1,2\}, \{2,3\}, \{3,4\}, \{4,5\}$.
    }
    \label{fig:lines_4-cyanopyrene}
\end{figure}

\newpage
\subsection{MCMC analysis and results}

\begin{table}[htb!]
    \caption{Priors that were used for the MCMC analysis, where $N(\mu,\sigma^2)$ denotes a normal (Gaussian) parameter distribution with mean, $\mu$, and variance, $\sigma^2$, and $U\{a,b\}$ denotes a uniform (unweighted) parameter distribution between $a$ and $b$. The minimum and maximum values for each distribution are reported. These priors were derived from the marginalized posterior for the MCMC analysis of 
    1-cyanopyrene~\cite{wenzelinreview}.}
    \centering
    \begin{tabular}{cccccccccc}
    \toprule
      Component &  $v_\mathrm{lsr}$	&	Size	&	$\mathrm{log_{10}}(N_T)$	&	$T_\mathrm{ex}$	&	$\Delta V$	\\
	No. & [$\mathrm{km\,s^{-1}}$] &[$^{\prime\prime}$]	&	[$\mathrm{cm}^{-2}$]	&	[$\mathrm{K}$]	&	[$\mathrm{km\,s^{-1}}$]\\
	\midrule
	1 & {$N(5.603,0.005)$} & \multirow{4}{*}{$N(50,1)$}	 &  	 \multirow{4}{*}
 {$U\{\mathrm{a,b}\}$} &
 \multirow{4}{*}{$U\{\mathrm{a,b}\}$}	 & 	 \multirow{4}{*}{$N(0.150,0.05)$}\\
	2 &	{$N(5.747,0.005)$} &	 & 	 & 	 & 	 \\
        3 & {$N(5.930,0.005)$} &	 & 	 &   & 	 \\
	4 &	{$N(6.036,0.005)$} &  &	 & 	 & 	 \\
 \midrule
        Min & $0.0$ & $5$ & $10.5$ & $4.0$ & $0.05$ \\
        Max & $10.0$ & $100$ & $13.0$ & $10.0$ & $0.25$ \\
	\bottomrule
\\
    \end{tabular}
    \label{tab:MCMCpriors}
\end{table}

\begin{table}[!htb]
    \centering
    \caption{MCMC fit parameters for 1-cyanopyrene as reported in ref.~\cite{wenzelinreview}, and 2- and 4-cyanopyrene (this work) using the priors reported in Supplementary Table~\ref{tab:MCMCpriors}.}
    \label{tab:MCMCresults_isomers}
    \begin{tabular}{cccccccccc}
    \toprule
    \multicolumn{6}{c}{\textbf{1-cyanopyrene}}\\
          Component &  $v_\mathrm{lsr}$	&	Size	&	$N_T$	&	$T_\mathrm{ex}$	&	$\Delta V$	\\
	No. & [$\mathrm{km\,s^{-1}}$] & [$^{\prime\prime}$]	&	[$10^{11}\,\mathrm{cm}^{-2}$]	&	[$\mathrm{K}$]	&	[$\mathrm{km\,s^{-1}}$]\\
    \midrule
	1 &	$5.603^{+0.012}_{-0.012}$	 & 	$50$	 & 	$4.15^{+0.62}_{-0.59}$	 & 	 \multirow{4}{*}{$7.87^{+0.43}_{-0.40}$}	 & 	 \multirow{4}{*}{$0.150^{+0.016}_{-0.013}$}\\
	2 &	$5.747^{+0.010}_{-0.011}$	 & 	$50$	 & 	$5.81^{+0.68}_{-0.61}$	 & 	 & 	 \\		
        3 & $5.930^{+0.016}_{-0.026}$	 & 	$49$	 & 	$3.75^{+0.70}_{-1.21}$	 & 	 & 	 \\
	4 &	$6.036^{+0.046}_{-0.042}$	 & 	$49$	 & 	$1.49^{+1.38}_{-0.71}$	 & 	 & 	 \\
	\midrule
		\multicolumn{6}{c}{$N_T(\mathrm{Total}):\;1.52^{+0.18}_{-0.16}\times 10^{12}\,\mathrm{cm}^{-2}$}\\
  \midrule
  \\
\multicolumn{6}{c}{\textbf{2-cyanopyrene}}\\
      Component &  $v_\mathrm{lsr}$	&	Size	&	$N_T$	&	$T_\mathrm{ex}$	&	$\Delta V$	\\
	No. & [$\mathrm{km\,s^{-1}}$] & [$^{\prime\prime}$]	&	[$10^{11}\,\mathrm{cm}^{-2}$]	&	[$\mathrm{K}$]	&	[$\mathrm{km\,s^{-1}}$]\\
	\midrule

	1 & $5.603^{+0.005}_{-0.005}$	 & 	$49$	 & 	$0.78^{+0.36}_{-0.30}$	 & 	 \multirow{4}{*}{$7.90^{+0.53}_{-0.48}$}	 & 	 \multirow{4}{*}{$0.191^{+0.018}_{-0.019}$}\\
	2 &	$5.751^{+0.005}_{-0.005}$	 & 	$49$	 & 	$3.52^{+0.52}_{-0.46}$	 & 	 & 	 \\		
        3 & $5.928^{+0.005}_{-0.005}$	 & 	$50$	 & 	$1.60^{+0.48}_{-0.51}$	 & 	 & 	 \\
	4 &	$6.038^{+0.005}_{-0.005}$	 & 	$50$	 & 	$2.51^{+0.50}_{-0.44}$	 & 	 & 	 \\
	\midrule
		\multicolumn{6}{c}{$N_T(\mathrm{Total}):\;8.41^{+0.94}_{-0.87}\times 10^{11}\,\mathrm{cm}^{-2}$}\\
  
\midrule
\\
\multicolumn{6}{c}{\textbf{4-cyanopyrene}}\\
      Component &  $v_\mathrm{lsr}$	&	Size	&	$N_T$	&	$T_\mathrm{ex}$	&	$\Delta V$	\\
	No. & [$\mathrm{km\,s^{-1}}$] & [$^{\prime\prime}$]	&	[$10^{11}\,\mathrm{cm}^{-2}$]	&	[$\mathrm{K}$]	&	[$\mathrm{km\,s^{-1}}$]\\
	\midrule
	1 &	$5.601^{+0.005}_{-0.004}$	 & 	$50$	 & 	$3.94^{+0.46}_{-0.42}$	 & 	 \multirow{4}{*}{$8.27^{+0.46}_{-0.44}$}	 & 	 \multirow{4}{*}{$0.179^{+0.016}_{-0.015}$}\\
	2 &	$5.752^{+0.005}_{-0.005}$	 & 	$50$	 & 	$3.49^{+0.53}_{-0.51}$	 & 	 & 	 \\		
        3 & $5.925^{+0.005}_{-0.005}$	 & 	$50$	 & 	$5.20^{+0.62}_{-0.56}$	 & 	 & 	 \\
	4 & $6.036^{+0.005}_{-0.005}$	 & 	$50$	 & 	$0.66^{+0.41}_{-0.25}$	 & 	 & 	 \\
	\midrule
		\multicolumn{6}{c}{$N_T(\mathrm{Total}):\;1.33^{+0.10}_{-0.09}\times 10^{12}\,\mathrm{cm}^{-2}$}\\

\bottomrule
    \end{tabular}
\end{table}

\begin{figure}[!ht]
    \centering
    \includegraphics[width=1\linewidth]{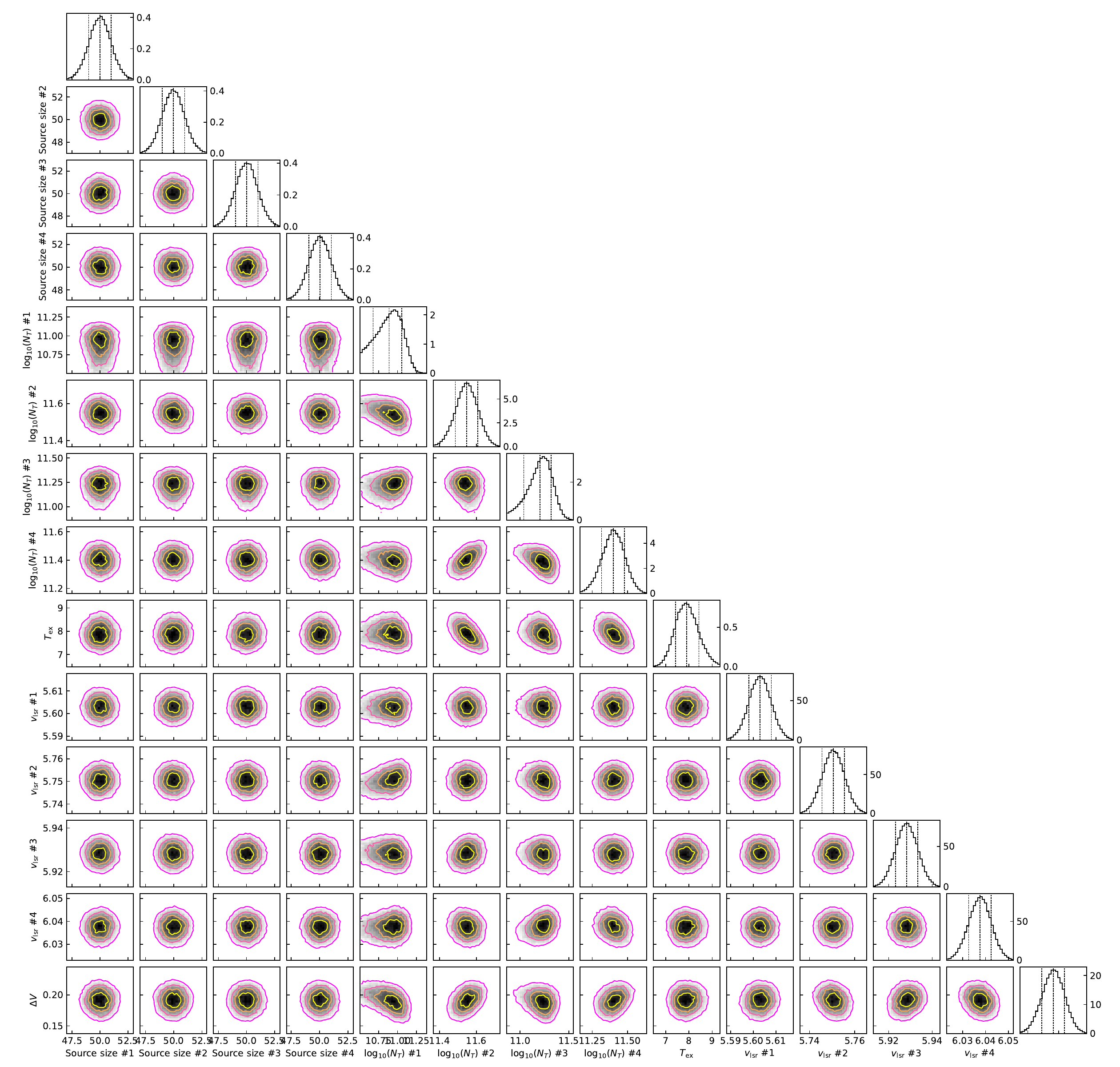}
    \caption{\textbf{Corner plot of MCMC analysis for 2-cyanopyrene.} This shows the parameter covariances on the off-diagonal and marginalized posterior distributions considered in the marginalized posterior distributions on-diagonal. The $16^\mathrm{th}$, $50^\mathrm{th}$, and $84^\mathrm{th}$ confidence intervals (corresponding to $\pm 1\,\sigma$ for a Gaussian posterior distribution) are shown as vertical lines on the diagonal.}
    \label{fig:corner2}
\end{figure}

\begin{figure}
    \centering
    \includegraphics[width=1\linewidth]{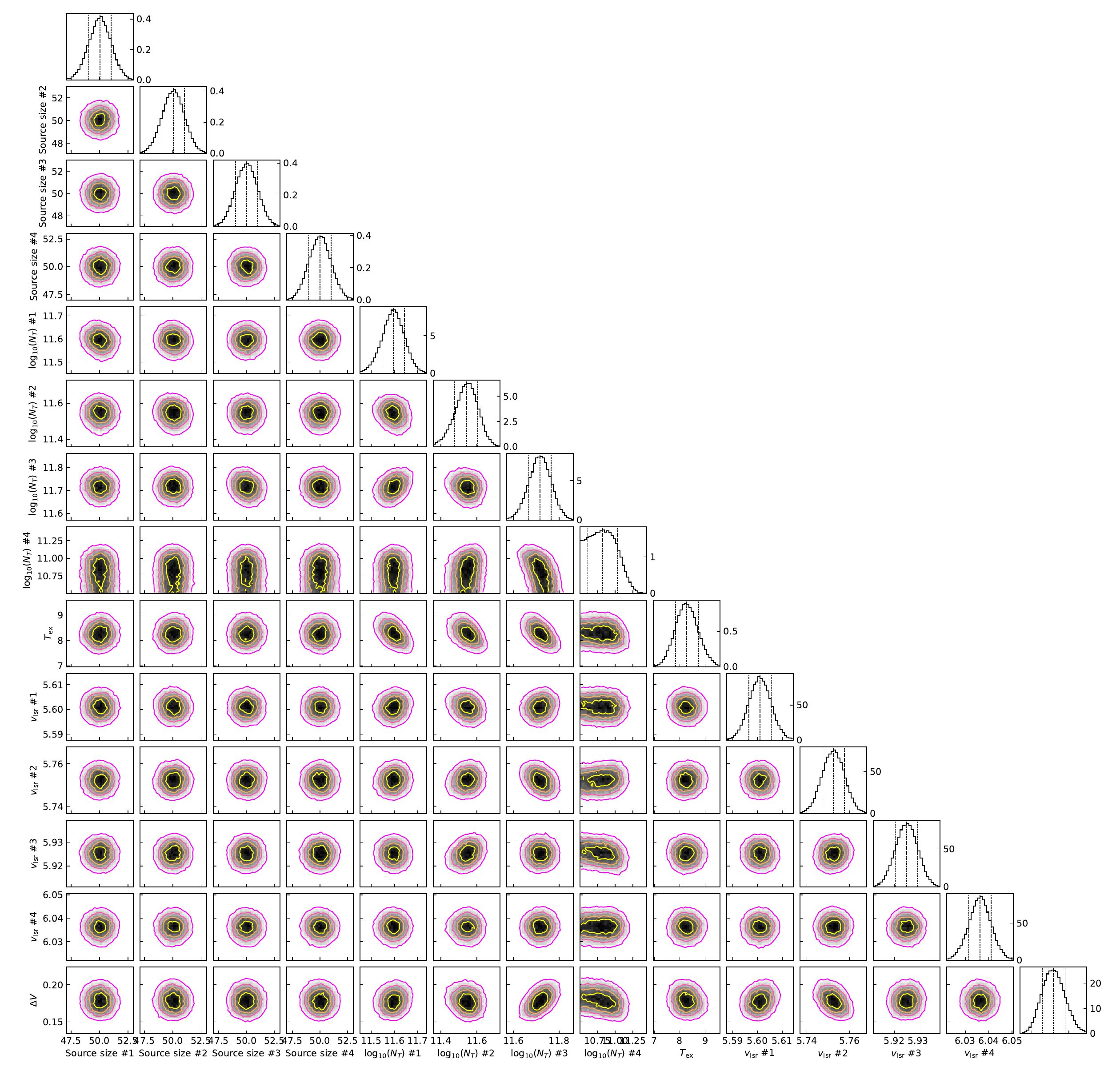}
    \caption{\textbf{Same as Supplementary Fig.~\ref{fig:corner2} for 4-cyanopyrene.}}
    \label{fig:corner4}
\end{figure}

\clearpage

\subsection{Partition functions for 2- and 4-cyanopyrene}

\begin{table}[htb!]
    \centering
    \caption{Partition functions for 2- and 4-cyanopyrene as determined by SPCAT and used in the MCMC analyses. The difference in partition function between the two isomers is due to the $C_{2v}$ symmetry of 2-cyanopyrene, whose rotational lines are spin-weighted at a relative intensity of 17:15 between \textit{ortho} and \textit{para} states.}
    \begin{tabular}{S[table-format=3.3]S[table-format=3.5]S[table-format=3.4]S[table-format=3.4]}
    \hline
     & & \multicolumn{2}{c}{Partition function} \\
    {Temperature [K]} & & {2-cyanopyrene} & {4-cyanopyrene} \\
        
    \hline
       & \\
1.0  &&   29554.8507 & 1806.3918 \\
1.5  &&   54226.4137 & 3313.8116 \\
2.0  &&   83433.7777 & 5098.3047 \\
2.5  &&  116557.6617 & 7122.0481 \\
3.0  &&  153179.9968 & 9359.5064 \\
3.5  &&  192993.7673 & 11791.9267 \\
4.0  &&  235761.0430 & 14404.7753 \\
4.5  &&  281290.3696 & 17186.3574 \\
5.0  &&  329423.3438 & 20126.9968 \\
5.5  &&  380026.0667 & 23218.5142 \\
6.0  &&  432983.4063 & 26453.8766 \\
6.5  &&  488194.9818 & 29826.9521 \\
7.0  &&  545572.2581 & 33332.3322 \\
7.5  &&  605036.3833 & 36965.2001 \\
8.0  &&  666516.5423 & 40721.2299 \\ 
8.5  &&  729948.6770 & 44596.5083 \\
9.375 && 845479.2656 & 51654.6444 \\
18.75 && 2390908.6578 & 146069.1269 \\
37.5  && 6758452.5511 & 412535.9667 \\
75.0  && 18683690.3920 & 1123913.3725 \\
150.0 && 45353238.5366 & 2584507.7670 \\
225.0 && 68471605.0991 & 3708796.1878 \\
275.0 && 81317164.7213 & 4283442.5837\\
300.0 && 87051710.1965 & 4529584.0918 \\
400.0 && 106225362.2531 & 5310351.5683 \\
500.0 && 120801263.2841 & 5864688.7714 \\        
        \hline
    \end{tabular}
    \label{tab:partfunc}
\end{table}

\clearpage
\subsection{DFT computations}
\label{sec:MESMER}

\begin{table}[!h]
    \centering
        \caption{Site-specific bimolecular rate coefficients and branching ratios (B.R.) for the addition of CN to pyrene leading to cyanopyrene (CNP) and H atoms, at the EP3//wB97XD4/def2-TZVPP level including error calculated by propagating a 10 kJ mol$^{-1}$ change in the barrier heights ($\pm\Delta$10).}
    \begin{tabular}{ccccccc}
    \toprule
         &  \multicolumn{2}{c}{EP3//$\omega$B97X-D4/def2-TZVPP }&  \multicolumn{2}{c}{$\omega$B97X-D4/def2-TZVPP}\\
         \midrule
         &  \textit{k} ($+ \Delta$10, $- \Delta$10) &  B.R. ($+ \Delta$10, $- \Delta$10) & \textit{k}  & B.R.\\
         & [$10^{-10}\,\mathrm{cm^3\,s^{-1}}$] & & [$10^{-10}\,\mathrm{cm^3\,s^{-1}}$] &\\
         \midrule
         1-CNP&  2.02 (0.50, 2.02)&  0.40 (0.32, 0.40)&  2.07& 0.40\\
         2-CNP&  1.01 (0.34, 1.01)&  0.20 (0.24, 0.20)&  1.03& 0.20\\
 4-CNP& 2.02 (0.63, 2.02) & 0.20 (0.44, 0.40)& 2.06&0.40\\
 \bottomrule
    \end{tabular}
    \label{tab:MESMER}
\end{table}

Here, the site-specific rate coefficients refer to the loss of CN via addition to pyrene that results in the prompt formation of cyanopyrenes and H atoms. The formation of thermalized 1- and 4-cyanopyrene adducts at 10\,K is not predicted to lead to the formation of their respective cyanopyrenes without additional reactions. Where the EP3//DFT-2 barrier heights were used in the MESMER calculations, no formation of thermalised adducts was predicted. This was also the case where the barriers to H atom elimination were lowered by 10\,kJ\,mol$^{-1}$. Where the barriers were raised by 10\,kJ\,mol$^{-1}$, formation of thermalised 1 and 4-cyanopyrene adducts became the dominant loss process for addition at the 1 and 4 positions ( $k_{1add} = 1.52 \times 10^{-10}\,\mathrm{cm^3\,s^{-1}}$,
$k_{4add} = 1.39 \times 10^{-10}\,\mathrm{cm^3\,s^{-1}}$). In contrast, H atom elimination was still the only pathway predicted following addition at the 2 position, albeit with a removal rate coefficient an order of magnitude below the collision limit.



\begin{figure}[htb!]
    \centering
    \includegraphics[width=1\linewidth]{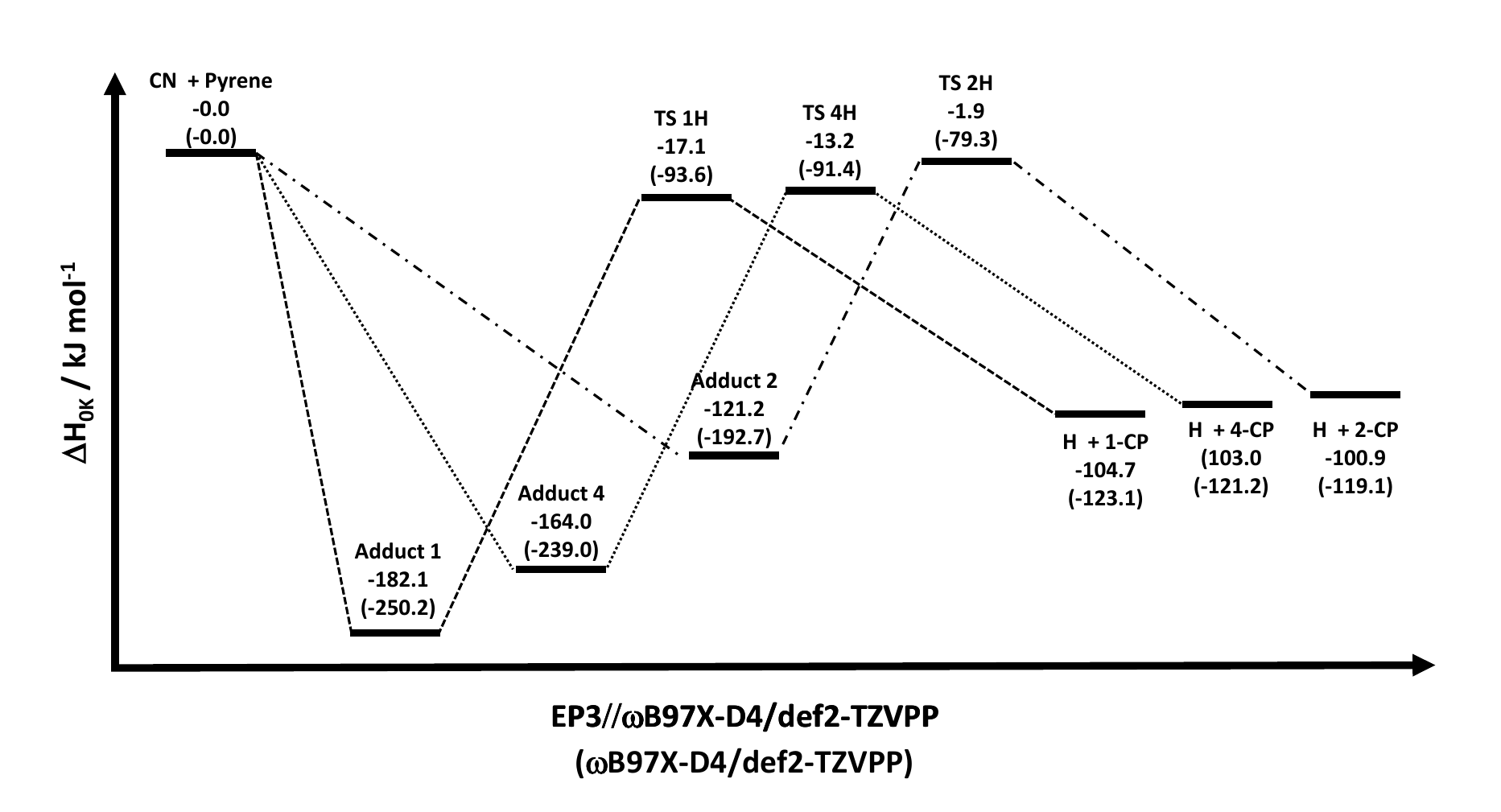}
    \caption{\textbf{Potential energy surface for the addition of CN to pyrene at the EP3//$\omega$B97X-D4/def2-TZVPP level with values prior to EP3 energy corrections given in parenthesis}}
    \label{fig:PES}
\end{figure}

\begin{table}[htb!]
    \caption{Energetics for the CN addition to Pyrene at the  EP3//$\omega$B97X-D4/def2-TZVPP level including scaled harmonic zero point energies (sZPE).}
    \centering
    \begin{tabular}{lccccc}
    \toprule
         &  DFT-2&  sZPE&  $\Delta$H$_{0 K}$ &  EP3//& $\Delta$H$_{0 K}$ \\
         & (Ha) & (Ha) & (kJ mol$^{-1}$) & (Ha) & (kJ mol$^{-1}$)\\
    \midrule
         CN + pyrene&  -709.04756859 &  0.20531153 &  0 & -707.51552912 & 0\\
         Adduct - 1 &  -709.14549706 &  0.20795853 &  -250.2&  -707.58753511
& -182.102
\\
         Adduct - 2&  -709.12292909 &  0.20727361 &  -192.7&  -707.56364325
& -121.2\\
         Adduct - 4&  -709.14131632 &  0.20802809 &  -239.0&  -707.58071359
& -164.0\\
         TS e1&  -709.07873496 &  0.20081908 &  -93.6&  -707.51756636
& -17.1\\
         TS e2&  -709.07323909 &  0.20078966 &  -79.3&  -707.51172060
& -1.9\\
         TS e4&  -709.07796105 &  0.20087818 &  -91.4&  -707.51613469
& -13.2\\
         H + 1-cyanopyrene &  -709.08858469 &  0.19944984 &  -123.1&  -707.54954531 & -104.7\\
         H + 2-cyanopyrene &  -709.08692346 &  0.19930482 &  -119.1&  -707.54795048 & -100.9\\
        H + 4 cyanopyrene & -709.08783089 & 0.19942930 & -121.2& -707.54887256 &-103.0\\
        \bottomrule
    \end{tabular}
    \label{tab:my_label}
\end{table}

\clearpage


\subsection{Updated Rate Coefficients}

\begin{table}[ht!]
    \centering
        \caption{Updated ion-neutral rate coefficients with benzonitrile and 2-cyanoindene derived using the Su-Chesnavich collision theory for \ce{H3+}, \ce{H+}, \ce{He+}, \ce{C+}, and \ce{HCO+}.}
    \begin{tabular}{l c c c c}
        \toprule
         &  \multicolumn{2}{c}{Benzonitrile}& \multicolumn{2}{c}{Cyanoindene}\\
         \midrule
         &   $\beta$ &  $\gamma$ & $\beta$ & $\gamma$ \\
         \hspace{0.1em}\vspace{-0.5em}\\
         \ce{H3+}
&  4.748$\times$10$^{-9}$&  4.53& 5.558 $\times$10$^{-9}$ &3.882\\
         \ce{H+}
&  8.146 $\times$10$^{-9}$ &  4.53& 9.604 $\times$10$^{-9}$ &3.882\\
         \ce{C+}
&  2.473 $\times$10$^{-9}$ &  4.53& 2.791 $\times$10$^{-9}$ &3.882\\
         \ce{He+} &  4.131 $\times$10$^{-9}$&  4.53& 4.818 $\times$10$^{-9}$ & 3.882\\
 \ce{HCO+} & 1.704 $\times$10$^{-9}$ & 4.53& 1.796 $\times$10$^{-9}$ &3.882\\
 \bottomrule
    \end{tabular}
    \label{tab:ionrxns}
\end{table}



\subsection{Synthesis}
\label{sec:synthesis}

\begin{figure}[htb]
    \centering
    \includegraphics[width=0.7\textwidth]{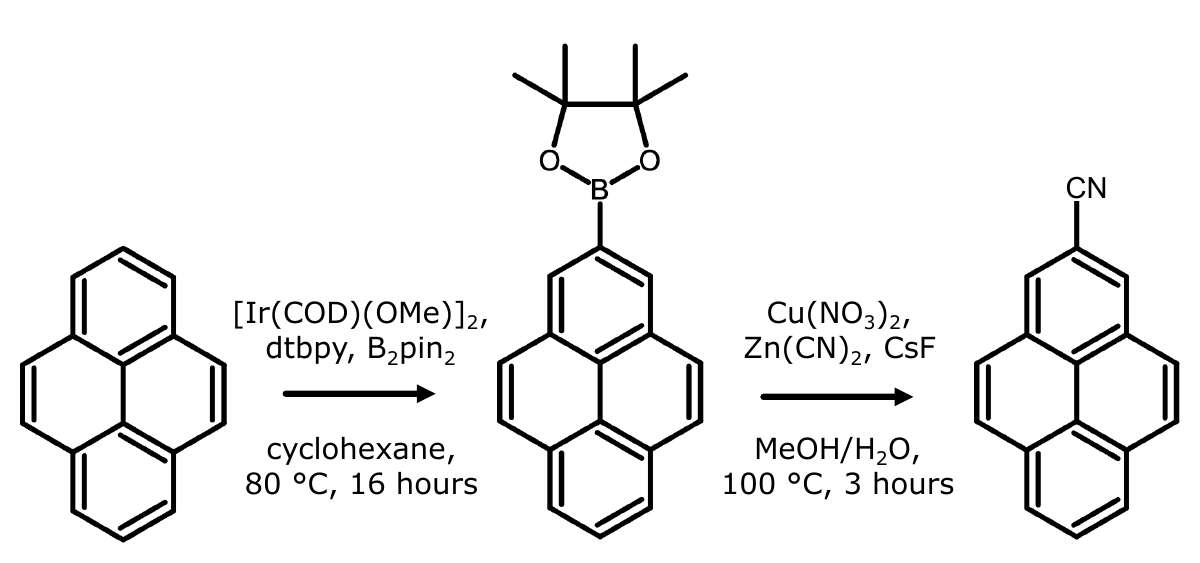}
    \caption{\textbf{Synthesis of 2-(Bpin)pyrene and 2-cyanopyrene}. Using pyrene as a starting material, first 2-(Bpin)pyrene was synthesized and further used to synthesize 2-cyanopyrene.}
    \label{fig:synthesis_2-cyanopyrene}
\end{figure}

\noindent \textbf{4,4,5,5-Tetramethyl-2-pyren-2-yl-[1,3,2]dioxaborolane (2-(Bpin)pyrene)} In a nitrogen-filled purge box, pyrene (5.0\,g, 24.74\,mmol), \ce{[Ir(COD)(OMe)]2} (164\,mg, 0.25\,mmol), 4,4'-di-\textit{tert}-butyl-2,2'-dipyridyl (dtbpy, 133\,mg, 0.50\,mmol) and bis(pinacolato)diboron (6.28\,g, 24.74\,mmol) were added to a 120\,mL pressure tube, cyclohexane (50\,mL) was then added and the tube was sealed. The reaction mixture was stirred at 80\,$^\circ$C for 16\,hours. Upon completion, the mixture was concentrated under vacuum, the residue was purified by column chromatography on silica gel (hexane/\ce{CH2Cl2} = 3/1) to give the 2-(Bpin)pyrene as a colorless solid (3.7\,g, 46\,\%). $^{1}$H NMR (400\,MHz, \ce{CDCl3}) $\delta$ 8.67 (s, 2H), 8.17 (d, $J$ = 7.6\,Hz, 2H), 8.12 (d, $J$ = 9.0\,Hz, 2H), 8.06 (d, $J$ = 9.0\,Hz, 2H), 8.02 (dd, $J$ = 8.1, 7.2\,Hz, 1H), 1.48 (s, 12H); $^{13}$C NMR (101\,MHz, \ce{CDCl3}) $\delta$ 131.77, 131.47, 130.54, 127.89, 127.39, 126.52, 126.46, 124.95, 124.74, 84.29, 25.15 ($^{13}$C-B not observed). The characterization of 2-(Bpin)pyrene is in agreement with the literature.~\cite{wang2021,dordevic2020,biswas2024}\\

\noindent\textbf{2-cyanopyrene} In a round-bottom flask fitted with a reflux condenser, 2-(Bpin)pyrene (20\,g, 60.9\,mmol) was dissolved in hot methanol (400\,mL), to which zinc cyanide (25\,g, 213.2\,mmol) and cesium fluoride (14\,g, 91.4\,mmol) were added. The suspension was heated to reflux with stirring in an oil bath, a solution of copper nitrate hydrate (30\,g, 121.9\,mmol) in water (160\,mL) was added dropwise. The temperature was then increased to 100\,$^\circ$C. The reaction mixture was stirred for 3\,hours. After cooling to room temperature, 1M NaOH aq. was added to adjust the pH to 10.0, and the suspension was extracted with \ce{CH2Cl2} (3 $\times$ 500\,mL). The organic layer was dried over \ce{Na2SO4}, filtered, and concentrated under vacuum, the residue was purified by column chromatography on silica gel (hexane/\ce{CH2Cl2} = 2/1) and then recrystallized from ethyl acetate to give the 2-cyanopyrene as colorless needles (7.6\,g, 55\,\%). $^{1}$H NMR (500\,MHz, \ce{CDCl3}) $\delta$ 8.40 (s, 2H), 8.27 (d, $J$ = 7.6\,Hz, 2H), 8.18 (d, $J$ = 8.9\,Hz, 2H), 8.12 (t, $J$ = 7.6\,Hz, 1H), 8.07 (d, $J$ = 8.9\,Hz, 2H); $^{13}$C NMR (126\,MHz, \ce{CDCl3}) $\delta$ 131.72,  131.43, 129.40,  127.70, 127.55, 126.72, 126.35, 126.35 (overlapping), 123.98, 119.81, 109.29. The characterization of 2-cyanopyrene is in agreement with the literature~\cite{ji2015}. \\

\begin{figure}[htb]
    \centering
    \includegraphics[width=0.7\textwidth]{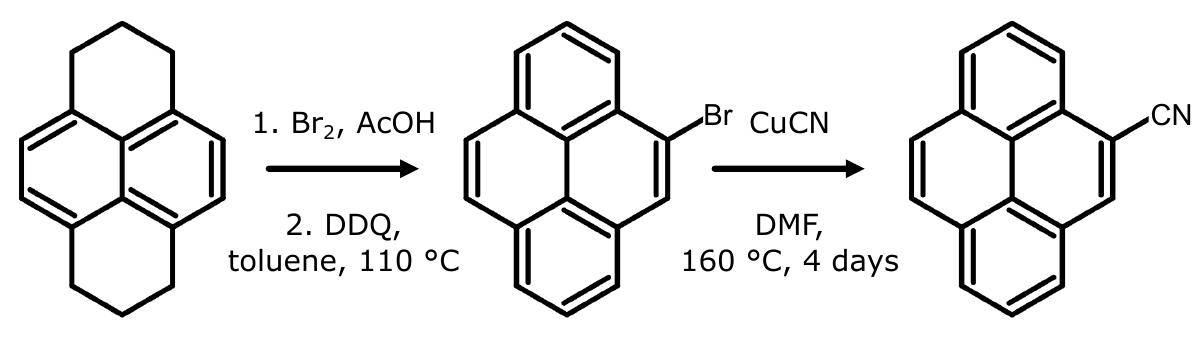}
    \caption{\textbf{Synthesis of 4-bromopyrene and 4-cyanopyrene}. Using hexahydropyrene as a starting material, first 4-bromopyrene was synthesized and further used to synthesize 4-cyanopyrene.}
    \label{fig:synthesis_4-cyanopyrene}
\end{figure}

\noindent\textbf{4-bromopyrene} Bromine (2.5\,mL, 48\,mmol) was mixed with 100\,mL of glacial acetic acid and the solution was added to an addition funnel, the above solution was added to a mixture of 1,2,3,6,7,8-hexahydropyrene (10\,g, 48\,mmol) in glacial acetic acid (200\,mL) dropwise. After 4\,hours, the addition was completed, the mixture was stirred for 12\,hours. The precipitated solid was filtered and washed with water (30\,mL). The resultant  4-bromo-1,2,3,6,7,8-hexahydropyrene was used without further purification.

In a 1000\,mL round-bottom flask, 4-bromo-1,2,3,6,7,8-hexahydropyrene (10\,g, 35\,mmol) and 2,3-dicyano-5,6-dichlorobenzoquinone (DDQ, 40\,g, 176\,mmol) were dissolved in toluene (300\,mL). The mixture was heated to 110\,$^\circ$C and stirred overnight in an oil bath. The mixture was cooled to room temperature and filtered, the filtrate was concentrated under vacuum and purified by column chromatography on silica gel (hexane, 100\,\%) to give 4-bromopyrene as a light-yellow solid (6.7\,g, 69\,\%).
$^1$H NMR (400\,MHz, \ce{CDCl3}) $\delta$ 8.60 (d, $J$ = 8.0\,Hz, 1H), 8.45 (s, 1H), 8.24 (dd, $J$ = 13.2, 7.7\,Hz, 2H), 8.15 – 8.07 (m, 4H), 8.02 (t, $J$ = 7.6\,Hz, 1H); $^{13}$C NMR (101\,MHz, \ce{CDCl3}) $\delta$ 131.48, 131.35, 131.00, 129.88, 127.60, 127.54, 126.62, 126.55, 126.39, 125.77, 125.56, 125.23, 125.09, 124.71, 123.92, 122.64. The characterization of 4-bromopyrene is in agreement with the literature~\cite{lu2020}.\\

\noindent\textbf{4-cyanopyrene} To a 150\,mL pressure tube, 4-bromopyrene (3.5\,g, 12.5\,mmol), copper(I) cyanide (2.24\,g, 25.0\,mmol) and anhydrous DMF (40\,mL) were added. The mixture was heated to 160 $^\circ$C in an oil bath and stirred for 3 days. Additional copper(I) cyanide (0.56\,g, 6.3\,mmol) was added, the mixture was stirred for another day. The resulting mixture was cooled to room temperature, diluted with water (150\,mL) and extracted with \ce{CH2Cl2} (3 $\times$ 300\,mL), the organic layer was dried over \ce{Na2SO4}, filtered, and concentrated under vacuum. The residue was purified by column chromatography on silica gel (hexane/\ce{CH2Cl2} = 2/1) and then recrystallized from ethyl acetate to obtain 4-cyanopyrene as pale green crystals (3.4\,g, 60\,\%).
$^{1}$H NMR (500\,MHz, \ce{CDCl3}) $\delta$ 8.60 – 8.53 (m, 2H), 8.33 (d, $J$ = 7.6\,Hz, 1H), 8.30 (d, $J$ = 7.7\,Hz, 1H), 8.26 (d, $J$ = 7.6\,Hz, 1H), 8.19 – 8.05 (m, 4H); $^{13}$C NMR (126\,MHz, \ce{CDCl3}) $\delta$ 136.32, 131.46, 131.24, 128.93, 128.40, 128.14, 127.96, 127.66, 127.14, 126.89, 126.84, 126.65, 125.49, 124.34, 123.22, 118.19, 110.25.
The characterization of 4-cyanopyrene is in agreement with the literature~\cite{bao-xi2019}.



\clearpage
\begin{figure}[htb!]
    \centering
    \includegraphics[width=1\linewidth]{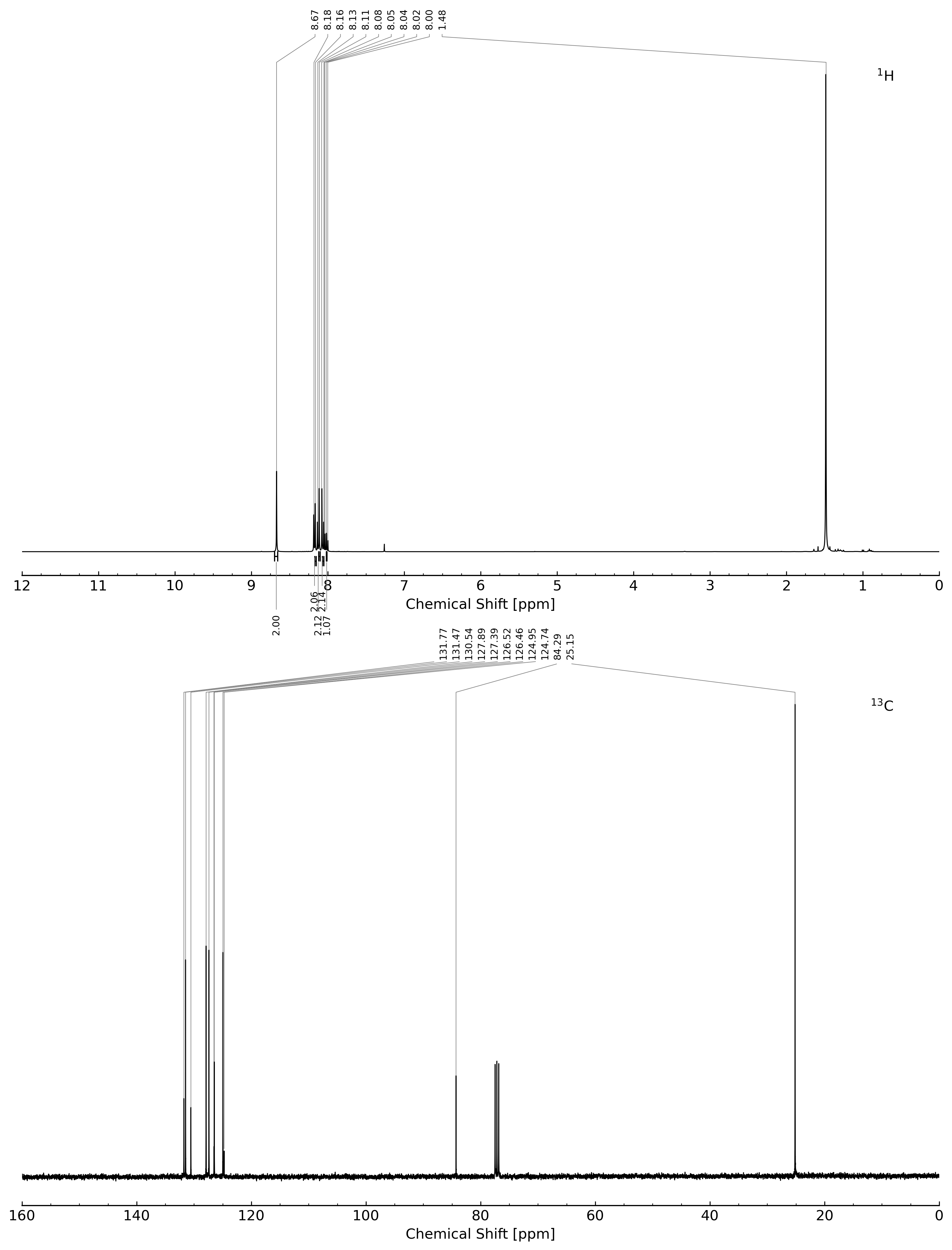}
    \caption{\textbf{Nuclear magnetic resonance (NMR) spectra of 2-(Bpin)pyrene.} $^{1}$H NMR (top panel) and $^{13}$C NMR (bottom panel) of 2-(Bpin)pyrene.}
    \label{fig:NMR_2-bpinpyrene}
\end{figure}

\clearpage
\begin{figure}[htb!]
    \centering
    \includegraphics[width=1\linewidth]{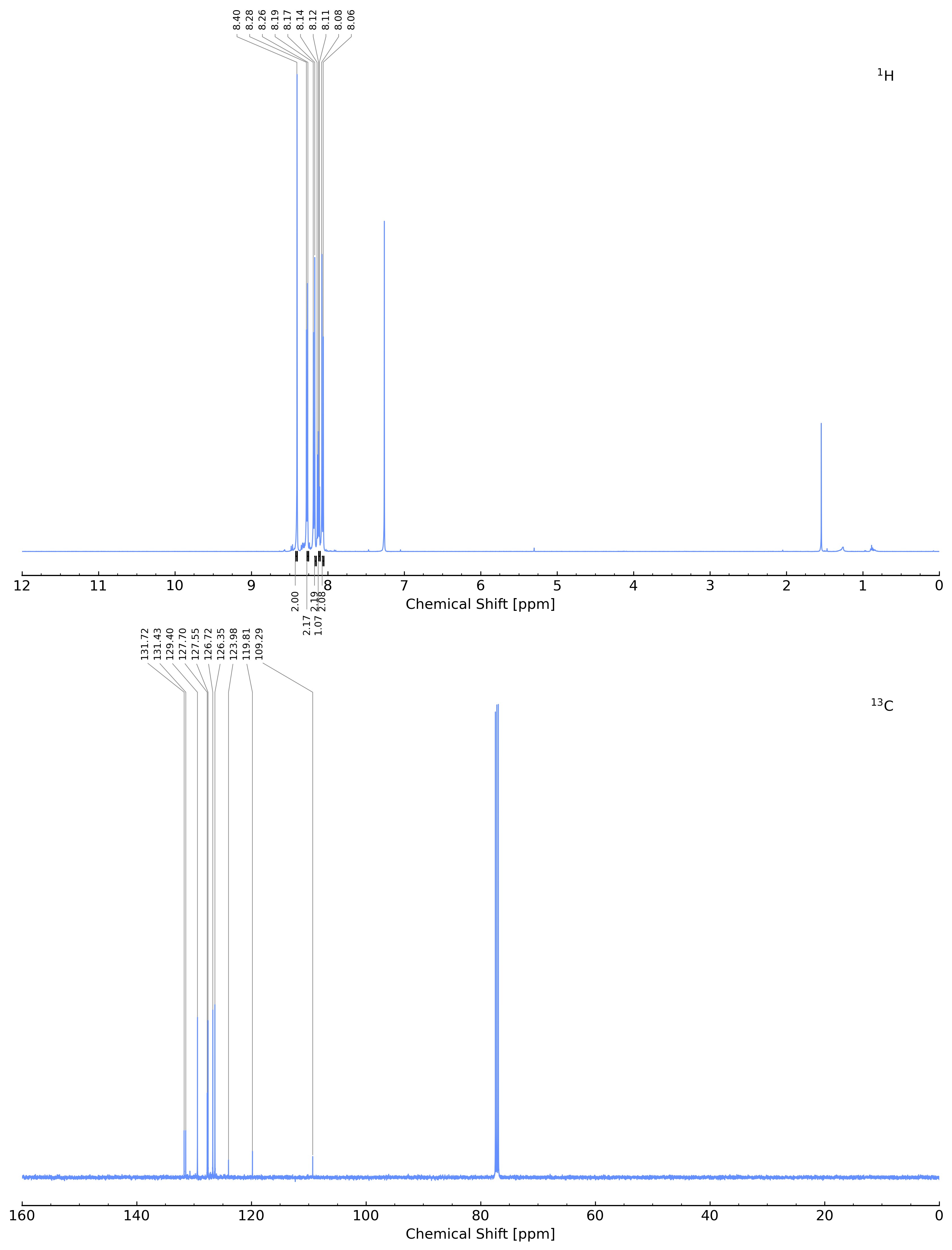}
    \caption{\textbf{Nuclear magnetic resonance (NMR) spectra of 2-cyanopyrene.} $^{1}$H NMR (top panel) and $^{13}$C NMR (bottom panel) of 2-cyanopyrene.}
    \label{fig:NMR_2-cyanopyrene}
\end{figure}

\clearpage
\begin{figure}[htb!]
    \centering
    \includegraphics[width=1\linewidth]{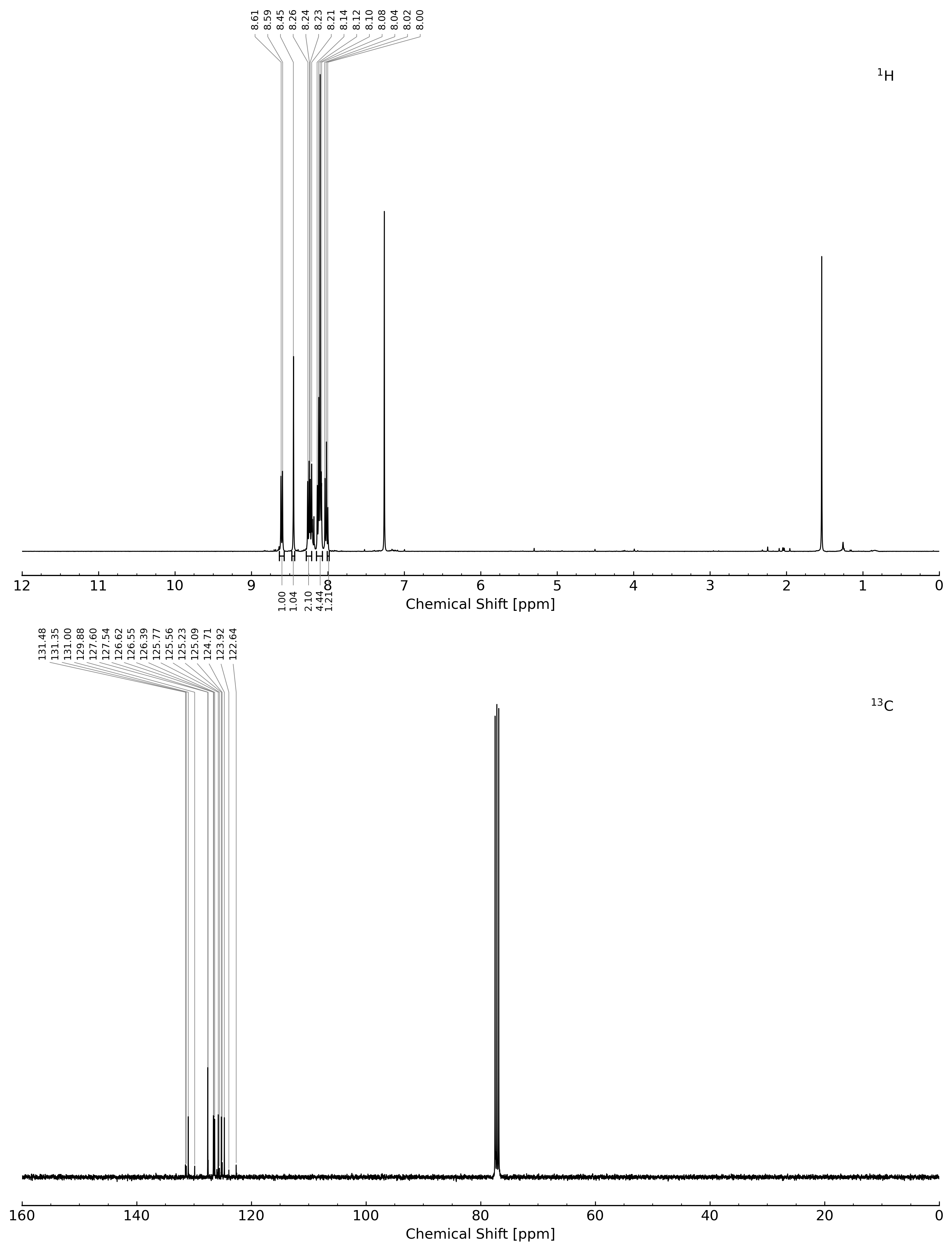}
    \caption{\textbf{Nuclear magnetic resonance (NMR) spectra of 4-bromopyrene.} $^{1}$H NMR (top panel) and $^{13}$C NMR (bottom panel) of 4-bromopyrene.}
    \label{fig:NMR_4-bromopyrene}
\end{figure}

\clearpage
\begin{figure}[htb!]
    \centering
    \includegraphics[width=1\linewidth]{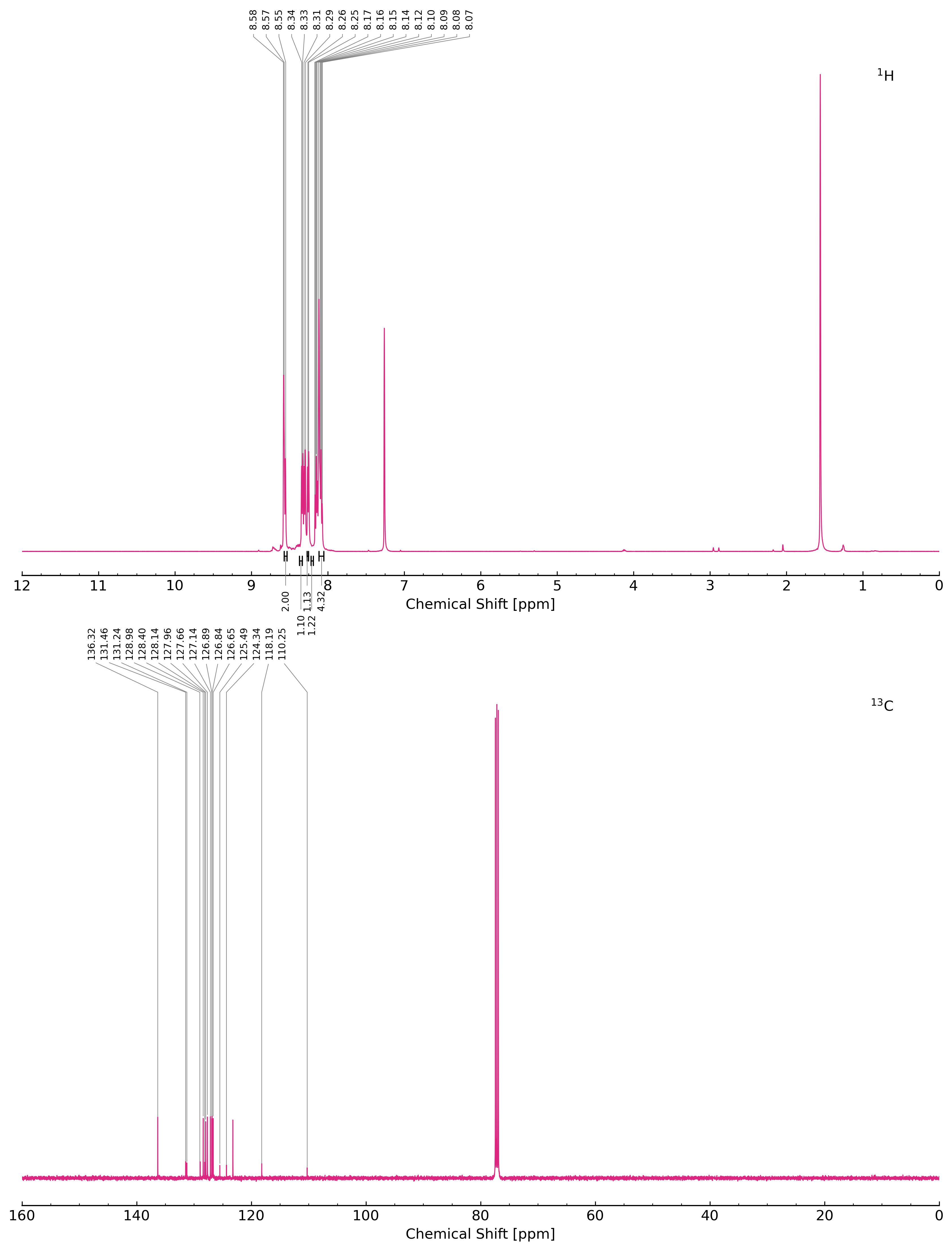}
    \caption{\textbf{Nuclear magnetic resonance (NMR) spectra of 4-cyanopyrene.} $^{1}$H NMR (top panel) and $^{13}$C NMR (bottom panel) of 4-cyanopyrene.}
    \label{fig:NMR_4-cyanopyrene}
\end{figure}




\end{appendices}


\end{document}